\documentclass[preprint,12pt,3p]{elsarticle}

\usepackage{amssymb}
\usepackage{amsthm}
\usepackage{amsmath}
\usepackage{aas_macros}
\usepackage{rotating}
\usepackage{url}
\usepackage{morefloats} 
\usepackage{float}
\usepackage{threeparttable}
\usepackage{rotating}
\usepackage{longtable}
\usepackage{array}
\usepackage{color}
\usepackage{graphicx}
\usepackage[utf8]{inputenc}
\usepackage[export]{adjustbox}
\usepackage{wrapfig}
\usepackage{lineno}
\usepackage{natbib}
\usepackage[table]{xcolor}

\definecolor{mygray}{gray}{0.9}

\journal{Icarus}

\newcommand{\fig}[1]{{fig.\,\ref{#1}}}
\newcommand{\tab}[1]{{table\,\ref{#1}}}
\newcommand{\eqn}[1]{{eq.\,\ref{#1}}}

\newcommand{\Hawaii}{{Hawai`i}}

\newcommand{\eg}{{\it e.g.}}
\newcommand{\ie}{{\it i.e.}}

\newcommand{\insitu}{{\it in situ}}

\newcommand{\dif}{\mathrm{d}}

\newcommand{\boldm}[1] {\mathversion{bold}#1\mathversion{normal}}

\newcommand{\xbar}{\bar x}

\newcommand{\qei}{(q,e,i)}

\newcommand{\digesttwo}{{\texttt{digest2}}}

\newcommand{\HtwoO}{{H$_2$O}}
\newcommand{\CO}{{CO}}
\newcommand{\COtwo}{{CO$_2$}}
\newcommand{\Otwo}{{O$_2$}}

\newcommand{\arcdeg}{{^{\circ}}}

\newcommand{\arcsec}{^{\prime\prime}}

\newcommand{\au}{\,\mathrm{au}}
\newcommand{\km}{\,\mathrm{km}}

\newcommand{\meter}{\,\mathrm{m}}

\newcommand{\um}{\,\mu \mathrm{m}}
\newcommand{\nm}{\,\mathrm{nm}}

\newcommand{\yr}{\,\mathrm{yr}}
\newcommand{\yrs}{\,\mathrm{yrs}}

\newcommand{\Day}{\,\mathrm{day}}

\newcommand{\mags}{\,\mathrm{mag}}

\newcommand{\kg}{\,\mathrm{kg}}

\newcommand{\Mearth}{\,\mathrm{M}_\oplus}

\newcommand{\Uone}{{1I/2017 U1 (‘Oumuamua)}}

\newcommand{\ips}{\ensuremath{i_{\rm P1}}}

\newcommand{\wps}{\ensuremath{w_{\rm P1}}}

\newcommand{\PS}{\protect \hbox {Pan-STARRS}}
\newcommand{\PSone}{\protect \hbox {Pan-STARRS1}}

\begin{document}

\begin{frontmatter}

\title{The Orbit and Size-Frequency Distribution of Long Period Comets\\ Observed by \PSone}

\author[label1]{Benjamin~Boe}
\author[label1]{Robert~Jedicke}
\ead{jedicke@hawaii.edu}
\author[label1]{Karen~J.~Meech}
\author[label2]{Paul~Wiegert}
\author[label1]{Robert~J.~Weryk}

\author[label1]{K.~C.~Chambers} 
\author[label1]{L.~Denneau} 
\author[label4]{N.~Kaiser} 
\author[label1,label3]{R.-P.~Kudritzki} 
\author[label1]{E.~A.~Magnier} 
\author[label1]{R.~J.~Wainscoat} 
\author[label1]{C.~Waters}





\address[label1]{Institute for Astronomy, University of Hawai`i, 2680 Woodlawn Drive, Honolulu, HI 96822, USA}
\address[label2]{The University of Western Ontario, London, Ontario, Canada}
\address[label3]{Munich University Observatory, Munich, Germany}
\address[label4]{{\'E}cole Normale Sup{\'e}rieure, Paris, France.}


\begin{abstract}
We introduce a new technique to estimate the comet nuclear size frequency distribution (SFD) that combines a cometary activity model with a survey simulation and apply it to 150 long period comets (LPC) detected by the \PSone\ near-Earth object survey.  The debiased LPC size-frequency distribution is in agreement with previous estimates for large comets with nuclear diameter $\gtrsim 1\km$ but we measure a significant drop in the SFD slope for small objects with diameters $<1\km$ and approaching only $100\meter$ diameter.  Large objects have a slope $\alpha_{big} = 0.72 \pm 0.09 (stat.) \pm 0.15 (sys.)$ while small objects behave as $\alpha_{small} = 0.07 \pm 0.03 (stat.) \pm 0.09 (sys.)$ where the SFD is $\propto 10^{\alpha H_N}$ and $H_N$ represents the cometary nuclear absolute magnitude.  The total number of LPCs that are $>1\km$ diameter and have perihelia $q<10\au$ is $0.46 \pm 0.15 \times 10^9$ while there are only $2.4 \pm 0.5 (stat.) \pm 2 (sys.) \times 10^9$ objects with diameters $>100\meter$ due to the shallow slope of the SFD for diameters $<1\km$.  We estimate that the total number of `potentially active' objects with diameters $\ge 1\km$ in the Oort cloud, objects that would be defined as LPCs if their perihelia evolved to $<10\au$, is $(1.5\pm1)\times10^{12}$ with a combined mass of $1.3\pm0.9\Mearth$.  The debiased LPC orbit distribution is broadly in agreement with expectations from contemporary dynamical models but there are discrepancies that could point towards a future ability to disentangle the relative importance of stellar perturbations and galactic tides in producing the LPC population.
\end{abstract}


\end{frontmatter}


\section{Introduction}
\label{s.Introduction}

\citet{Oort1950} first proposed the existence of a comet cloud around the Solar System based on the observation of an excess of 19 long-period comets (LPC) on orbits with large semimajor axes ($a> 20,000\au$).  He argued that they ``belong very definitely to the solar system'' and couldn't be captured from an interstellar population but did not propose  a formation mechanism for the Oort cloud except to hypothesize it was material lost from the asteroid belt or possibly from a disrupted planet \citep[see note 2 in][]{Oort1950}.  We now understand that LPCs were injected into the Oort cloud through a two-step process. First, icy planetesimals were scattered from the inner proto-solar nebula onto high-semimajor axis, high-eccentricity orbits by Uranus and Neptune \citep{Duncan1987}. Second, their perihelion distances were raised through the effects of distant encounters with passing stars or torques exerted by the Milky Way's tidal field \citep{Heisler1986} so that most of them would never return to near-Sun space. Subsequent torques by the Galactic tide or passing stars can reduce some of their perihelia and return the planetesimals to the inner Solar System as LPCs billions of years later. As a result, LPCs are tracers of both of the dynamical evolution occurring during planet formation and the influences acting at the extreme edge of our solar system. In this work we measure the size and orbit distribution of LPCs detected by \PSone\ in order to illuminate these processes.

The size distribution of the objects in any population provides insight into their formation and evolutionary processes. Comets have been proposed to be either primordial rubble piles \citep{Davidsson2016,Weidenschilling1997,Weissman1986} or collisional fragments from larger parent bodies, and may have grown incrementally and slowly by accumulating smaller objects \citep[\eg][]{Jutzi2015-Science-CometaryNuclei,Hartmann1968} or been rapidly ``born big'' through \eg\ streaming instabilities \citep[\eg][]{Blum2017,Johansen2007-Nature,Morbidelli2009}. Spectacular high-resolution ALMA sub-mm images of exo-planetary systems provide evidence of the growth of dust into pebbles and planets \citep{Zhang2015} but cannot yet reveal the details of the formation process of the planetesimals that form the building blocks of planetary construction. The size-frequency distribution (SFD) of the small body populations in our solar system can bridge this gap between our understanding of dust and planets in exoplanetary systems.  A ``simple'' power-law SFD could imply that the population reached a stable self-similar collisional cascade \citep[\eg][]{OBrien2003,Dohnanyi1969} whereas ``breaks'' or ``waviness'' in the distribution, as observed for main belt asteroids \citep[\eg][]{Ivezic2001-SDSS-MainBelt,Jedicke1998}, have implications for the internal strength of the objects and their formation mechanisms \citep[\eg][]{Bottke2005,OBrien2005,Durda1998-MB-SFD}.

Attempts to directly measure the sizes of the nuclei of members of the LPC population have been frustrated by their comae. Dynamically new comets entering the inner solar system for the first time (traditionally considered to be those with $a>10,000\au$ but \citet{Krolikowska2017} suggest that they must have $a>20,000\au$) are typically bright even when they are far from the Sun because of significant coma activity driven by sublimation of \CO\ or \COtwo.  Historically, reports of nuclear sizes referred to ``the part of the comet that is star-like in appearance'' \citep{Roemer1967} or the ``central condensation''.  This was effectively a measure of the combined scattered light from the nucleus plus the steep core of the coma surface brightness distribution. In a major campaign to photographically measure cometary properties using the U.S. Naval Observatory in Flagstaff, \citet{Roemer1968} concluded that nearly parabolic LPCs had a flatter size distribution than short-period comets extending out to large diameters.  While this was state-of-the-art at the time, there was almost certainly a contribution from the coma in these measurements.  Indeed, \citet{Hui2018} suggest that unless the nuclear signal is $\ge10$\% of the total signal that ``there is probably no way to debias results from this technique''.

With in-situ results from space missions we now know that most known, active cometary nuclei are relatively small, low albedo (few percent) bodies with a mean diameter of $\sim2.8\km$ \citep{Meech2017}. Since sublimation of water-ice can begin to lift optically detectable dust from comet's surface even at $\sim 6\au$ \citep{Meech2004}, and other more volatile ices can create a dust coma throughout much of the solar system, large telescopes or specialized techniques are needed to make direct measurements of the nuclei of LPCs.  For this reason there are perhaps only a dozen or so reliable LPC nucleus size measurements \citep{Meech2017-BeforeRosetta} obtained from radar observations \citep[\eg][]{Harmon1997-HyakutakeRadarDetection}, or where it was possible to model and remove the coma because of high resolution Hubble Space Telescope data \citep{Meech2004}, or space-based thermal-IR measurements \citep{Bauer2017,Bauer2015}.  \citet{Bauer2015} suggested on the basis of measurements from the WISE observatory that the known, active LPCs were on average about twice as large as the short-period comets, but they did not have a large sample and the measurements were not corrected for observational selection effects.

Thus, while every comet discovery is exciting, and each new object provides fresh opportunities for scientific inquiry, the aspect of comets that makes them interesting and beautiful, their time-varying and unpredictable comae and tails, makes it difficult to accurately quantify their orbit and size-frequency distribution (SFD).  Furthermore, while 2,657 LPCs\footnote{The total number of objects on the JPL Small-Body Database Search Engine (\protect\url{https://ssd.jpl.nasa.gov/sbdb_query.cgi}) with either (orbital period $\ge200\yrs$ and $q<10\au$) or (hyperbolic or parabolic) orbital classes.\protect\label{footnote.JPL-LPCs}} are known as of 18 January 2019, they were discovered over the course of centuries with a wide range of instruments and few individual telescopic surveys have ever discovered or detected a large sample.  Thus, techniques that have been developed to `debias' high-statistics samples of asteroids detected by astronomical surveys \citep[\eg][]{Jedicke2016} cannot be readily applied to comets because it has been difficult to quantify the comet detection efficiency.  While an asteroid's apparent brightness is well-characterized as a function of its absolute magnitude ($H$), topocentric ($\Delta$) and heliocentric ($r$) distance, and phase function \citep[\eg][]{Muinonen2010,Bowell1988}, a comet's apparent brightness also depends on the the amount of dust in its coma which is related to the fraction of its surface that is active ($f$). This fraction is a function of heliocentric distance and the types of volatile ices present in the objects \citep[\eg][]{Meech1986}.  Furthermore, while asteroids are routinely and automatically detected by software algorithms whose detection efficiency can be readily measured \citep[\eg][]{Jedicke2016,Bannister2016,Gladman2009}, automated discovery of comets in all their morphological phases has, as yet, not been achieved.  The reliance on humans in the comet discovery chain introduces an aspect of discovery that has been difficult to quantify \citep[\eg][]{Hsieh2015}.

In this work we debias the population of LPCs discovered and detected by the \PSone\ survey to determine the population's true orbit and size-frequency distribution.  We measure the system's comet detection efficiency by simulating the survey's ability to detect a synthetic LPC population.  To do so we have to assume that all the synthetic LPCs behave as `average' comets and invoke the use of a photometric model to calculate their apparent brightness.  We do so with full awareness that every comet is special but, at this time, with the data set at hand, it is necessary and sufficient for our analysis of the LPC orbit and size-frequency distributions.

\section{Method}
\label{s.Method}

Our technique for measuring the LPC SFD is straightforward in principle --- we use the set of 150 LPCs detected by the \PSone\ sky survey combined with our measurement of the system's absolute LPC detection efficiency to correct for selection effects in the detected sample and obtain the true SFD.  

The primary complication, and the new technique that we introduce for this analysis, is our method of determining the LPCs' nuclear absolute magnitudes ($H_N$).  Our starting point is the $H_N'$ provided by JPL's Small-Body Database Search Engine (\url{https://ssd.jpl.nasa.gov/sbdb_query.cgi}) because these values are calculated in a consistent way but from a large number of observations provided by many different observers.  We did not restrict ourselves to \PSone\ data because it typically obtains only a few measurements of individual objects whereas the data available to JPL are all reported values for all objects.  The problem is that JPL's $H_N'$ have systematic errors in almost all the values due to strong contamination from the LPCs' coma.  Even though the error for each LPC is different, because it is activity-dependent, we let the average systematic error (offset) for all 150 LPCs in our sample be represented by $\Delta H_{JPL}$ so that $H_N = H_N' + \Delta H_{JPL}$.

We measure \PSone's LPC detection efficiency ($\epsilon$) using 1) a synthetic LPC population, 2) a thermodynamic sublimation model that calculates the expected brightness of an object from the nuclear albedo, absolute magnitude and a small number of fixed input parameters ($\bar x$), $V(H_N,p_V,\bar x)=V(H_N' + \Delta H_{JPL},p_V,\bar x)$, and 3) a \PSone\ survey simulator.  This process makes {\it our calculated detection efficiency and our debiased population estimates functions of} $\Delta H_{JPL}$!  Thus, and finally, we determine $\Delta H_{JPL}$, and our final debiased SFD and orbit distributions, by minimizing the difference between the actual observed perihelion and time of perihelion distributions and their simulated counterparts.

\subsection{\PSone}
\label{ss.PanSTARRS}

The Panoramic Survey Telescope \& Rapid Response System \citep[\PS;][]{Kaiser2004} was developed by the Institute for Astronomy at the University of \Hawaii\ for the purpose of discovering near-Earth objects \citep[NEO;][]{Schunova-Lilly2017}.  The \PS\ prototype telescope, \PSone, has been surveying the sky since early 2010 and is currently the leading discovery system for both asteroids and comets \citep{Chambers2016}.  The \PSone\ image processing pipeline's \citep[IPP;][]{Magnier2016} source data stream is automatically processed to identify moving objects using their moving object processing system \citep[MOPS;][]{Denneau2013} but the most interesting objects, particularly newly discovered NEOs and comets, are vetted by a human `czar' before submitting the observations to the Minor Planet Center (MPC).

\ \PSone\ surveys the sky using six filters \citep{Tonry2012} but the majority (53.1\%) of field pointings with at least three visits used the wide passband ($\wps$) spanning the range of wavelengths from about $550\nm$ to $920\nm$. During the bright period of a lunation \PSone\ uses its $\ips$ filter with a bandpass from about $820\nm$ to $920\nm$ to limit sky-background from the Moon, accounting for 38.3\% of the three-visit field pointings.   

The typical survey mode that resulted in moving object detections involved acquiring four sequential images (visits), each of them about 20 minutes apart, at the same telescope boresight.  MOPS then required that at least one source in each of 3 or 4 images in the sequence be consistent with detections of the same moving object to create a `tracklet' --- a candidate detection of a moving object.  Known objects are automatically attributed to tracklets \citep{Milani2012} while tracklets that have `unusual' rates of motion, apparent magnitude, or morphology at their observed location on the sky are flagged for individual examination to ensure that they are composed of real detections and check for cometary activity.    

One of the primary features of the \PSone\ MOPS is that it is capable of processing synthetic data using the same software used for processing actual data \citep{Denneau2013}.  This capability allows the determination of the survey detection efficiency for any asteroid or comet population if the synthetic population and simulation are sufficiently representative of reality (\S\ref{ss.SimulationVerisimilitude}).  Thus, we simulated the detection of a synthetic LPC population (\S\ref{sss.SyntheticLPCOrbitDistribution}) by \PSone\ using the system's actual pointing history in 230,591 $\wps$ band exposures and 192,573 $\ips$ band exposures acquired by the system from 23 February 2010 through 22 December 2016. MOPS determines which of the synthetic objects appeared in each field and provides their apparent positions, rate and direction of motion, apparent magnitude \citep[$V_0$, assuming $H_N=0$ and using the now out-dated $H$-$G$ photometric system;][]{Bowell1988}, and the heliocentric and topocentric distance.  These data are then post-processed as described below to determine the LPC detection efficiency as a function of their orbital parameters and size.

\subsection{LPC orbit distribution}
\label{ss.LPCOrbitDistribution}

LPCs are commonly defined as active objects with orbital periods greater than 200 years that enter the planetary region ($q<10\au$) of the solar system \citep[\eg][]{Dones2015}.  We will also use the LPC initialism for `potentially active' objects from the Oort cloud that {\it would} become active if they entered the inner solar system.  To determine \PSone's LPC detection efficiency our analysis requires an LPC orbit distribution for which we resort to using a theoretical model.  In this subsection we describe two contemporary LPC models that we will compare to our resulting debiased distributions.  The second model was used to generate our synthetic input LPC distribution (\S\ref{sss.SyntheticLPCOrbitDistribution}).

\subsubsection{\citet{Fouchard2017b}}
\label{sss.FouchardLPCOrbitDistribution}

\citet{Fouchard2017b} developed their Oort cloud beginning with a population of planetesimals in a scattered disk beyond, but interacting with, Uranus and Neptune.  They tracked the evolution of objects transferred to large orbits, \ie\ the Oort cloud, and then back into the planetary system under the influence of the galactic tide and the giant planets and, in some of their models, passing stars.  Their inclusion of passing stars is the first of two significant features that differ from the model we used as the basis of our efficiency determination (\S\ref{sss.WiegertLPCOrbitDistribution}), the second being that they do not introduce a fading function to account for the reduced activity of comets with each perihelion passage.  \citet{Fouchard2017b} found that their models with and without passing stars produce similar, though not identical, Oort clouds to that of \citet{Wiegert1999} which we adopted for our efficiency analysis.  The \citet{Fouchard2017b} models specifically exclude strong LPC `showers' induced by the passing stars to better mimic the LPC steady state orbital population.

\subsubsection{\citet{Wiegert1999}}
\label{sss.WiegertLPCOrbitDistribution}

We generated a set of synthetic LPCs from the \citet{Wiegert1999} steady state comet population model.  \citet{Wiegert1999} integrated orbits of cometary bodies from the Oort cloud into the solar system including dynamical perturbations from the galactic tide and the four giant planets.  The objects were followed until they collided with the Sun or a giant planet, or were ejected from the solar system on a hyperbolic orbit.  At each time step they recorded the perihelion distance ($q$), eccentricity ($e$), and inclination ($i$) of each object, thereby constructing a `residence time distribution' in $\qei$-space that is identical to the steady state distribution of orbital elements for objects evolving out of the Oort cloud {\it without} any morphological evolution.  

The LPC steady state $\qei$ distribution is derived from the integrations by introducing a simple `fading' law to account for objects that are `lost' due to physical processes such as break up or loss of volatiles during perihelion passage.  Objects contribute to the sample with a weight $n_q^{\beta}$ where $n_q$ is the number of perihelion passes in the inner solar system for a given comet. The model is agnostic to the physical mechanism that causes the loss of LPCs but the fitted value of $\beta=-0.6$ reproduces the observed LPC distribution well while introducing only one free parameter. In effect, the model assumes that LPCs are lost due to thermal heating and/or planetary/tidal effects associated with perihelion passage and that the process is less likely as comets age due, perhaps, to a build up of an insulating dust layer on LPCs that survive many passes through the inner solar system.  The fraction of the final LPC steady state orbital element distribution within $(\dif q, \dif e, \dif i)$ of $(q,e,i)$ is represented by $f(q,e,i) \; \dif q \; \dif e \; \dif i$.

\subsubsection{Synthetic}
\label{sss.SyntheticLPCOrbitDistribution}

We then generated $10^5$ synthetic LPCs according to the \citet{Wiegert1999} $\qei$ model.  Using a multidimensional distribution accounts for the interdependence between the $\qei$ parameters imposed by  gravitational perturbations.  The orbits were generated with perihelia in the range $0\au < q \le 10\au$, eccentricities with $0.4 \le e < 1$, and inclinations satisfying $0\arcdeg \le i \le 180\arcdeg$.   Each orbit's longitude of the ascending node ($\Omega$) and argument of perihelion ($\omega$) was generated in the range $[0\arcdeg,360\arcdeg)$ according to the model's respective one dimensional probability distribution function (PDF) since they did not vary more than $\pm10$\% for different choices of the $\qei$.  

The final cometary orbital element, the time of perihelion ($t_p$), was generated in a manner to reduce the number of synthetic objects that would never be detected because they were too far from perihelion (or, equivalently, with an out-of-range mean anomaly).  Letting $T_j$ represent the $j^{th}$ object's orbital period and $n_j$ its mean angular rate in deg/day, we generate a random mean anomaly for the object ($M_j$) in the range $[-M_{limit,j},+M_{limit,j}]$ where $M_{limit,j} = n_j \times 3,000\Day$,  \ie\ a mean anomaly within 3,000~days of the time of perihelion (about 8.2 years).  The time of perihelion for the object is then
\begin{equation}
t_{p,j} = t_{PS1} - M_j \sqrt{{\frac{a_j^3}{G M_\odot}}},
\end{equation}
where $t_{PS1}$ is the mid-point of the \PSone\ survey considered herein (2013-07-24), $a_j$ is the object's semi-major axis, $G$ is the gravitational constant, and $M_\odot$ is the mass of the Sun.  This procedure results in a uniform distribution of times of perihelion in the range $t_{PS1}\pm$3,000~days.  To correct for the limited mean anomaly range we then weight the $j^{th}$ synthetic LPC by $w_j = 180 / M_{limit,j}$. Since the minimum orbital period is 200~years the minimum weight is $\sim12$ \ie\ a single 200~year period comet in our generated model represents $\sim12$ LPCs in the population model because our 6,000~day window is about 1/12$^{th}$ of the orbital period.

\begin{figure}[htbp]

	\includegraphics[width=0.5\textwidth]{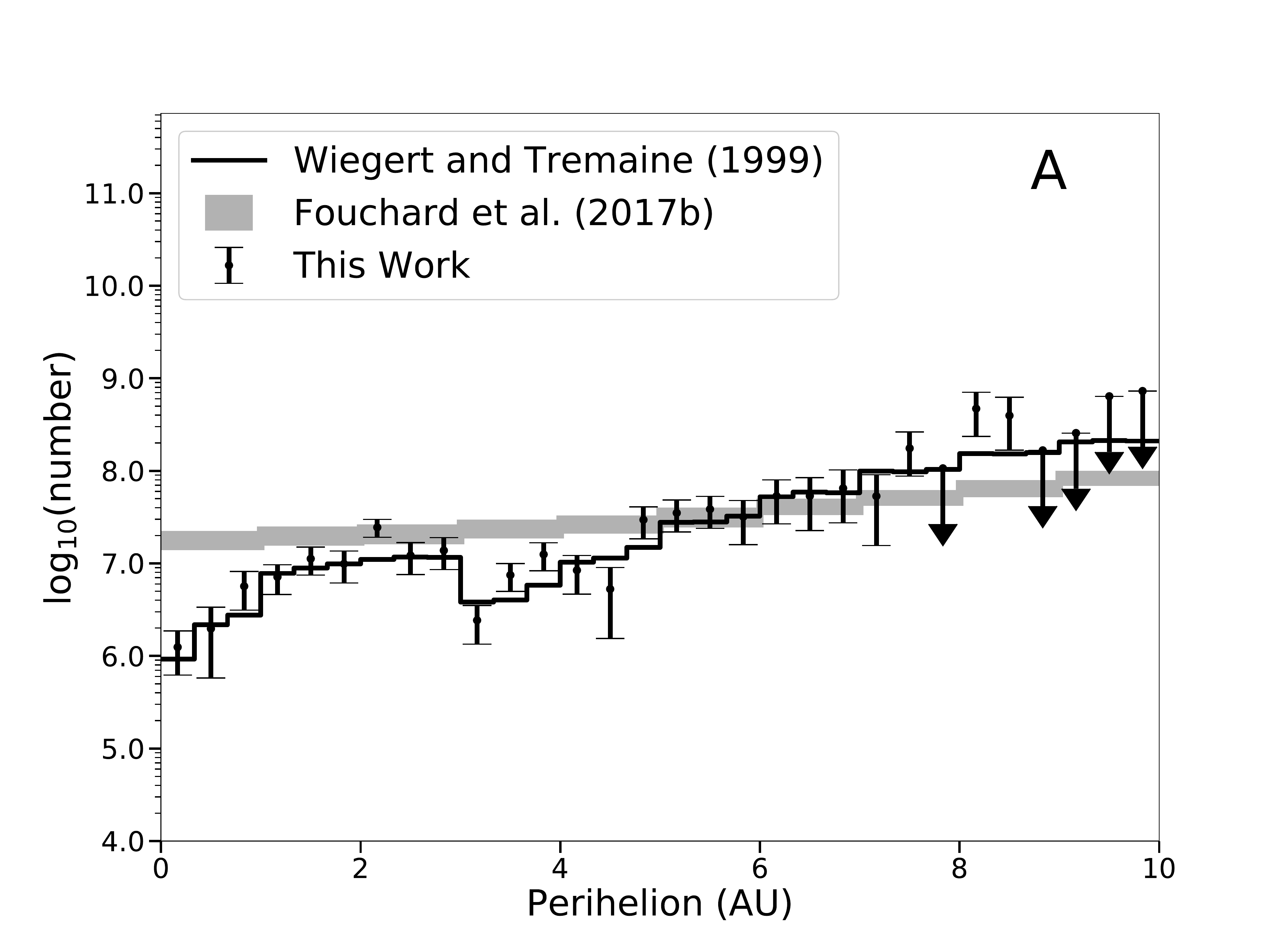}
	\includegraphics[width=0.5\textwidth]{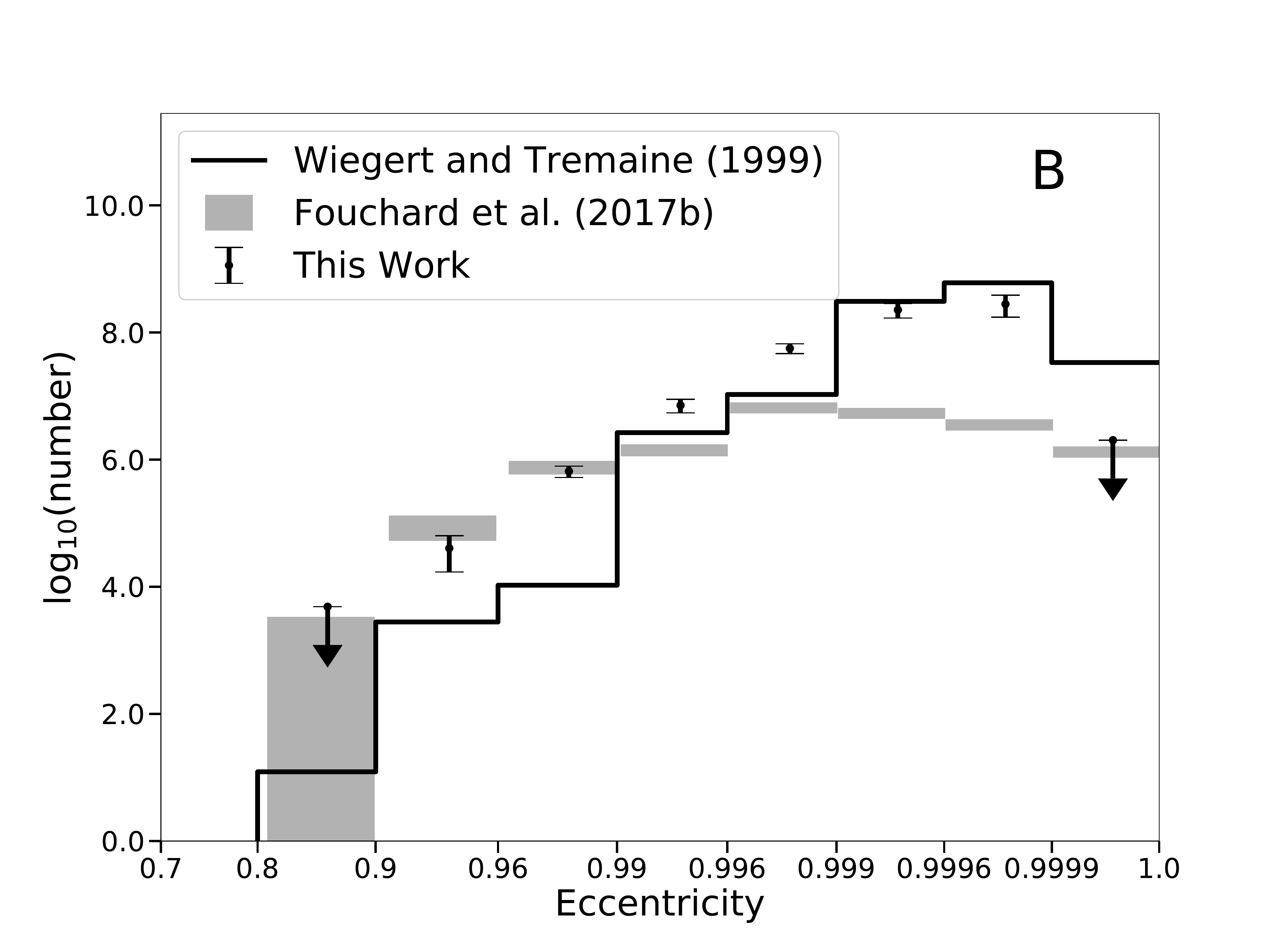}
	\includegraphics[width=0.5\textwidth]{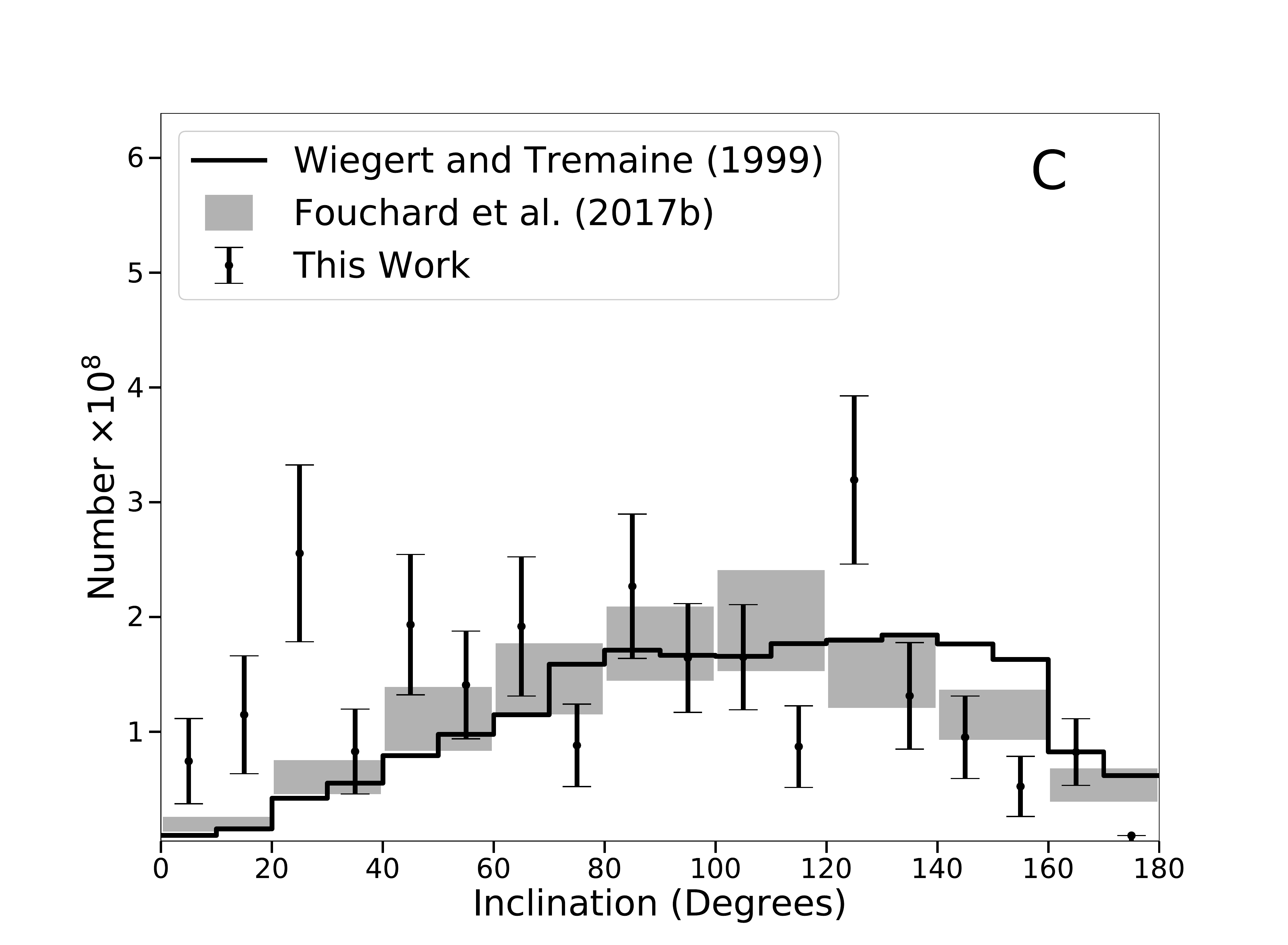}
	\includegraphics[width=0.5\textwidth]{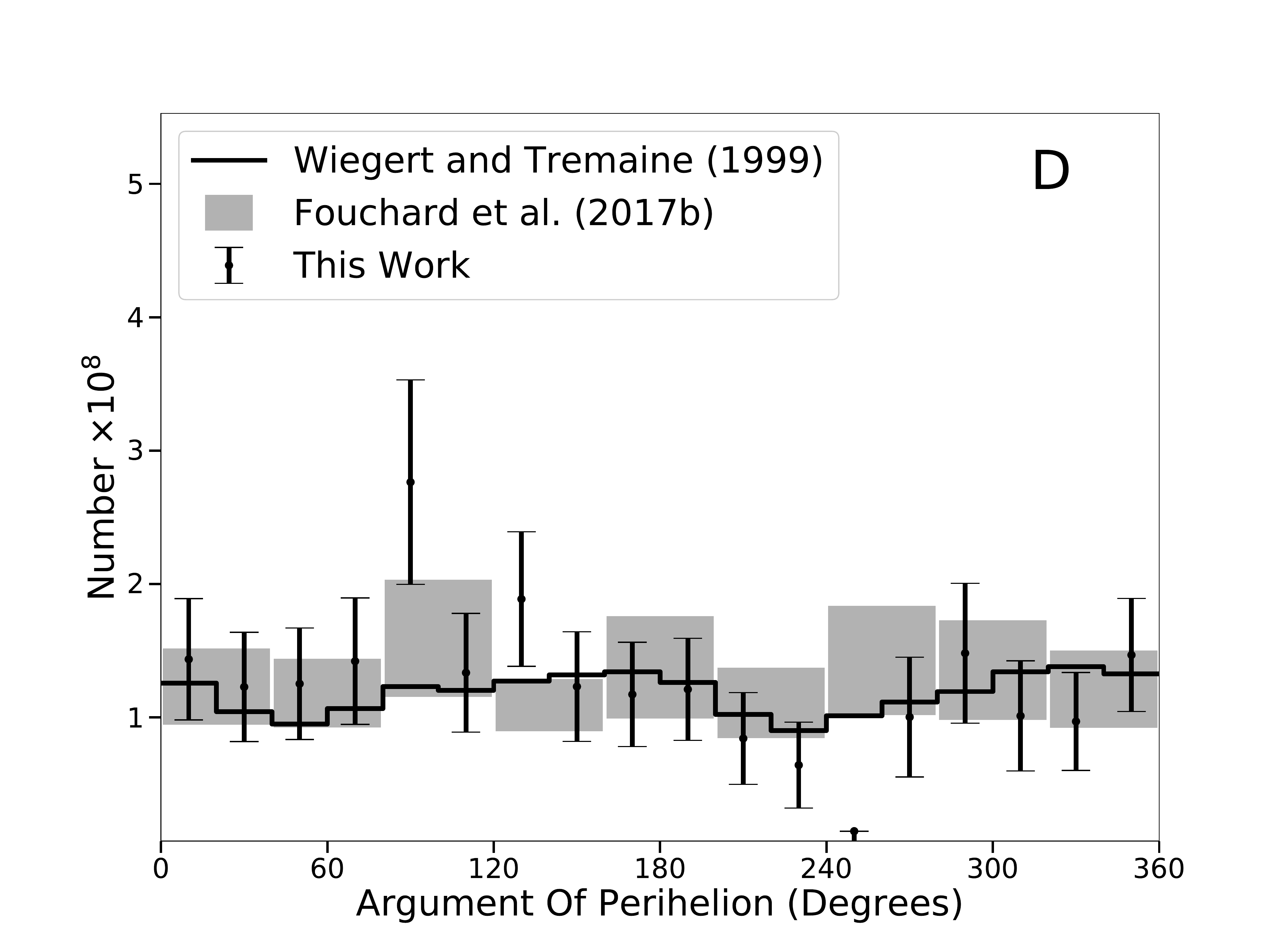}
	\includegraphics[width=0.5\textwidth]{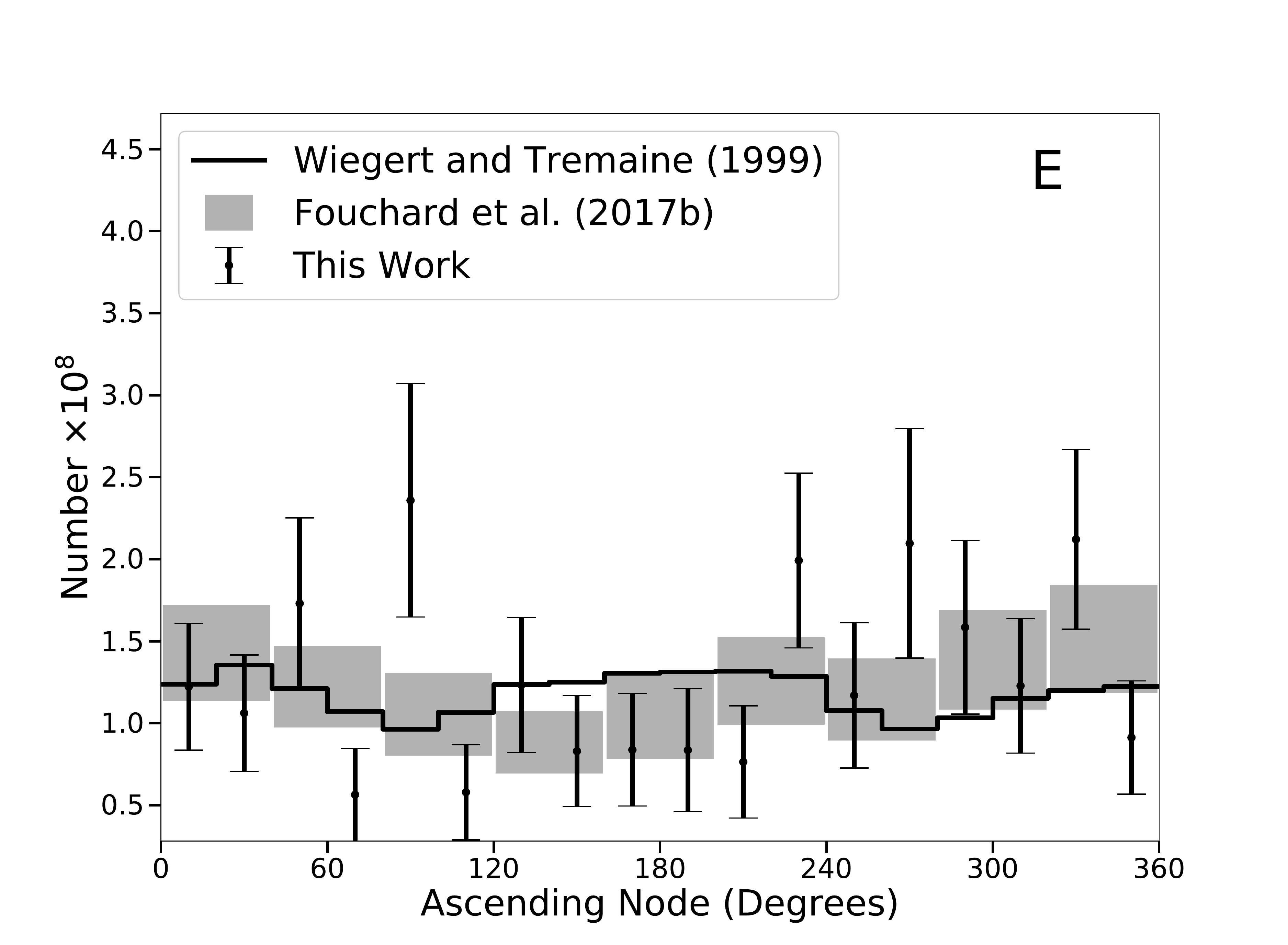}
	\includegraphics[width=0.5\textwidth]{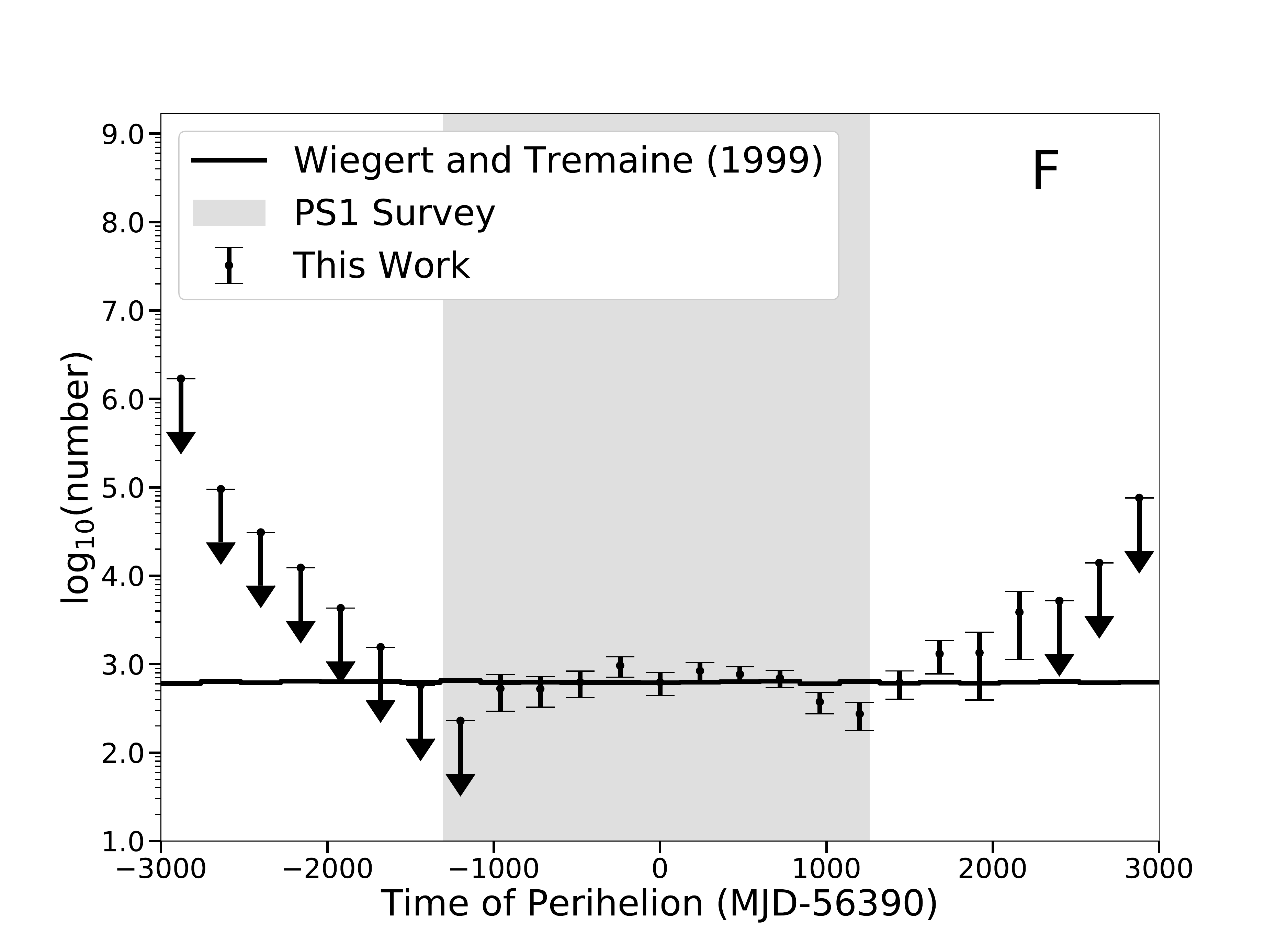}
	
	\caption{
	Orbital element distributions for synthetic long period comets (solid) and the \PSone\ corrected (debiased) population (data points with uncertainties).  The gray bands in panels A-E represent the $\pm1$-$\sigma$ ranges of the various LPC simulations of \citet[\eg][]{Fouchard2017b}. The gray region in the time of perihelion panel (F) in the range $[-1225,+1269]$\;days represents the time range of the \PSone\ survey fields used for our MOPS simulation (\S\ref{sss.SyntheticLPCOrbitDistribution}). The corrected distributions represent our estimate for the true number of comets in the Oort cloud within the tested parameter space.  All panels except for the eccentricity distribution (panel B) contain the values for 150 LPCs.  The eccentricity panel contains only 102 values because the other LPCs have measured eccentricities $>1$. Both LPC model distributions have been normalized to the debiased population. 
	}

	\label{fig.Synthetic-vs-Corrected-Population}
	
\end{figure}

Our LPC model's incremental perihelion number distribution increases roughly linearly with perihelion distance (\fig{fig.Synthetic-vs-Corrected-Population}). Though the distribution of perihelion distances injected from the Oort cloud would have a flat incremental distribution \citep[\eg][]{Wiegert1999} it is significantly modified by cometary fading, planetary ejection dynamics, and whether a comet's perihelion `creeps' or `jumps' across the Jupiter-Saturn barrier \citep{KaibQuinn2009,Fouchard2017b,Rickman2014}. The \citet{Fouchard2017b} perihelion distribution is shallower (\fig{fig.Synthetic-vs-Corrected-Population}), presumably due to ignoring the physical loss (fading) of LPCs with perihelion passage so that more objects survive passage deep into the solar system. 

The LPC eccentricity distribution is strongly peaked near $e\lesssim1.0$ because any object with an aphelion in the range of the Oort cloud and with $q<10\au$ must have high eccentricity.  Indeed, 99\% of the model LPCs have $e>0.9908$. The dynamical origin of the distribution stems from the fact that the effects that push Oort cloud comets back into the inner solar system are more efficient with longer lever arms, that is, larger semi-major axis. The low eccentricity tail of the distribution is the set of objects that have had sufficient time to dynamically evolve their aphelion distances to smaller values because they live long enough despite fading or because they suffered a chance encounter(s) with a giant planet(s) that modified their orbit.  The \citet{Fouchard2017b} LPC eccentricity distribution is also peaked at a high $e$ or $\sim0.997$ but is broader than the \citet{Wiegert1999} model.

Both the \citet{Fouchard2017b} and \citet{Wiegert1999} LPC models exhibit roughly the $\sin(i)$ distribution expected from phase space considerations but are skewed to retrograde orbits ($i>90\arcdeg$) (\fig{fig.Synthetic-vs-Corrected-Population}C).  The skew, which is more obvious in the \citet{Wiegert1999} model, occurs because objects on retrograde orbits are less efficiently scattered than those on prograde orbits so that objects on orbits with $i>90\arcdeg$ survive longer and therefore have greater representation in the steady-state model.

The \citet{Fouchard2017b} and \citet{Wiegert1999} LPC models show variations in the number of objects as a function of the argument of perihelion and ascending node (\fig{fig.Synthetic-vs-Corrected-Population}D \& E).  The differences are presumably due to the manner in which the two models consider the importance of stellar perturbations and the galactic tide in producing LPCs, \ie, the two models predict different angular element distributions in $(i,\omega,\Omega)$ so a sufficiently accurate debiased LPC model could distinguish between stellar and galactic tidal processes in the dynamical production of LPCs.

\subsubsection{LPC diameters, absolute and apparent magnitudes}
\label{sss.LPC-DHV}
  
Our fundamental parameter to characterize the `size' of a cometary nucleus is the absolute nuclear magnitude ($H_N$) \ie\ the absolute magnitude of a cometary nucleus without any contribution from a coma where \citep{Russell1916}: 
\begin{equation}
\frac{D(H_N,p_V)}{\mathrm{meters}} 
   = \frac{1.342\times 10^6}{\sqrt{p_V}} 10^{-H_N/5}
\label{eqn.HSize}
\end{equation}
where $p_V$ represents the nucleus's albedo. Most asteroids have albedos in the range $0.03 \lesssim p_V \lesssim 0.20$ \citep{Mainzer2011} while comet nuclei have $p_V\sim0.04$ \citep[\eg][]{Li2013,Lamy2004}.  We used $p_V=0.04$ when calculating the diameters for both the synthetic and real LPCs since we are attempting to characterize the SFD of a large population of objects with unknown albedos.

We used the nominal values of the \citet{Meech2004} LPC incremental nuclear size frequency distribution 
\begin{eqnarray}
    n(D) \propto \Bigl( \frac{D}{\mathrm{meter}} \Bigr)^{-2.45\pm0.05} &&   2\km \le D <  4\km, \;                           10\km \le D < 20\km \\
\nonumber
    n(D) \propto \Bigl( \frac{D}{\mathrm{meter}} \Bigr)^{-2.91\pm0.06} &&   4\km \le D < 10\km,
    \label{eqn.MeechSFD}
\end{eqnarray}
as the first step in determining the \PSone\ detection efficiency (\S\ref{sss.LPCDetectionEfficiency}) and extended it to both smaller and larger objects. This LPC SFD is relatively steep for comet nuclei in the  $4\km$ to $10\km$ diameter range and shallower at both smaller and larger diameters (\fig{fig.HEfficiency+HDistribution}).

Modeling cometary apparent magnitudes ($V$) is difficult due to the large number of parameters including, but not limited to, the dust-to-gas ratio, the types of volatiles, the active surface area, the grain sizes and density distribution, thermal properties of the surface materials and the diameter of the nucleus \citep{Prialnik1992}.  All these factors contribute to a complex relationship between a comet's intrinsic brightness and solar distance.  As comets approach the Sun the increasing solar radiation heats the surface, driving sublimation of volatiles within the comet that drag dust from the surface to create their characteristic comae and tails.  The total apparent brightness of a comet is usually dominated by light reflecting off dust particles in the coma and tail, swamping the contribution from the nucleus, because the light scattering surface area of ejected dust is orders of magnitude larger than the nucleus's surface area.  The LPCs show a wide range of activity levels and can be active even beyond Saturn indicating that they contain highly volatile species such as \COtwo\ and \CO~\citep{Meech2017}. Ground- and space-based spectroscopic investigations commonly identify \COtwo\ and \HtwoO\ while \CO\ is present only in a small fraction of LPCs \citep{Meech2017,Womack2017}.  Furthermore, there is a large variation in relative volatile abundances between comets \citep{Meech2017-BeforeRosetta}.  Thus, defining the characteristics and modelling the behavior of an `average' comet is complicated.

We used the \citet{Meech1986} thermodynamic sublimation model to calculate cometary apparent total magnitudes that include the effect of the parameters discussed above.  This model can account for the behavior of different volatile species and has been successfully applied to many individual comets \citep[\eg][]{Meech2013,Snodgrass2013,Meech2017-K2,Meech2017-ER61}. To model the entire population of LPCs detected by \PSone\ we employed the thermodynamic sublimation model using accepted nominal values for each parameter (\tab{tab.CometParameters}) and the \citet{Bobrovnikoff1954} dust velocity formulation.  Since our goal was to model the entire LPC population we used typical \HtwoO\ and \COtwo\ active areas (\tab{tab.CometParameters}) and ignored the \CO\ contribution.

For the purpose of the \PSone\ survey simulation all the synthetic LPCs had their absolute magnitudes set to zero ($H_N = 0 \equiv H_0$), roughly the size of Pluto and therefore much larger than any expected real object, so that we could determine the fields in which the synthetic objects might appear and ensure that they are always brighter than the system's limiting magnitude.  We then employed the sublimation model to calculate the object's total apparent magnitude inside an aperture one arcsecond in radius, $V_0(H_0,r,\Delta,t)$, given the nominal cometary parameters, the LPC's diameter as calculated from its absolute nuclear magnitude ($D(H_N)$), and its heliocentric ($r$) and geocentric distances ($\Delta$) at the times ($t$) of observation (from the survey simulation). Since all our synthetic LPCs had $H_N=0$ we could re-assign a specific synthetic LPC any absolute nuclear magnitude ($H_N$) and quickly determine its apparent magnitude at every observation time \ie\ $V(H_N,r,\Delta,t) = V_0(H_0,r,\Delta,t) + H_N$. 

\begin{table}[htbp]
	\centering
	\begin{tabular}{ | l | c | l |}
		\hline
		{\bf Physical Parameter}  & {\bf Value}           & {\bf Source}                    \\ \hline
		Grain radius              &    $1\um$             & \citet{Hanner2003}              \\ \hline
		Grain density             & $1,000\kg\meter^{-3}$ & \citet{Bradley1988,Fulle2015}   \\ \hline
		Nucleus density           &  $400\kg\meter^{-3}$  & \citet{AHearn2011,Sierks2015}   \\ \hline
		Active area (\HtwoO)      &    $4$\%              & \citet{AHearn1995}              \\ \hline
		Active area (\COtwo)      &  $0.1$\%              & see note$^\dag$                 \\ \hline
		Nucleus albedo            &    $4$\%  			  & \citet{Li2013}                  \\ \hline
		Emissivity                &    $0.9$              & \citet{Fernandez2013}           \\ \hline
		Nuclear phase coefficient &  $0.04$ mag/deg  	  & \citet{Li2013}                  \\ \hline
		Coma phase coefficient    &  $0.02$ mag/deg    	  & \citet{Meech1987}               \\ \hline
	\end{tabular}
	\caption{Cometary physical parameters adopted for this work. $^{\dag}$There exists a range of abundances of \COtwo\ and \CO\ relative to water from $<$1\% to 30\% \citep{Mumma2011}. Sublimation from these two volatile species is typically $<1\%$ so we use a canonical value that is representative of other comets \citep{Meech2017-K2,Meech2017-ER61}.}
	\label{tab.CometParameters}
\end{table}


\subsubsection{Post-processing of synthetic LPC detections}
\label{sss.MOPSPostProcessing}

The MOPS simulation returns detections of the synthetic LPCs that then require several more steps to increase the simulation's verisimilitude.  While we have attempted to be as realistic as possible we stress that LPCs are legendary for their unpredictable and individualistic behavior.  

First we required that synthetic tracklets have $\geq 3$ detections.  The \PSone\ surveying typically obtained 4 images at the same boresight over the course of an hour but some of those sequences would be aborted and there were other surveying modes that acquired more than 4 images in a sequence, particularly in the \ips\ filter.

Next, we eliminated any tracklets for objects that are moving too fast or too slow.  Since the LPCs are typically discovered at large heliocentric distances (average$\pm\sigma$=$5.4\pm2.5\au$) their apparent rates of motion are slow and peak at values more typical of main belt asteroids than NEOs.  \citet{Schunova-Lilly2017} has shown that trailing loss, the loss in SNR due to an object leaving a trail on an image rather than looking more like a point source, begins at about 1.5~deg/day and only an insignificant fraction of LPCs have rates larger than this threshold at the time of discovery.  We did require that tracklets have apparent rates of $<5$~deg/day but it has essentially no effect on the analysis because so few tracklets have rates $>1.5$~deg/day.  In the czaring process objects are prioritized and reviewed based mostly on their apparent rate of motion and those moving slower than 0.1~deg/day are considered `slow movers'. The slow moving objects are typically not systematically reviewed by the czars due to time constraints and the large number of false-positive tracklets at slow rates. We therefore place a lower limit threshold at 0.1~deg/day on the detectable rate of motion. 

Third, the \PSone\ focal plane has about a $\sim75$\% fill-factor due to CCD defects, chip gaps, and other issues \citep{Denneau2013}.  This is one of the reasons why the \PSone\ asteroid survey employs four sequential images and then requires that at least 3 detections are present to create a tracklet.  The blending of the fill-factor losses with the survey detection combinatorics resulted in an overall tracklet detection efficiency of $\sim78$\% \citep{Denneau2013} that has improved slightly over time.  This led us to randomly remove 20\% of the synthetic tracklets to mimic real \PSone\ operations. 

Fourth, each object would be assigned an absolute magnitude and each of its detection's apparent  magnitudes would be calculated as described above (\S\ref{sss.LPC-DHV}).  The important quantity for determining whether an object is detectable is the apparent total brightness within a typical PSF ($V_{PSF}$), not the apparent nuclear magnitude.  We used the \citet{Meech1986} sublimation model to calculate $V_{PSF}$ within a $1\arcsec$ diameter aperture and detections with $V_{PSF}>21.8$ for \wps\ and $V_{PSF}>21.0$ for \ips\ were discarded because they are fainter than the \PSone\ limiting magnitudes for the respective bands \citep{Denneau2013}, \ie, we assumed that the \PSone\ limiting magnitude was constant throughout time and we assumed that the source detection efficiency was 100\% for detections brighter than the limiting magnitude and 0\% for fainter sources.  Neither of these assumptions is correct but we think the impact on our results is negligible in comparison to other known systematic offsets, such as the $5.5\mags$ `JPL' offset discussed in \S\ref{ss.PS1LPCDetectionEfficiency}, and other systematic errors (\eg\ \S\ref{sss.SystematicUncertainties}).  For comparison, \citet{Veres2017-LSST-Performance} have shown that a more accurate treatment of the system performance near the limiting magnitude ($V_{lim}$) marginally increases the number of detected objects near the system limit and allows detections only a few tenths of a magnitude fainter than the nominal $V_{lim}$.  Since all the detections in a tracklet are typically acquired within about an hour the apparent magnitudes of each of the detections within a tracklet are essentially identical because we also ignore the possibility of any light curve variation due to an object's rotation. 

At this point in the simulation we have a set of synthetic LPC tracklets that were detected by the MOPS but we have not yet determined whether those tracklets would actually be identified as LPCs.  This final step typically requires additional observations to identify an LPC by its orbital elements and/or by its morphology, \ie\ by identifying cometary activity, and we consider this post-processing step the most difficult to model because detecting comets has been as much art as science.

Inactive or low-level activity comets, like those typically discovered by the czars, will be nearly PSF-like in the \PSone\ images so the IPP will identify the sources with essentially the same efficiency as asteroids.  The MOPS czars may then flag the tracklet as interesting for one of two reasons:  1) it has an unusual rate and/or direction of motion or 2) it has a slightly non-PSF appearance.  

The first reason can be partially quantified using the Minor Planet Center's `\digesttwo' score that provides a metric that an object is interesting based on its sky-plane location, rates of motion, and apparent brightness \citep[\eg][]{Jedicke1996}.  For instance, objects are interesting if they appear to have main belt-like rates of motion on the ecliptic but their apparent brightness suggests that they are unknown, or if they have main belt-like rates but are far from the ecliptic, or if they have unusual rates of motion even if they are embedded in the ecliptic.  The \digesttwo\ score is effectively the fraction of orbits consistent with the tracklet that correspond to a near-Earth object (NEO).  Each \PSone\ tracklet's \digesttwo\ score is calculated and those with scores $<65$ are typically not vetted by the \PSone\ czars unless the object has a non-PSF morphology that might indicate a cometary coma.  Real objects with scores in the range from 65 to 90 are submitted to the MPC as being interesting but they are usually not confirmed by other sites because there are too many objects to followup.  Thus, any LPC with a \digesttwo\ score greater than 90 would likely be identified as interesting regardless of whether they exhibited any cometary behavior.  Subsequent followup by other observers would reveal the cometary nature of the orbit and thus we consider any synthetic LPC with \digesttwo$>$90 to be an LPC discovery. 

\begin{figure}[htbp]

	\includegraphics[width=\textwidth]{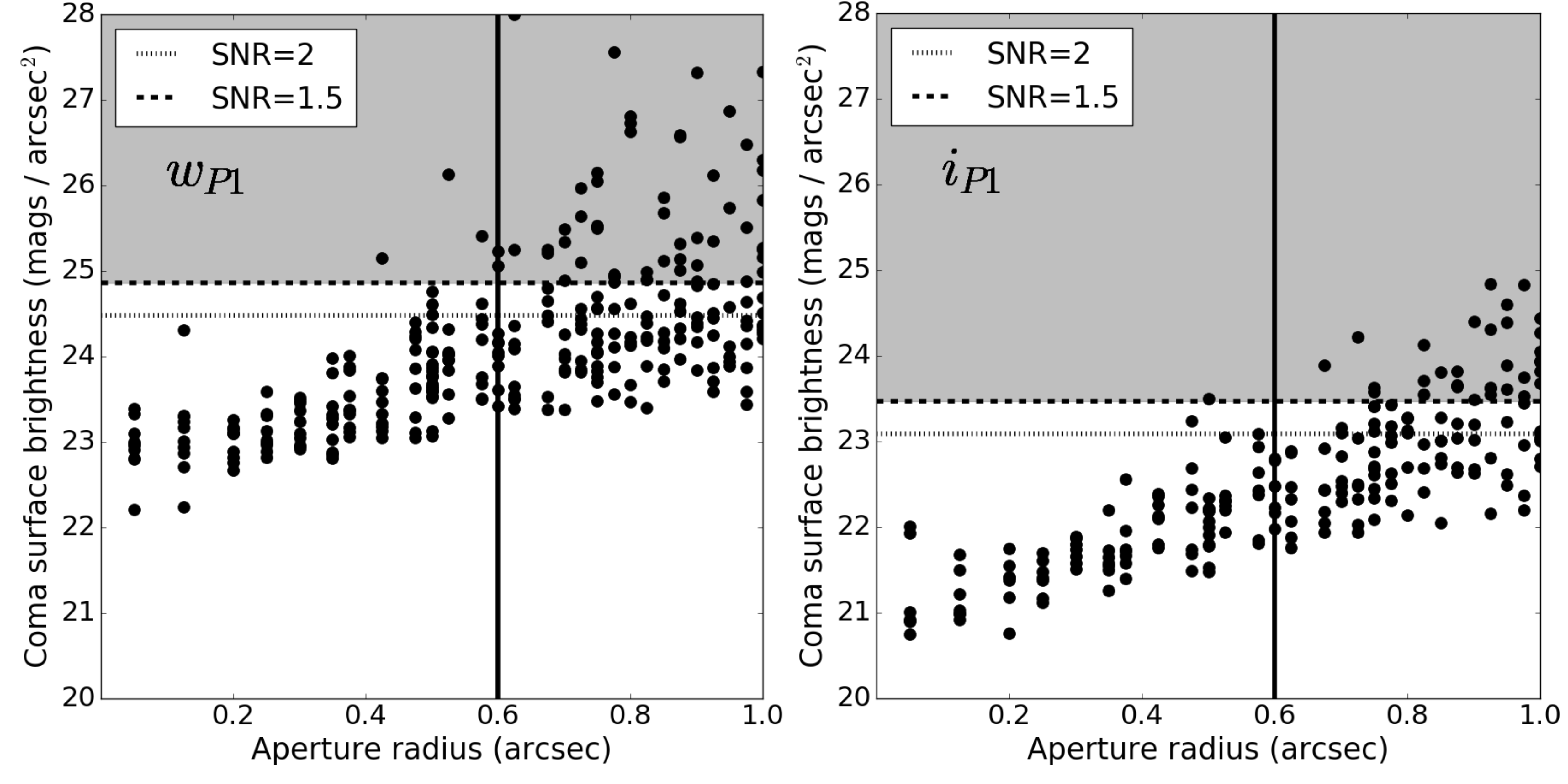}
	
	\caption{Measured total coma brightness in (left) $\wps$ and (right) $\ips$ bands as a function of aperture radius for faint, low-activity comets near the human detection threshold, representative of the morphology of the faintest \PSone\ comet discoveries.  We used 11 comets in $\wps$ and 7 comets in $\ips$.  The vertical line at 0.6$\arcsec$ represents an aperture radius slightly greater than the typical PSF radius in \PSone\ images, the distance at which any PSF-extension due to coma would be barely detectable by a human czar.  The horizontal lines represent the 1.5 and 2-$\sigma$ SNR above the typical sky background level. 
	}
	
	\label{fig.ComaBrightness}
	
\end{figure}

The second reason relies on the experience of the \PSone\ czars who have shown tremendous skill in identifying barely non-PSF morphologies.  Some of the comets discovered by \PSone\ czars have morphologies with PSF FWHM only 20\% larger than nearly stars with typical FWHMs of $\sim1.0\arcsec$. Given that \PSone\ primarily identifies comets using a human czaring process, our model must determine the detection threshold for comets that mimics the actual procedure. We selected the first \PSone\ observations of 18 comets successfully identified by the czars near the detection threshold (11 comets in \wps\ and 7 in \ips\ spanning the survey) and measured the typical surface brightness of these comets to determine the LPC detection limits. The IPP provides processed images with the sky background already subtracted by differencing successive images at the same location.  Each comet image was fit to a Gaussian PSF with an additional $1/r$ coma profile \citep{Jewitt1987} which was then used to measure the flux in successive apertures to determine the surface brightness profile as a function of radius, $\rho$, from the centroid (\fig{fig.ComaBrightness}).  Typical stellar \PSone\ PSFs have a diameter of about $1\arcsec$ so a comet will be detectable if there is detectable surface brightness above the sky background outside a radius of $\sim0.6\arcsec$ from the nucleus.  Based on actual discoveries of low-activity comets with barely detectable comae (\fig{fig.ComaBrightness}), and the well-known human ability to detect signal, we set our LPC coma detection threshold at SNR=1.5 at a radius of $0.6\arcsec$ corresponding to $\rho_{\wps}=24.8$~mags/arcsec$^2$ in the \wps\ band and $\rho_{\ips}=23.5$~mags/arcsec$^2$ in the \ips\ band.  \ie\ if the calculated coma-brightness of a synthetic LPC at $0.6\arcsec$ from the nucleus exceeds the sky brightness at SNR=1.5 in that band then we consider the synthetic tracklet to be an LPC discovery. 
 
Finally, assuming that an LPC's coma flux drops as $1/\rho$, where $\rho$ is the distance from the nucleus \citep{Jewitt1987}, we used the \citet{Meech1986} sublimation model to calculate the expected apparent coma surface brightness at $0.6\arcsec$ from the nucleus.

\subsubsection{\PSone\ LPC detection efficiency and debiasing}
\label{sss.LPCDetectionEfficiency}

We use `debiasing' to refer to the process of correcting an observed distribution for intrinsic selection effects induced by the experiment.  The process depends upon an accurate estimate of the system's detection efficiency with respect to all the relevant parameters \citep[\eg][]{Jedicke2002}.  This is usually impossible or extremely difficult so most debiasing is performed on a subset of the relevant parameters and, often, on a distribution that is collapsed into a single dimension (\eg\ diameter or absolute magnitude).  Debiasing in a limited set of parameters can be pernicious because the measured detection efficiency in each of those parameters is implicitly dependent on the `hidden' parameters.  Doing so introduces systematic errors into the debiased population that are often difficult to quantify.

Our primary goal is to measure the LPC nuclear SFD so our main requirement is an accurate determination of the \PSone\ detection efficiency as a one dimensional (1-d) function of nuclear absolute magnitude, $\epsilon(H_N)$.  In this case we could generate a synthetic population with {\it any} $H_N$ distribution, $N_{syn}(H_N)$, and run that population of objects through a survey simulator to generate the observed synthetic population, $n_{syn}(H_N)$ so that $\epsilon_{syn}(H_N) = n_{syn}(H_N) / N_{syn}(H_N)$.  The efficiency will be a good representation of the system detection efficiency if the survey simulator {\it and} the orbit distribution model are realistic.  

More generally, we used a synthetic LPC population model with a size-frequency distribution that is independent of the orbital element distribution through the MOPS simulator (\ie\ $N_{syn}(H_N) \; f(q,e,i,\xbar)$).  The simulator's output was a synthetic observed LPC population, $n_{syn}(H_N,q,e,i,\xbar)$, where $\xbar$ represents the set of other angular orbital elements.  The orbital elements and absolute nuclear magnitudes are no longer independent in the synthetic observed population because of correlations between cometary activity, heliocentric distance, and other observation biases.   The synthetic efficiency is then
\begin{equation}
    \epsilon_{syn}(H_N,q,e,i,\xbar) 
       = \frac{n_{syn}(H_N, q,e,i,\xbar)}
              {N_{syn}(H_N) f(q,e,i,\xbar)}
    \label{eqn.SyntheticEfficiency}
\end{equation}
where the `syn' subscript on the efficiency specifies that the result is our synthetic estimate of the system detection efficiency (presumably representative of the actual system detection efficiency).  In practice, we then integrate over all the dimensions except the one of interest, and even in that dimension we integrate over a bin due to limited computational time and statistics, so that the LPC detection efficiency for LPCs with nuclear absolute magnitudes in the range $[H_N^{min},H_N^{max}]$ is
\begin{equation}
    \epsilon_{syn}(H_N^{min},H_N^{max}) 
       = \frac{\int_{H_N^{min}}^{H_N^{max}} \int n_{syn}(H,   q,e,i,\xbar) \dif \; H \; \dif q \; \dif e \; \dif i \; \dif \; \xbar}
              {\int_{H_N^{min}}^{H_N^{max}} \int N_{syn}(H) f(q,e,i,\xbar) \; \dif H \; \dif q \; \dif e \; \dif i \; \dif \; \xbar}.
    \label{eqn.SyntheticEfficiencyGeneral}
\end{equation}

Then, with the assumption that we have accurately modeled the LPC orbit distribution and system performance the debiased $H$ distribution is given by
\begin{equation}
    N'(H_N^{min},H_N^{max}) = \frac{      n_{real}(H_N^{min},H_N^{max})}
                              {\epsilon_{syn}(H_N^{min},H_N^{max})}
\end{equation}
where we use the prime on $N'$ to indicate that it is our estimate of the LPC SFD.

A major issue is that the synthetic LPCs have `true' nuclear magnitudes but the real objects have measured nuclear magnitudes that are significantly in error.  Since our analysis includes a large population of detected LPCs and `average' LPC behavior we also assume that there is a constant error in the reported JPL nuclear magnitude\footnote{\protect\label{JPL_H_N}JPL nuclear magnitudes were extracted from the JPL Small-Body Database Search Engine \protect\url{https://ssd.jpl.nasa.gov/sbdb_query.cgi}} and fit the data to determine the offset, $\Delta H_{JPL}$.  \ie\ each real LPC designated by the subscript $i$ with absolute nuclear magnitude $H_{N,i}$ is set to $H_{N,i} + \Delta H_{JPL}$ so that the number of objects in the range $[H_N^{min},H_N^{max}]$ depends on $\Delta H_{JPL}$ as does the debiased number of objects:
\begin{equation}
    N'(H_N^{min},H_N^{max},\Delta H_{JPL}) = \frac{      n_{real}(H_N^{min},H_N^{max},\Delta H_{JPL})}
                                              {\epsilon_{syn}(H_N^{min},H_N^{max})}.
\end{equation}
We understand that each JPL nuclear magnitude is uniquely in error in its own way but this analysis assumes that the error can be characterized by an average error over the large population of \PSone\ LPCs.  We will present evidence to support this assumption below when we fit for the value of $\Delta H_{JPL}$ (\fig{fig.Likelihood_vs_JPLOffset} and \S\ref{ss.PS1LPCDetections}).

At this point we have our own debiased LPC size-frequency distribution that may be different from the original distribution employed in \eqn{eqn.SyntheticEfficiency}.  We use it to generate an improved synthetic input LPC population model by substituting $N'$ for $N_{syn}$ in \eqn{eqn.SyntheticEfficiencyGeneral} and then repeat the efficiency determination process to determine $\epsilon_{syn}^*(x_{min},x_{max},\Delta H_{JPL})$ where $x$ represents any of the orbital parameters or $H_N$ (at this point, debiasing the $H_N$ distribution returns our derived $H_N$ distribution).  

We then calculate the debiased 1-d distributions in all the orbital elements.  Following the $H_N$ derivation above, the debiased number of objects in the range is 
\begin{equation}
    N^*(x_{min},x_{max},\Delta H_{JPL}) = \frac{        n_{real}(x_{min},x_{max},\Delta H_{JPL})}
                                               {\epsilon_{syn}^*(x_{min},x_{max})}.
\end{equation}

In our final debiasing step we fit the real observed distributions, $n_{real}(x)$, in perihelion distance ($q$) and time of perihelion ($t_q$) to the synthetic observed distributions, $n_{syn}(x)$.  Letting $p_{KS}(x)$ represent the $p$-value of the Kolmogorov-Smirnov test \citep[KS;][]{Kolmogorov1933,Smirnov1948}  for orbital element $x$ we then maximize
\begin{equation}
    p_2(\Delta H_{JPL}) =  p_{KS}(q;\Delta H_{JPL}) \; p_{KS}(t_q;\Delta H_{JPL})
    \label{eqn.KSproduct}
\end{equation}
as a function of $\Delta H_{JPL}$.  These two orbital parameters are most sensitive to $\Delta H_{JPL}$ because they respectively determine how close the LPC approaches the Sun (and Earth) and how far away an object may be detected.  We tested introducing other orbital parameters in the product but they had little to no effect on $\Delta H_{JPL}$ and the resulting debiased distributions.


\section{Results and Discussion}
\label{s.ResultsAndDiscussion}

\subsection{\PSone\ LPC detections}
\label{ss.PS1LPCDetections}
Our observed population of 150 LPCs is a subset of the 2,657 known LPCs$^{\protect\ref{footnote.JPL-LPCs}}$ as of 2019-01-18 that were detected by \PSone\ in the $\wps$ and $\ips$ bands from 2010 to the end of 2016 (\fig{fig.Synthetic+Real-Observed-LPCOrbitalElements}).  A total of 229 LPCs were detected by {\it all} telescopes during the same time period implying that \PSone\ detected about 70\% of all known LPCs during the time range under consideration \citep[in agreement with][]{Hsieh2015}. 

The observed LPC orbital element distributions are explicable in terms of our current understanding of their true orbital element distributions and the survey system's operations (\fig{fig.Synthetic+Real-Observed-LPCOrbitalElements}).  The perihelion distribution peaks at about $2\au$ because LPCs at this distance are close and bright.  Objects with smaller perihelion distance are less efficiently detected by surveys like \PSone\ that concentrate their efforts towards opposition, not in the direction of the Sun.  LPCs with larger perihelion distances are simply more distant and more difficult to detect.  The detected eccentricities are strongly peaked near $1$ for the simple reason that these objects are most likely to enter the inner solar system and be detected by \PSone\ (note the unusual binning in the eccentricity panel, \fig{fig.Synthetic+Real-Observed-LPCOrbitalElements}B).  The apparent discrepancy in the eccentricity distribution that is exaggerated by our bin selections is likely due to our excluding the 48 LPCs with $e>1$ in that panel {\it only} since it only extends to $e=1$.  Those objects must have suffered a dynamical interaction in their most recent passage through the solar system that caused them to become hyperbolic so assigning them to any other bin in the histogram is problematic.  (For the purpose of most of our analysis we assign objects with $e\ge1$ a new eccentricity $e'$ given by $e'=(Q-q)/(Q+q)$ where $Q=10^5\au$ is a large aphelion distance.)  The observed inclination distribution very roughly follows a $\cos(i)$ distribution as expected from phase space arguments.  The enhancements as $i \rightarrow 0\arcdeg$ and $i \rightarrow 180\arcdeg$ are likely due to \PSone\ favoring surveying in the ecliptic plane thereby increasing the detection efficiency for objects with $i\sim0\arcdeg$ and $i\sim180\arcdeg$ relative to $i\sim90\arcdeg$ objects.  The observed distribution in the argument of perihelion is essentially flat with the exception of the dip at $\omega\sim250\arcdeg$ that will be discussed below.  The ascending node distribution is essentially flat as would be expected with a survey like \PSone\ that images most of the night sky each lunation and 3/4 of the entire sky during the year.  The skewed distribution in the perihelion dates to the latter half of the \PSone\ survey window is mostly due to a dramatic increase in the amount of time devoted to solar system observations as the survey progressed and also due to an increase in the detection efficiency as the survey strategy evolved and the MOPS czars gained experience.  

\begin{figure}[htbp]
\begin{center}

    \includegraphics[width=0.7\columnwidth]{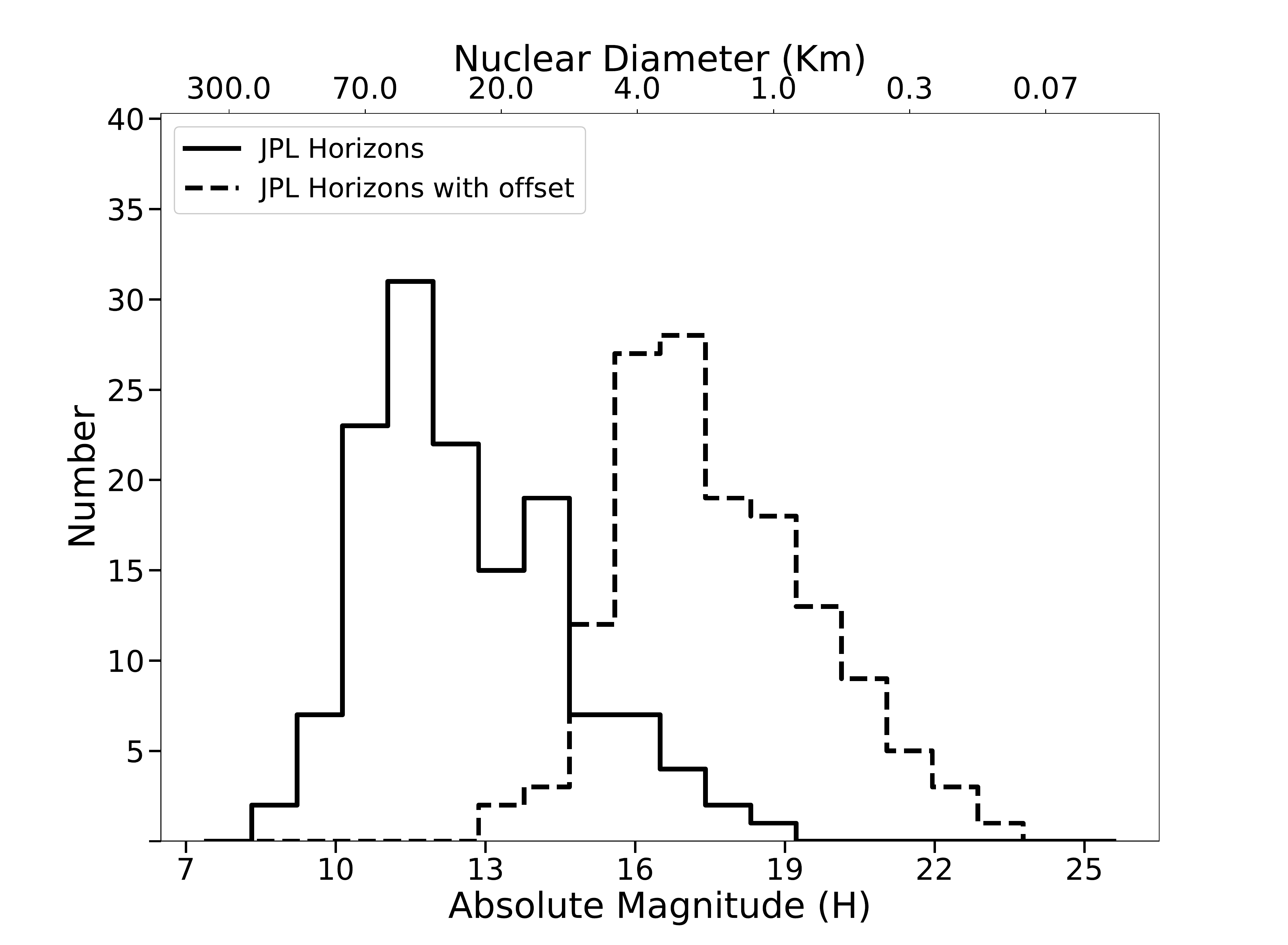}
    
    \caption{Nuclear absolute magnitude distribution for the 140 LPCs which had reported nuclear magnitudes from JPL.  The top scale shows the equivalent nuclear diameter assuming that $p_V=0.04$.  The distributions with and without the JPL offset ($H^{JPL}_{offset}=5.5$) are not identical because the bins in this histogram have widths of $1\mags$ but the offset is non-integral.
    }

    \label{fig:JPL-H-distribution}
    
\end{center}
\end{figure}

Of the 150 LPCs detected by \PSone\ and used in this study a subset of 140 have reported nuclear absolute magnitudes by JPL. The JPL LPC nuclear magnitudes range from $7 < H_N < 20$ that naively correspond to diameters in the range $1\km < D < 260\km$ with an assumed cometary albedo of 0.04 (\fig{fig:JPL-H-distribution}).  The mode of the diameter distribution is $\sim 40\km$, significantly larger than the $0.5\km$ to $10\km$ \insitu\ values for cometary nuclei measured by spacecraft missions and through remote observations when the LPCs are at large heliocentric distances and at zero to low activity levels \citep[\eg][]{Meech2004}. The tremendous discrepancy between the JPL nuclear magnitudes and those that are carefully measured for selected objects by a small number of researchers is due to JPL's use of observations reported by many different observers for a large number of objects.  We will correct for this issue using an average offset ($H^{JPL}_{offset}$) to the reported JPL nuclear magnitudes that provides the best fit between our LPC model and the observed \PSone\ LPCs (as described in \S\ref{sss.LPCDetectionEfficiency}).

\begin{figure}[htbp]
\begin{center}
	\includegraphics[width=0.7\columnwidth]{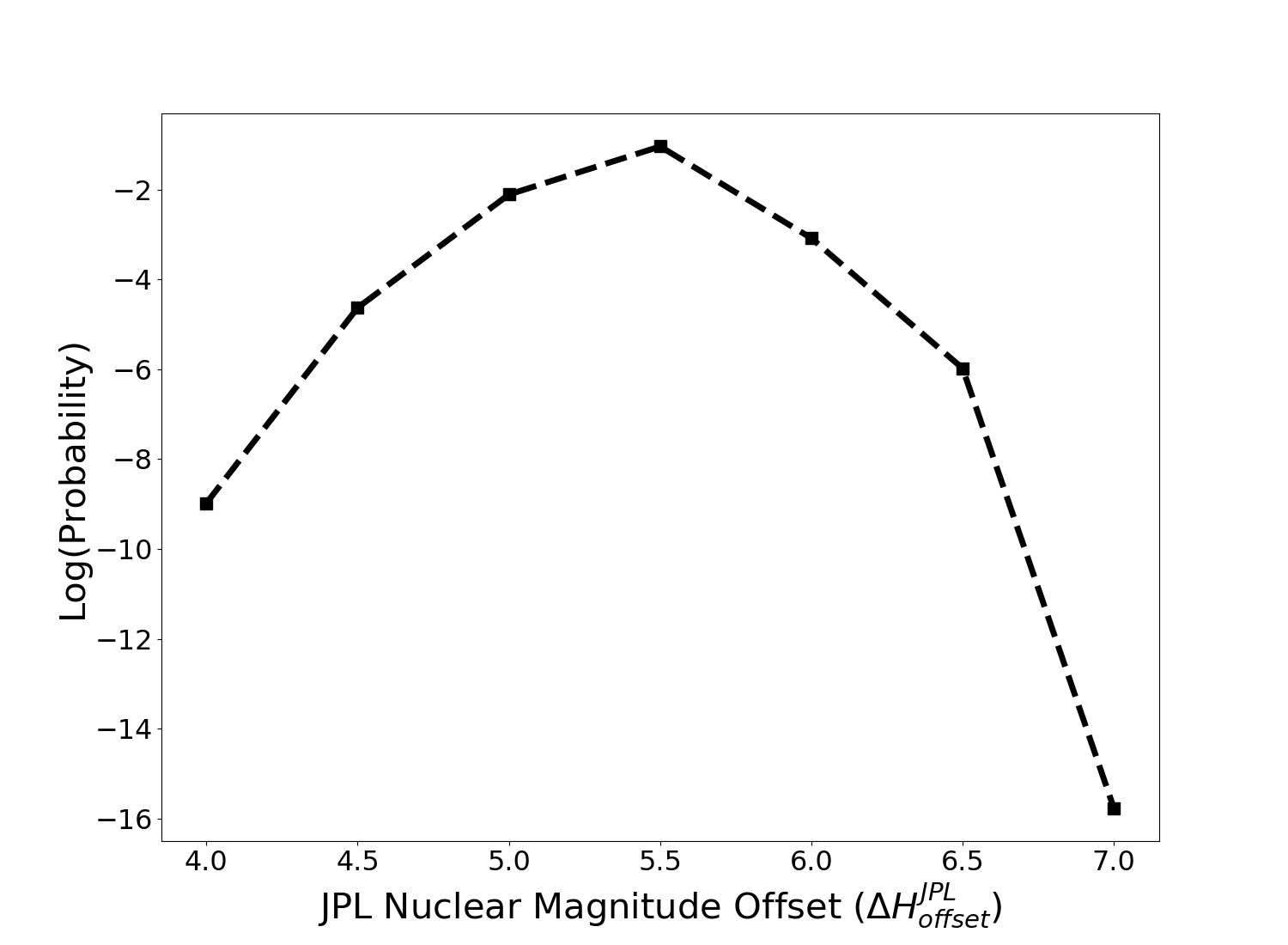}
	\caption{
	    The KS-probability product ($p_2$; \eqn{eqn.KSproduct}) as a function of the JPL nuclear magnitude offset ($H^{JPL}_{offset}$).  
	}
	\label{fig.Likelihood_vs_JPLOffset}
\end{center}
\end{figure}

\subsection{\PSone\ LPC detection efficiency}
\label{ss.PS1LPCDetectionEfficiency}

The calculation of the \PSone\ LPC detection efficiency depends on the JPL nuclear magnitude offset ($H^{JPL}_{offset}$, \fig{fig.Likelihood_vs_JPLOffset}) due to the iterative procedure described in \S\ref{sss.LPCDetectionEfficiency}.  There is a clear trend to a much better fit as $H^{JPL}_{offset} \rightarrow 5.5$ and a modest drop in the fit quality for larger values.  All the following discussion uses our nominal fitted value of $H^{JPL}_{offset} = 5.5$.

\begin{figure}[htbp]

	\includegraphics[width=0.5\textwidth]{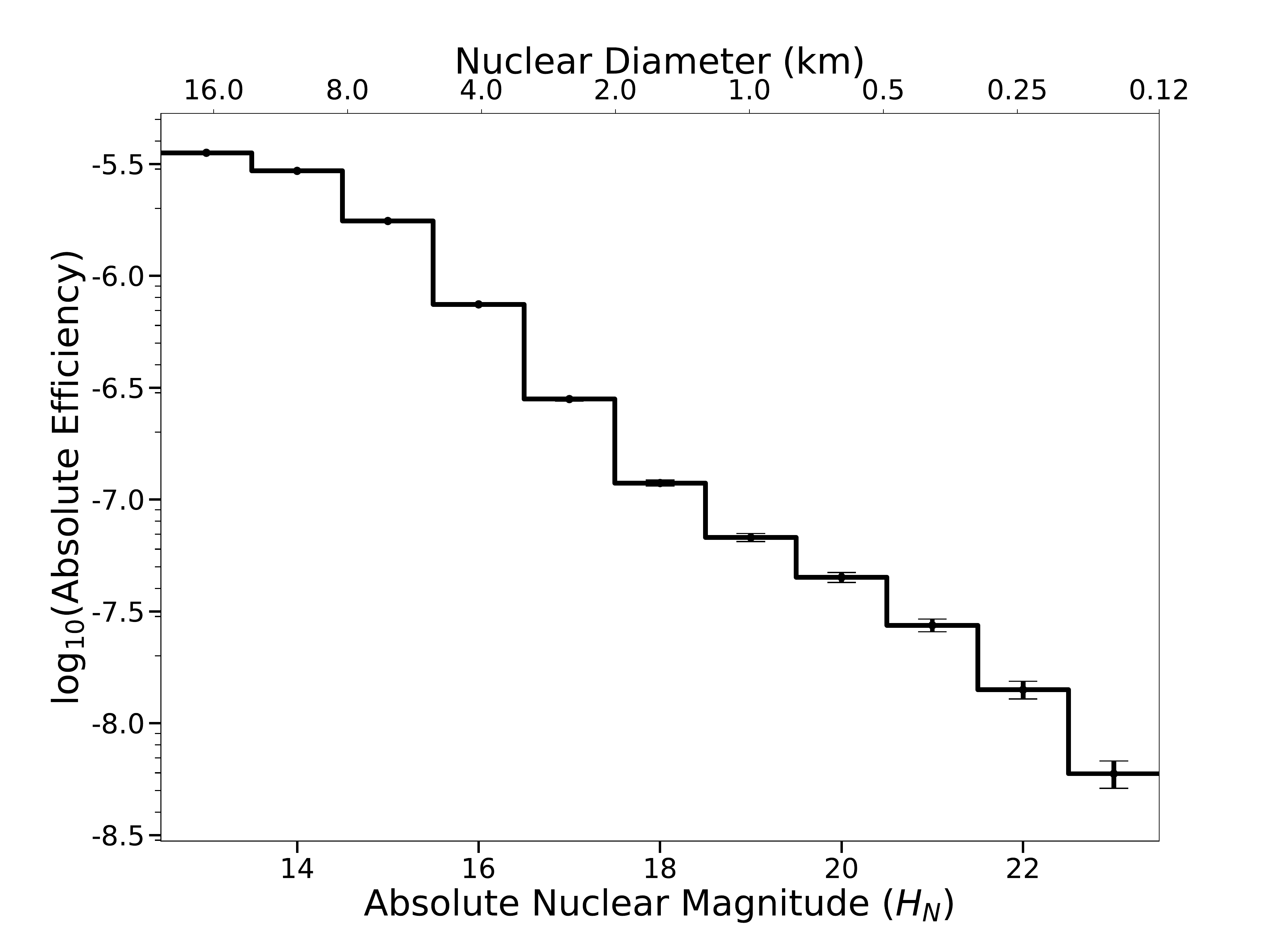}
	\includegraphics[width=0.5\textwidth]{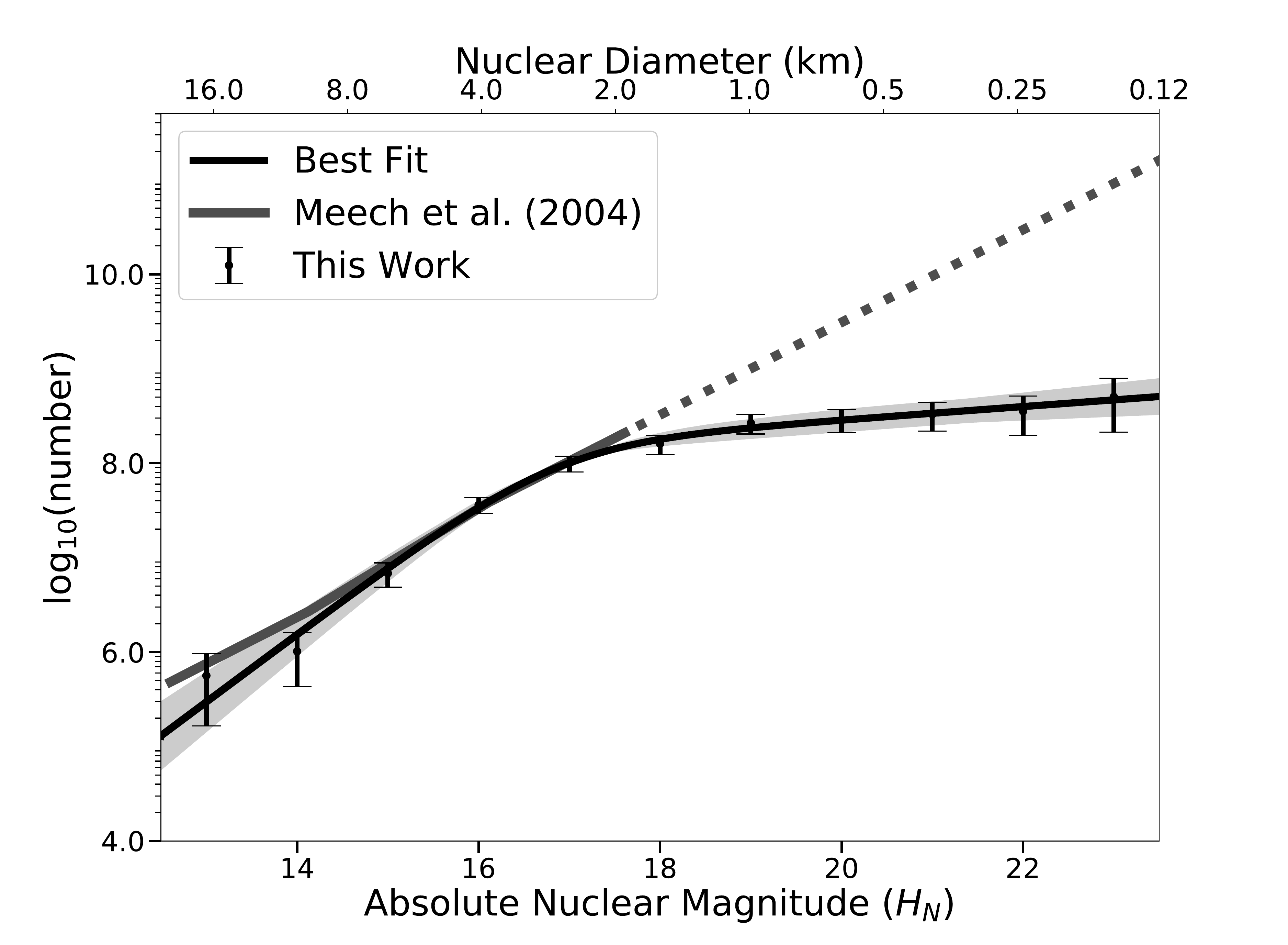}

	\caption{
	  (left) \PSone\ system absolute LPC detection efficiency as a function of absolute nuclear magnitude (\ie\ corrected for the JPL $H_N$ offset described in \S\ref{sss.LPCDetectionEfficiency}).  The nuclear diameter (top axis) assumes a typical cometary nuclear albedo of 4\%.  (right) Bias corrected incremental absolute nuclear magnitude distribution for LPCs.  The shaded gray region represents the range of solutions resulting from the least squares fit to the data. The \citet{Meech2004} function was normalized to our debiased distribution at $H_N = 16$ (to $N = 3.8 \times 10^7$) and the dotted portion for $D<2\km$ diameter is a simple extrapolation of their result to smaller sizes. 
	  \label{fig.HEfficiency+HDistribution}
	}

\end{figure}

The \PSone\ absolute detection efficiency for LPCs is very small because their orbital period is long and the observations used in this analysis encompass only $\sim6.8$ years.  The efficiency drops by $\gtrsim 10\times$ for objects from $10\km$ to $1\km$ diameter (\fig{fig.HEfficiency+HDistribution}) and remains non-zero for objects $\gtrsim100\meter$ diameter.

\begin{figure*}[htbp]

	\includegraphics[width=0.5\textwidth]{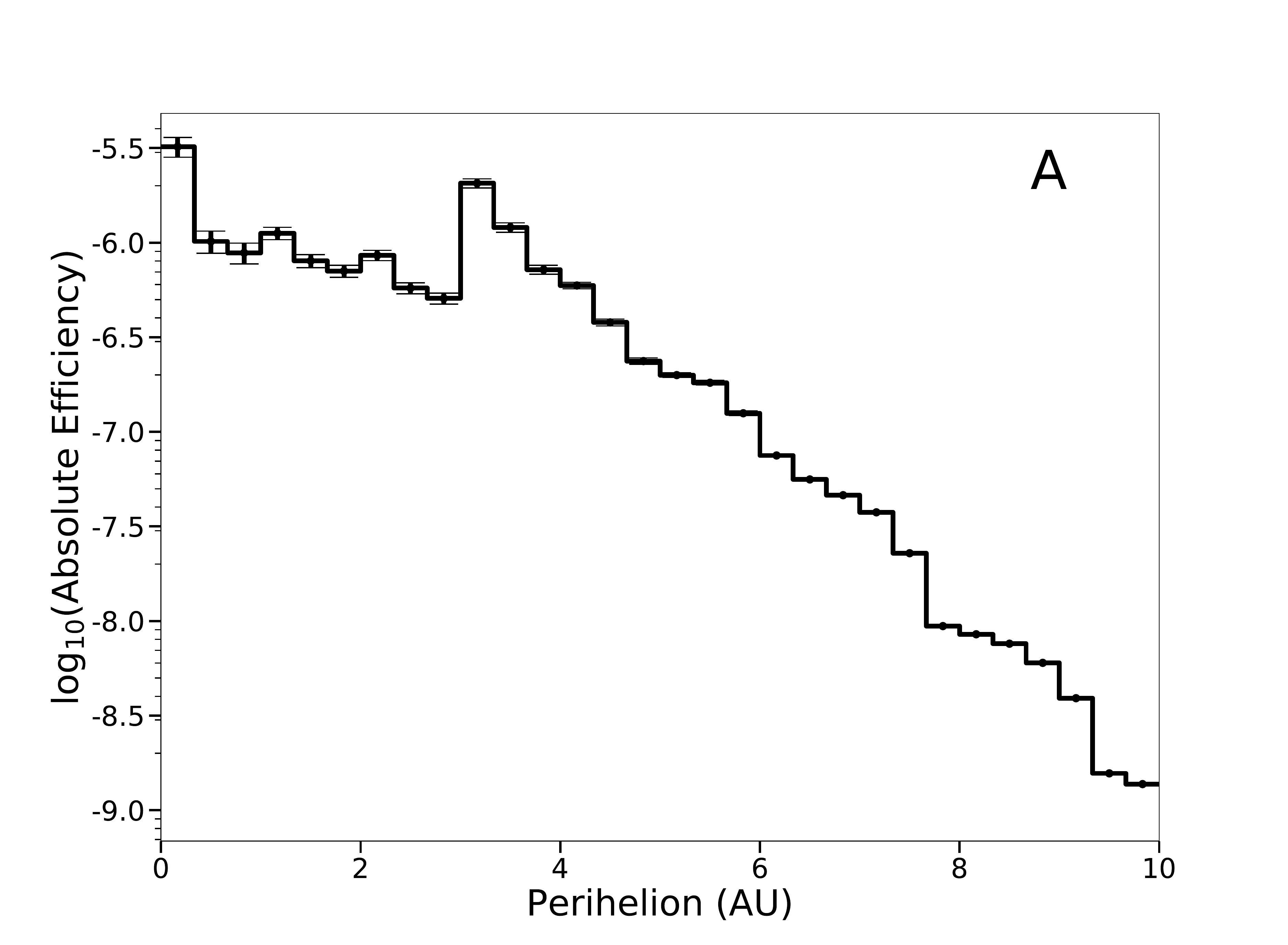}
	\includegraphics[width=0.5\textwidth]{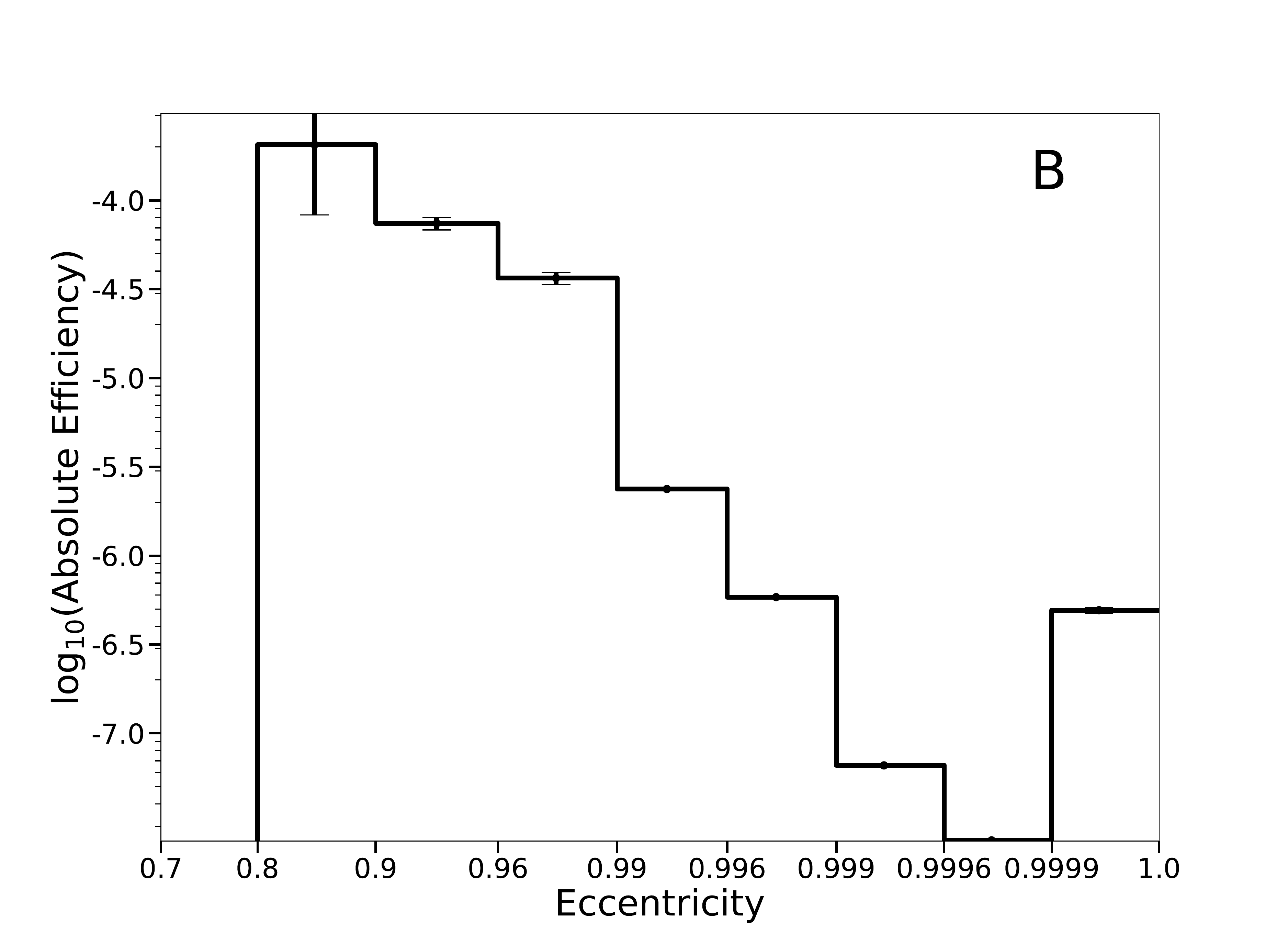}
	\includegraphics[width=0.5\textwidth]{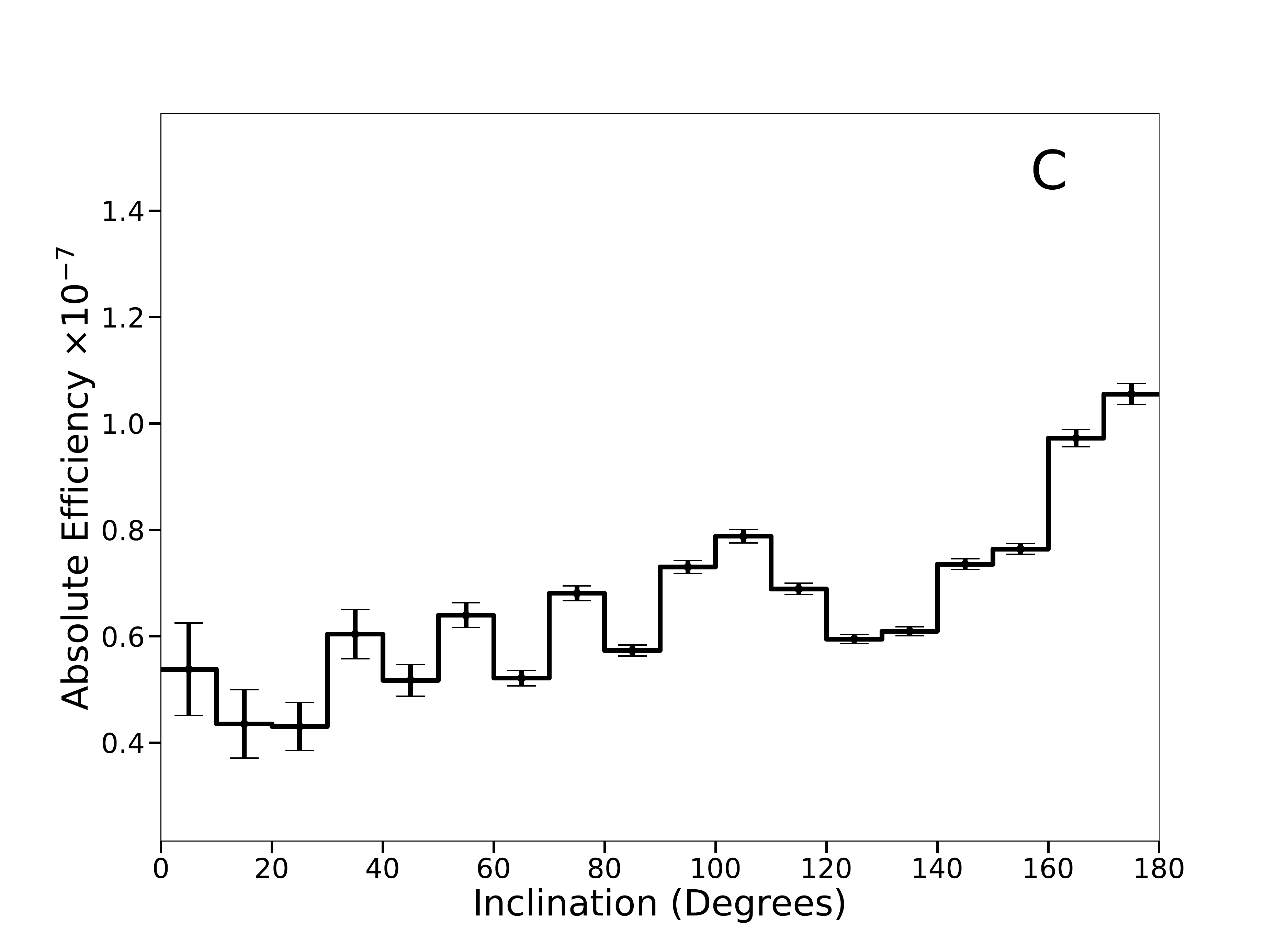}
	\includegraphics[width=0.5\textwidth]{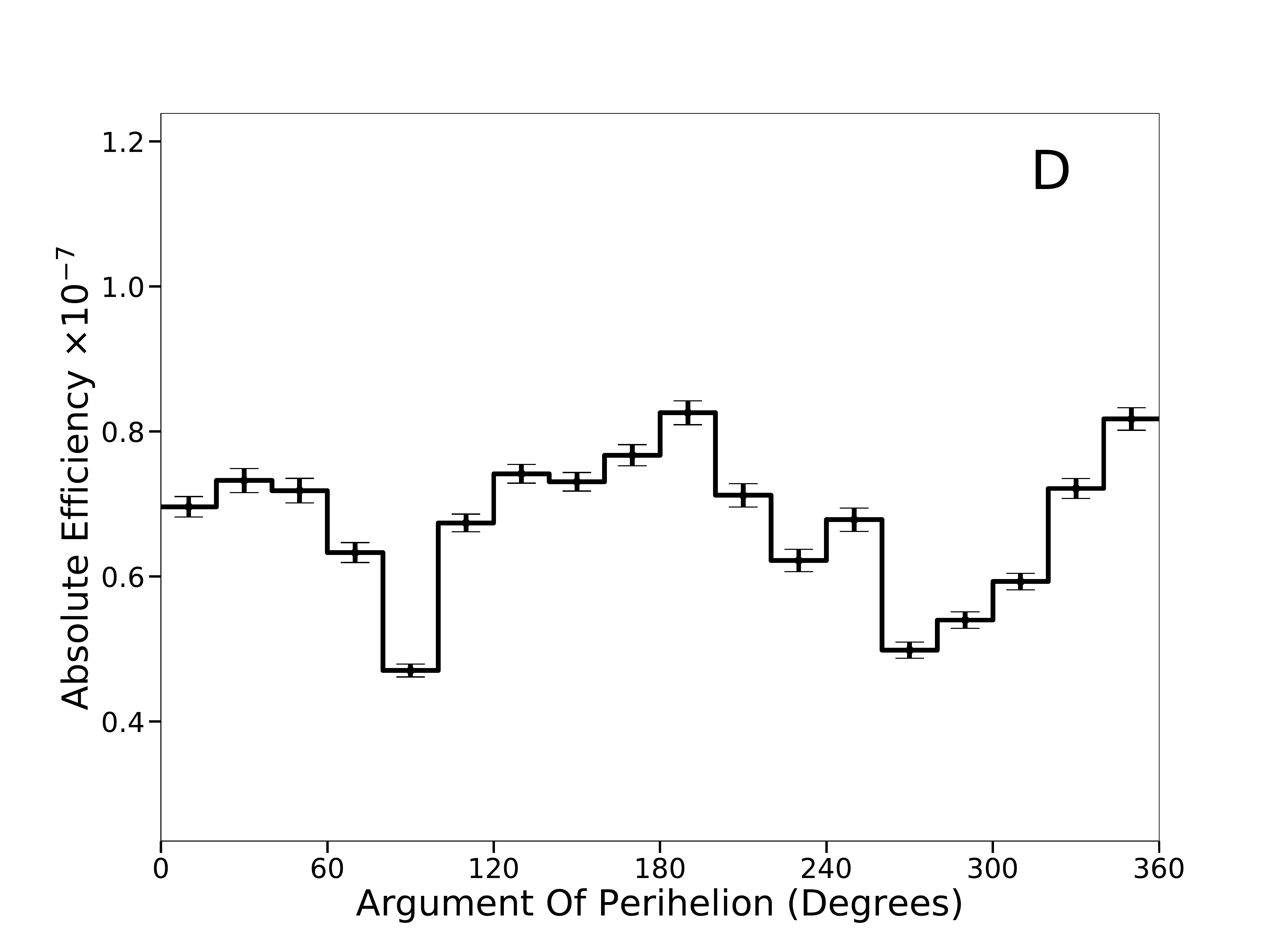}
	\includegraphics[width=0.5\textwidth]{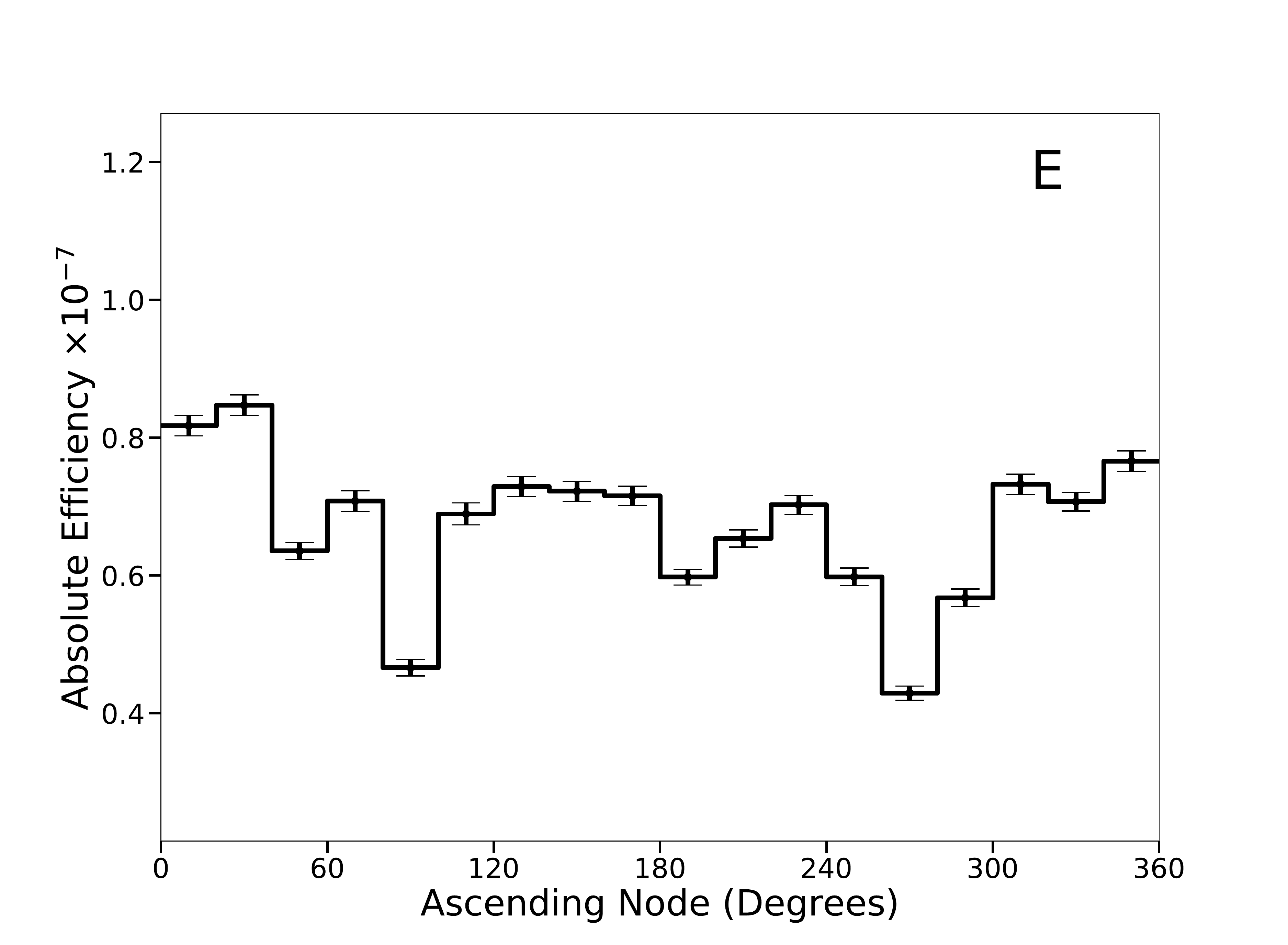}
	\includegraphics[width=0.5\textwidth]{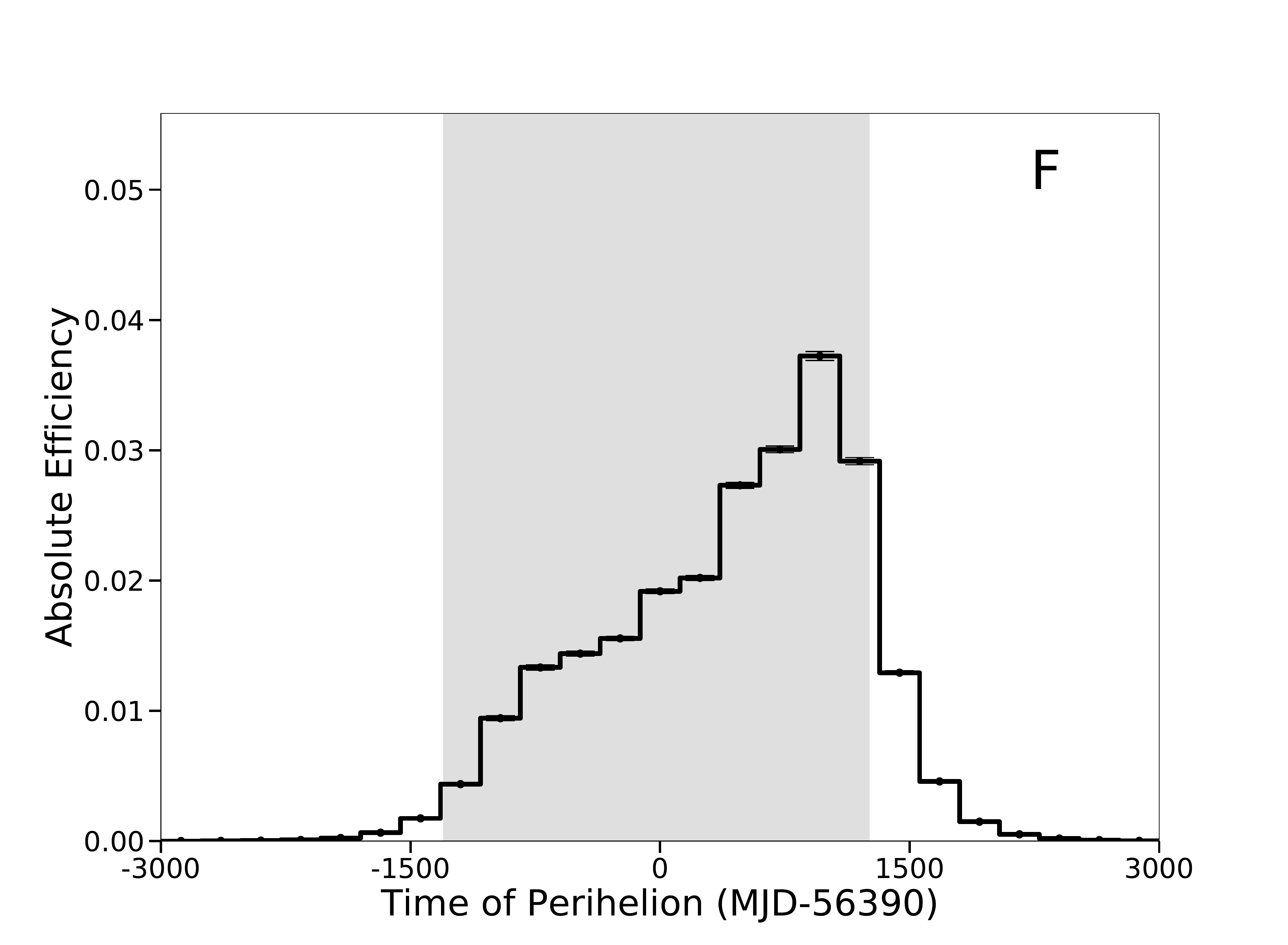}
	
	\caption{\PSone\ LPC absolute detection efficiency as a function of the six orbital elements.  Note the odd binning in eccentricity. The gray band in the time of perihelion panel represents the time period of the \PSone\ survey considered here.}
	\label{fig.LPCOrbitElementEfficiency}
	
\end{figure*}

The absolute LPC detection efficiencies as a function of their orbital elements are well behaved with, perhaps, one exception (\fig{fig.LPCOrbitElementEfficiency}).  The efficiency decreases roughly smoothly with perihelion distance in the $[\sim0.5\au,\sim10\au]$ range except for an enhancement between about $3\au$ and $4\au$.  We have not explored details of the origin of the enhancement but we are not surprised at the behaviour because there are several competing effects that combine to produce the overall system response.  In particular, our use of the MPC's \digesttwo\ score to select LPCs could introduce this behavior as it mostly measures the probability that a newly discovered object is not a main belt asteroid and 1) the outer edge of the belt is at about $3.3\au$ and 2) the region of the belt at and beyond $\sim3\au$ is i) distant and ii) dominated by dark objects.  The efficiency remains non-zero beyond about $6\au$ due to (limited) water sublimation at those heliocentric distances and because there are still some large objects that can be detected (recall that these efficiencies incorporate the corrected LPC SFD).  

Higher eccentricity LPCs are less efficiently detected than lower eccentricity objects by a few orders of magnitude primarily due to the fact that the smaller $e$ objects can only achieve those values through interaction with one of the planets in the solar system.  Interestingly, the LPC detection efficiency increases roughly linearly with inclination due to the \digesttwo\ score cuts since objects with high inclination are more likely to appear NEO-like and retrograde objects are more obviously interesting than prograde ones.  The average efficiency for $0\arcdeg \le i < 90\arcdeg$ is $5.5 \times 10^{-8}$ compared to $7.7 \times 10^{-8}$ for those objects with $i \ge 90\arcdeg$.  Both the argument of perihelion and ascending node distributions show reduced efficiency where the ecliptic passes through the winter and summer Milky Way ($\sim 90\arcdeg$ and $\sim 270\arcdeg$ respectively) but are otherwise roughly flat.  

The interpretation of the detection efficiency with respect to the LPC's time of perihelion requires some explanation.  In the case of a single-night all-sky survey this histogram would have a Gaussian shape centered on the survey night because it would have roughly equal probability of detecting LPCs $\pm N$ nights from their time of perihelion (ignoring the difference in pre- and post-perihelion activity levels).  The \PSone\ survey can broadly be divided into two periods before and after 2014 March 28 (MJD 56744) corresponding respectively to part-time and nearly-dedicated surveying for moving objects.  During the two time periods the LPC discovery efficiency was different, but roughly constant, due to the amount of surveying time.  Thus, \fig{fig.LPCOrbitElementEfficiency}F can be understood as the convolution of the Gaussian single-night efficiency with a double-plateau flat efficiency function corresponding to the two different time periods.  The shape of the realized efficiency with respect to time of perihelion is further complicated by the continual but modest increase in detection efficiency due to ongoing software, hardware, and operational improvements.

\begin{figure}[htbp]

	\includegraphics[width=0.5\textwidth]{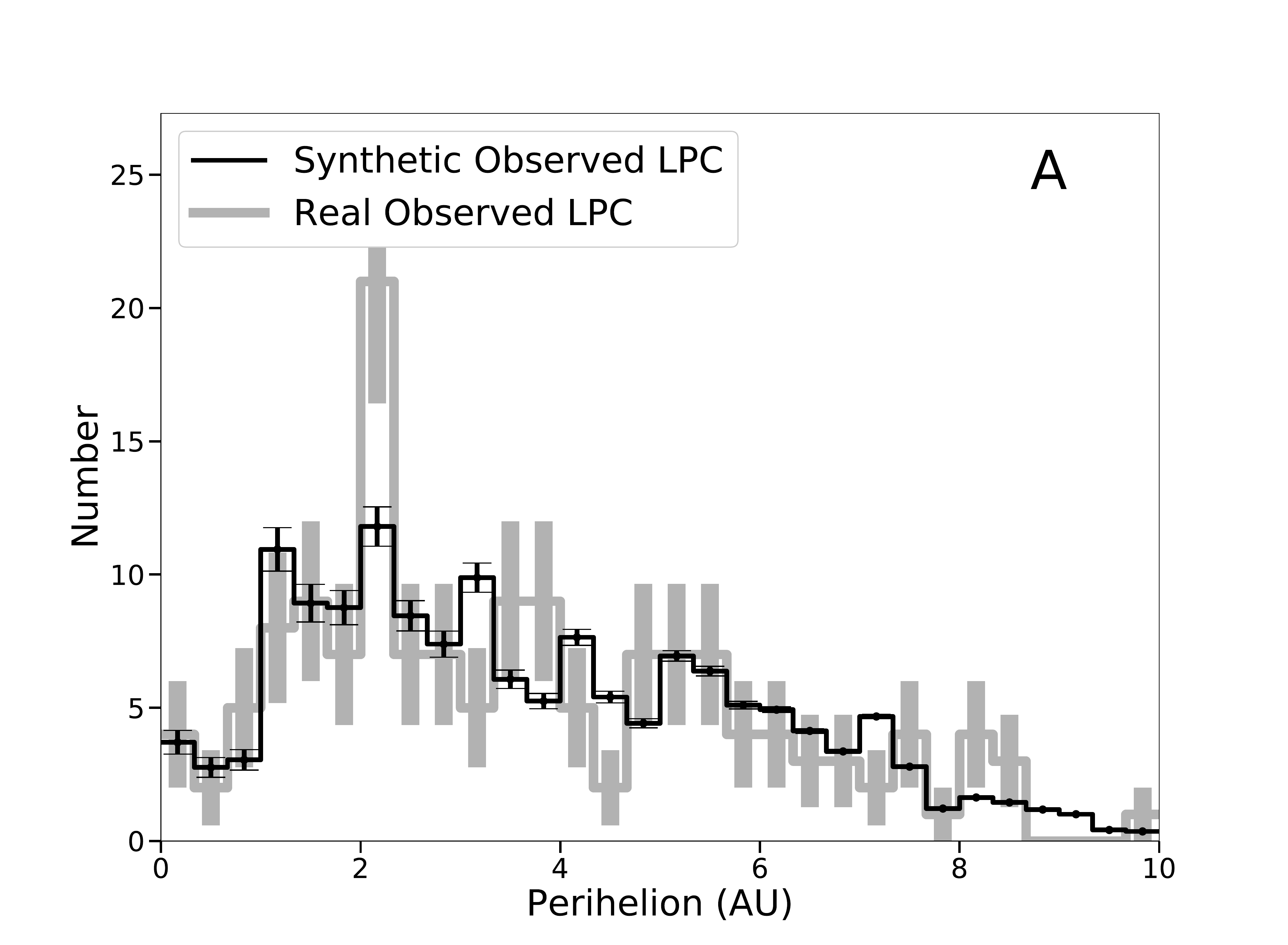}
	\includegraphics[width=0.5\textwidth]{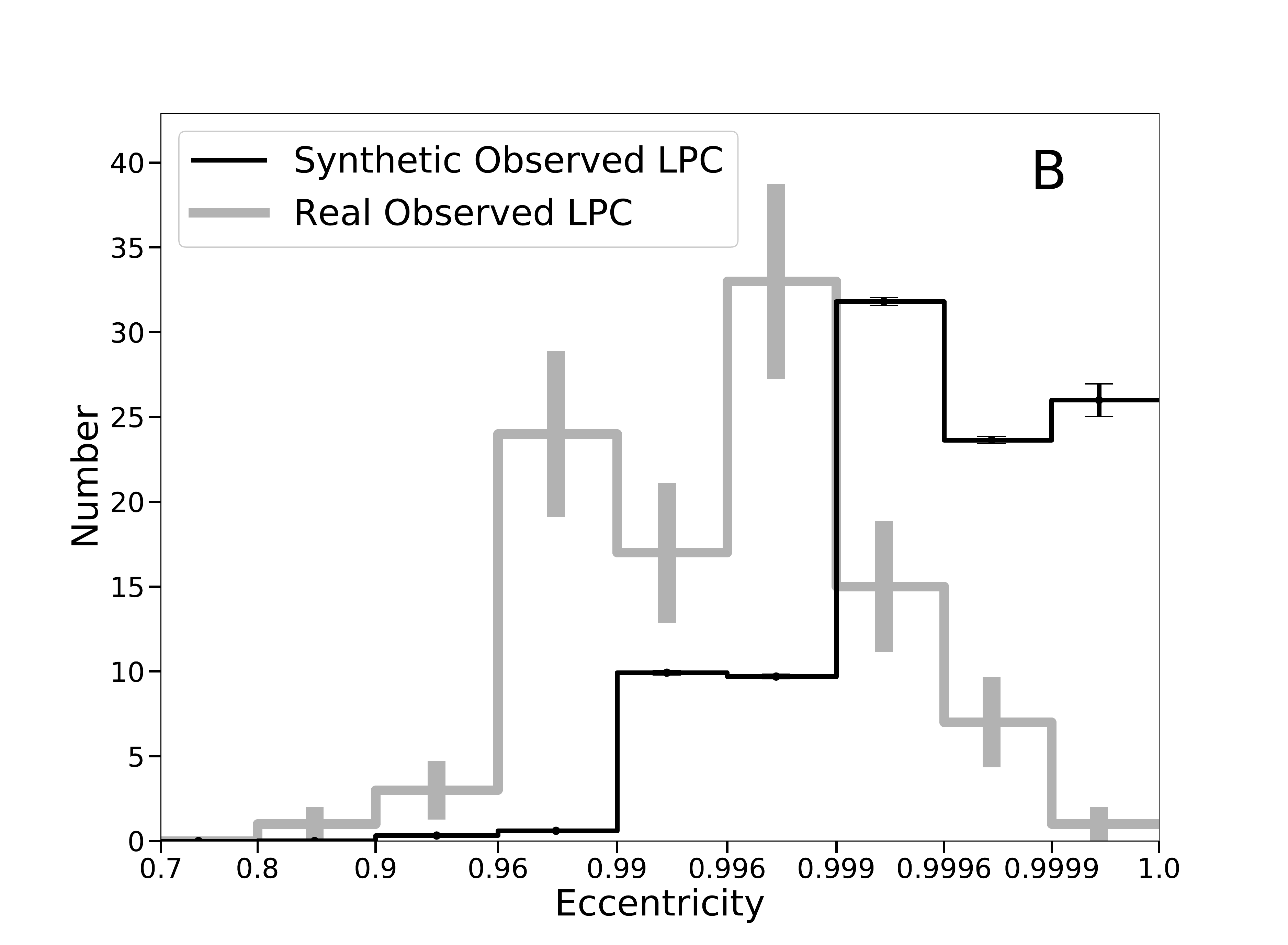}
	\includegraphics[width=0.5\textwidth]{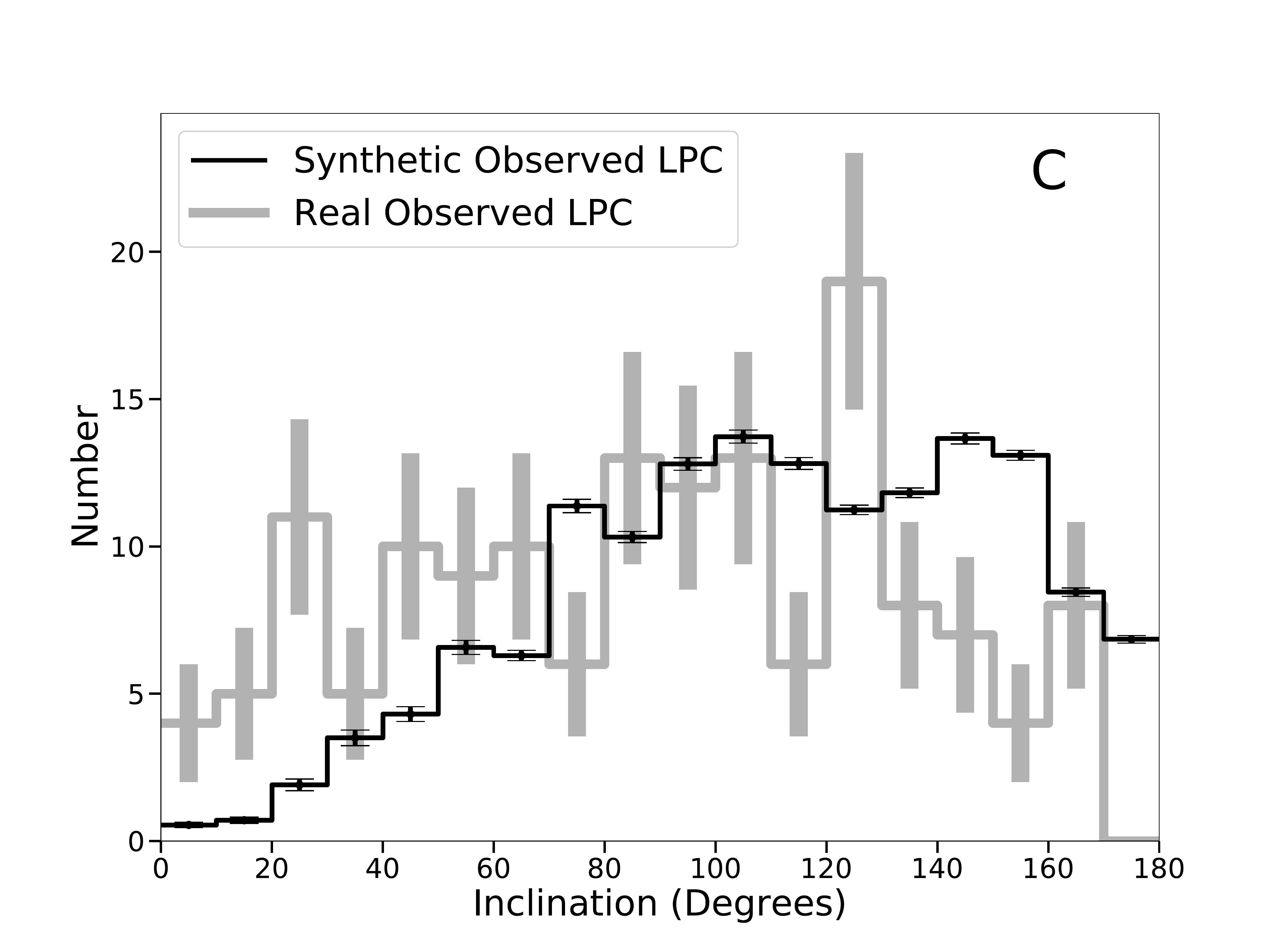}
	\includegraphics[width=0.5\textwidth]{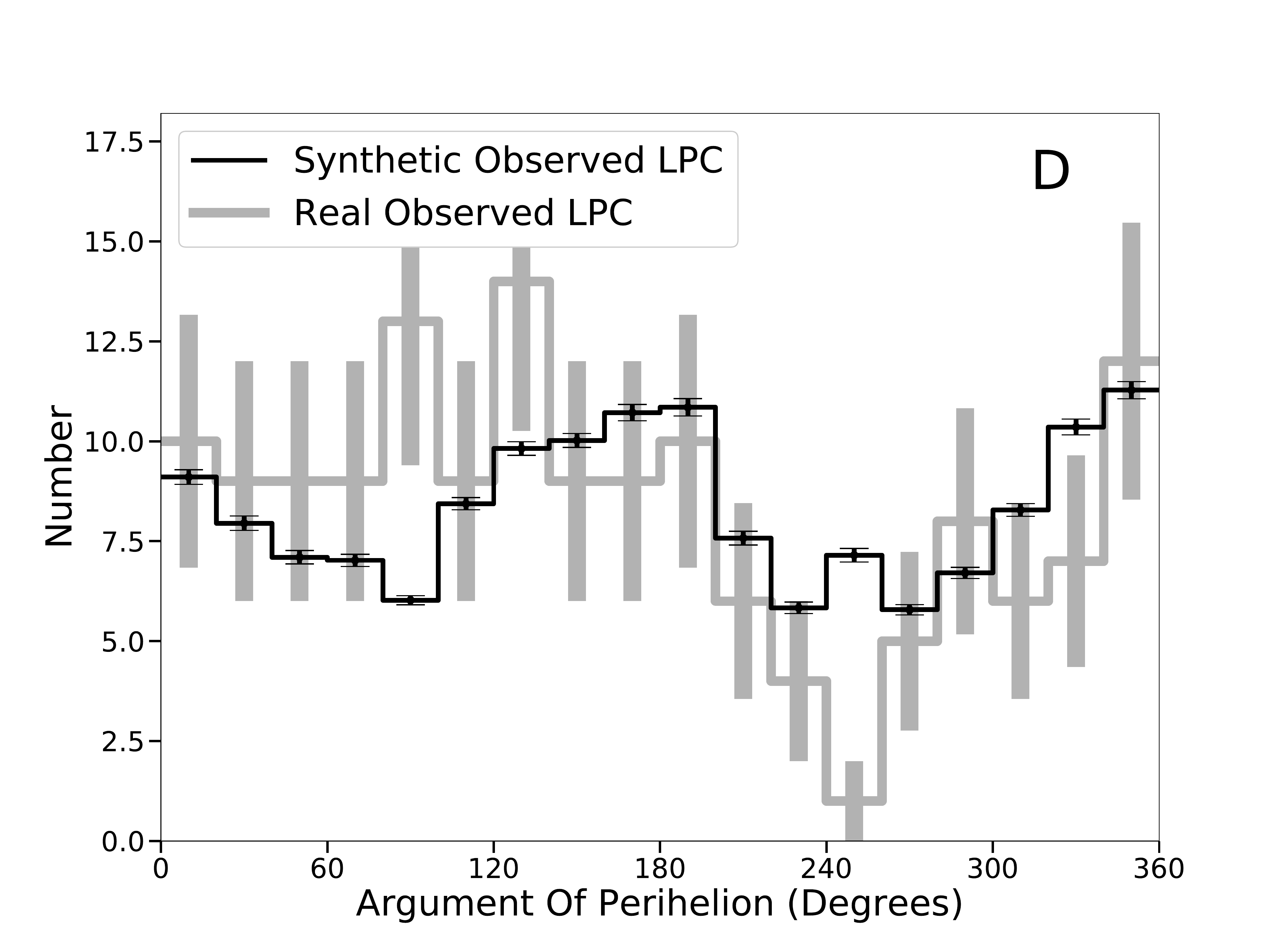}
	\includegraphics[width=0.5\textwidth]{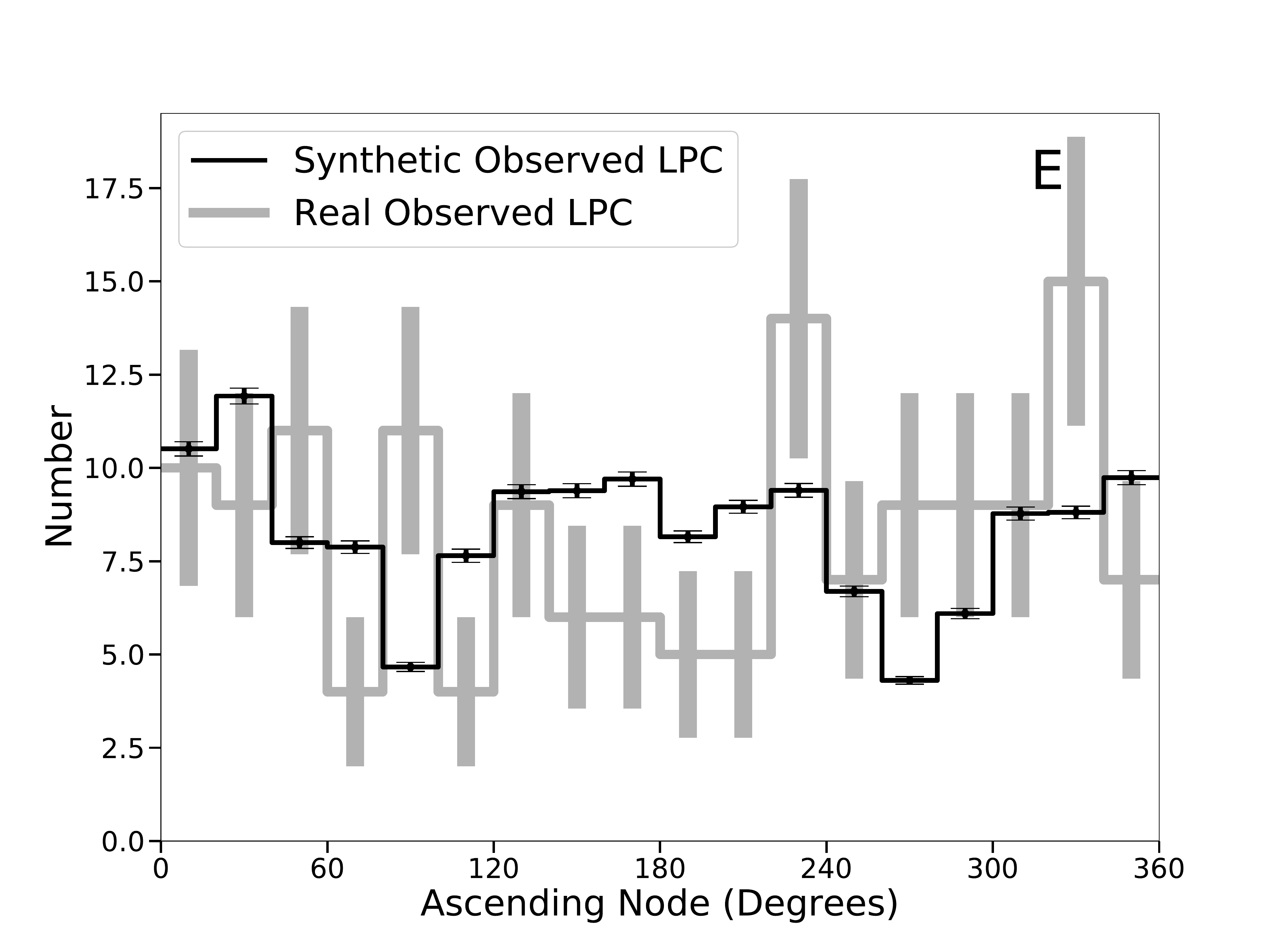}
	\includegraphics[width=0.5\textwidth]{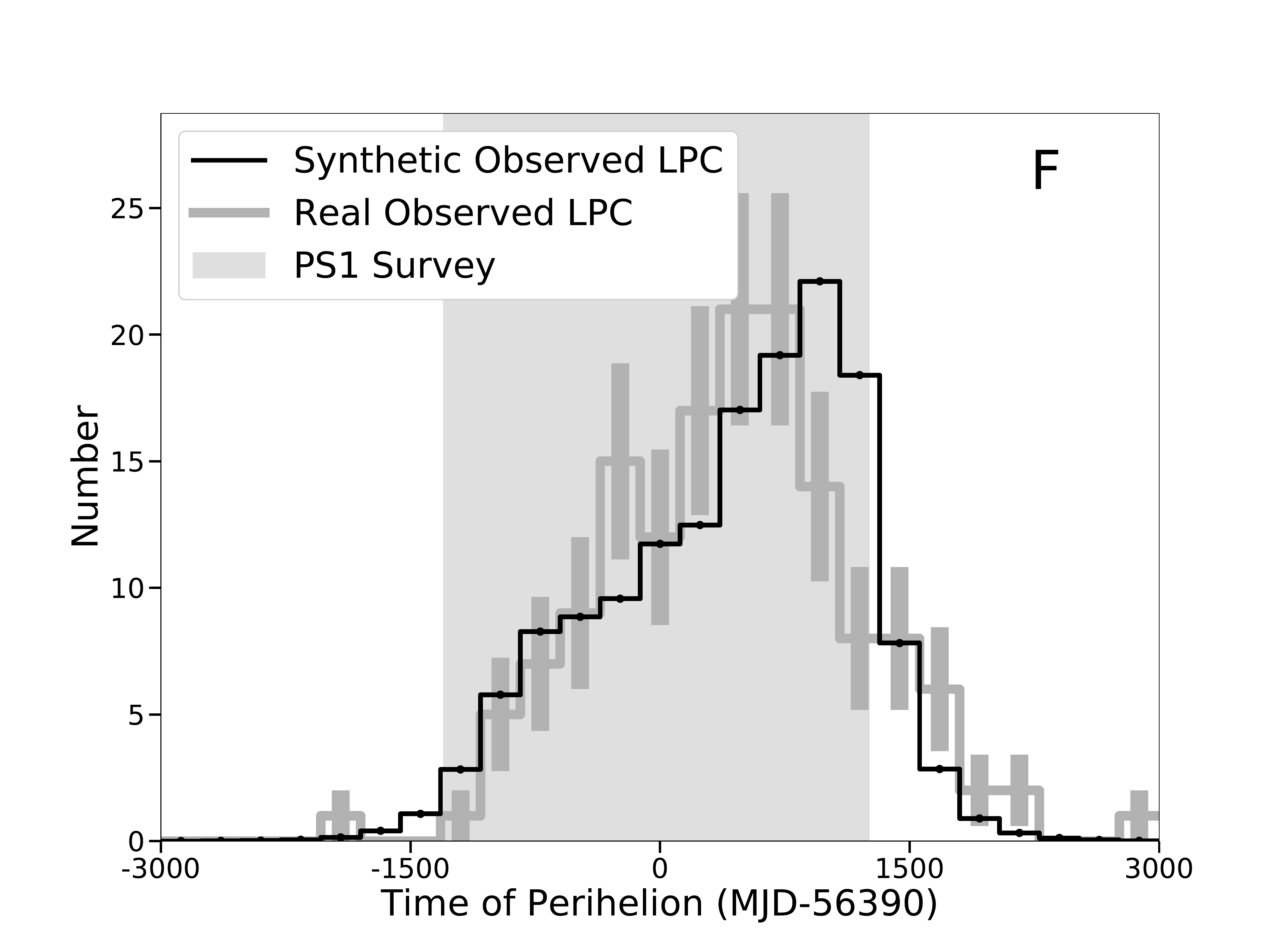}

	\caption{(gray line) Distribution of orbital elements for LPCs detected by \PSone\ from 2010-02-23 through 2016-12-22 and used in this analysis.  (dark line) Distribution of orbital elements for synthetic LPCs detected in the simulation after `fitting' and iterating on the SFD as described in \S\ref{sss.LPCDetectionEfficiency}.  All panels except for the eccentricity distribution (panel B) contain the values for 150 LPCs.  The eccentricity panel contains only 102 values because the other LPCs have measured eccentricities $>1$.  Note the unusual binning in the eccentricity that intentionally highlights the details of the $e\lesssim1$ distribution.
	}
	
	\label{fig.Synthetic+Real-Observed-LPCOrbitalElements}
	
\end{figure}

The efficacy of the efficiency calculation is intimately connected to the agreement between the real observed LPC orbital element distributions and the synthetic \PSone\ LPC observations (\fig{fig.Synthetic+Real-Observed-LPCOrbitalElements}).  We consider the match to be good considering the difficulty in modeling an `average' LPC and the long term operations of an evolving NEO survey.  (We provide formal metrics for the agreement below.)  That being the case, the perihelion ($q$) and time of perihelion ($t_q$) distributions are in good agreement.  Similarly, the argument of perihelion ($\omega$) and ascending node ($\Omega$) distributions are also in good agreement, both being nominally flat except for the explicable dips near the winter and summer Milky Way as discussed above.  The eccentricity distribution (\fig{fig.Synthetic+Real-Observed-LPCOrbitalElements}B) may appear to be in disagreement due to the unusual histogram binning.  In fact, both the model and the data are in good agreement as virtually all the real and synthetic LPCs have $e>0.9$ and any apparent disagreement occurs in bins with $e \gtrsim 0.96$.  The apparent discrepancy is likely due to excluding 48 LPCs with $e>1$ from the panel as described above.  

The orbital element that is clearly not in good agreement is the inclination (\fig{fig.Synthetic+Real-Observed-LPCOrbitalElements}).  The real observed LPCs exhibit a $\cos(i)$-like distribution despite observational selection effects while the synthetic LPC detections are skewed towards retrograde objects.  In the real data $51 \pm 7$\% of the objects are retrograde compared to $78.7 \pm 0.4$\% in the synthetic data.  It is possible that the discrepancy is due to unmodeled observational selection effects or to our input synthetic LPC model.  Indeed, the \citet{Fouchard2017b} LPC model (\fig{fig.Synthetic-vs-Corrected-Population}) would probably provide a better match to the observed LPC inclination distribution but not as good a match to other orbital parameters.  

We formally characterize the agreement between the model and the data in the individual orbital parameters and the multi-dimensional model in the next section.

\subsection{Simulation verisimilitude}
\label{ss.SimulationVerisimilitude}

\begin{figure*}[htbp]

	\includegraphics[width=0.5\textwidth]{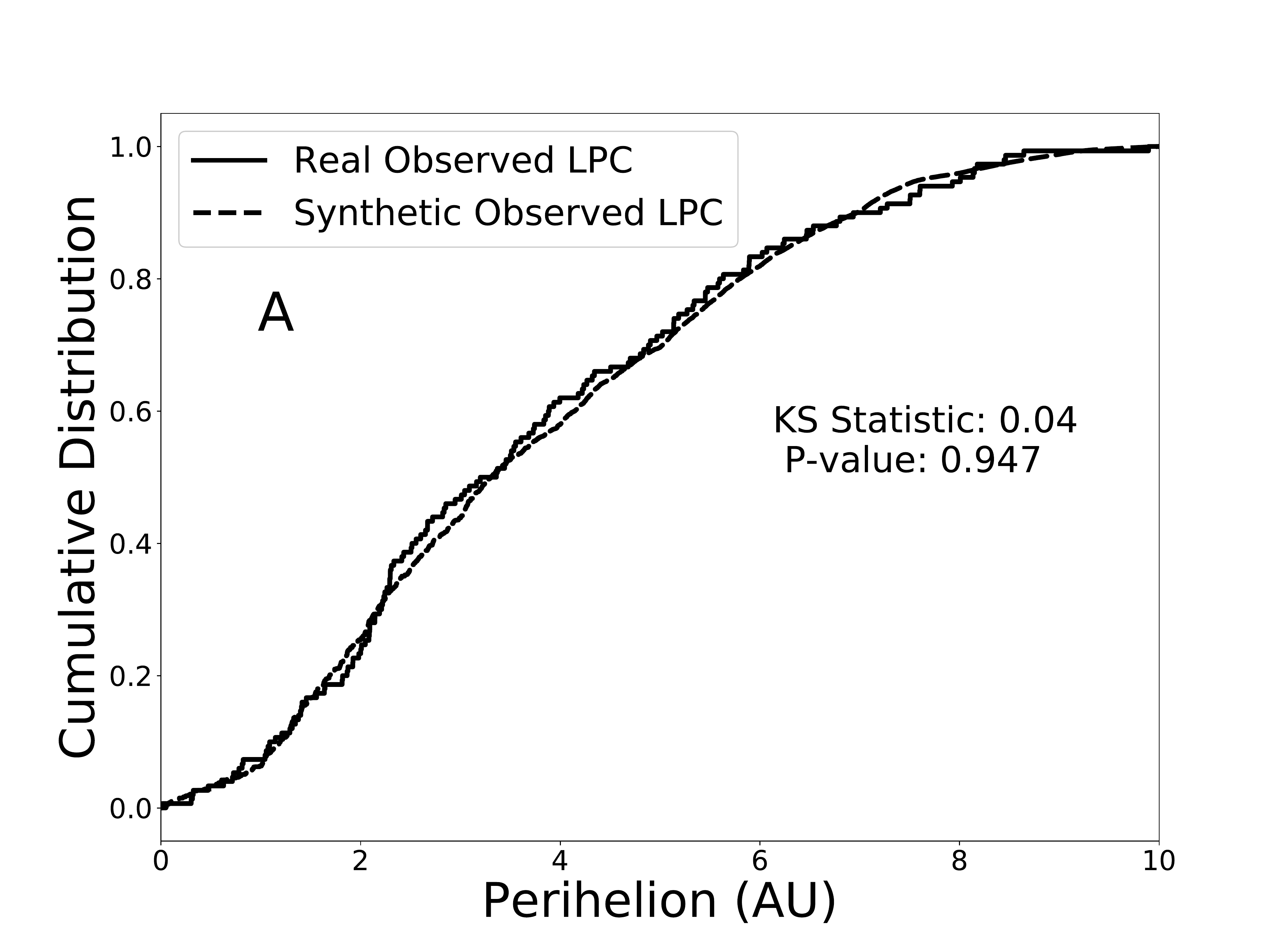}
	\includegraphics[width=0.5\textwidth]{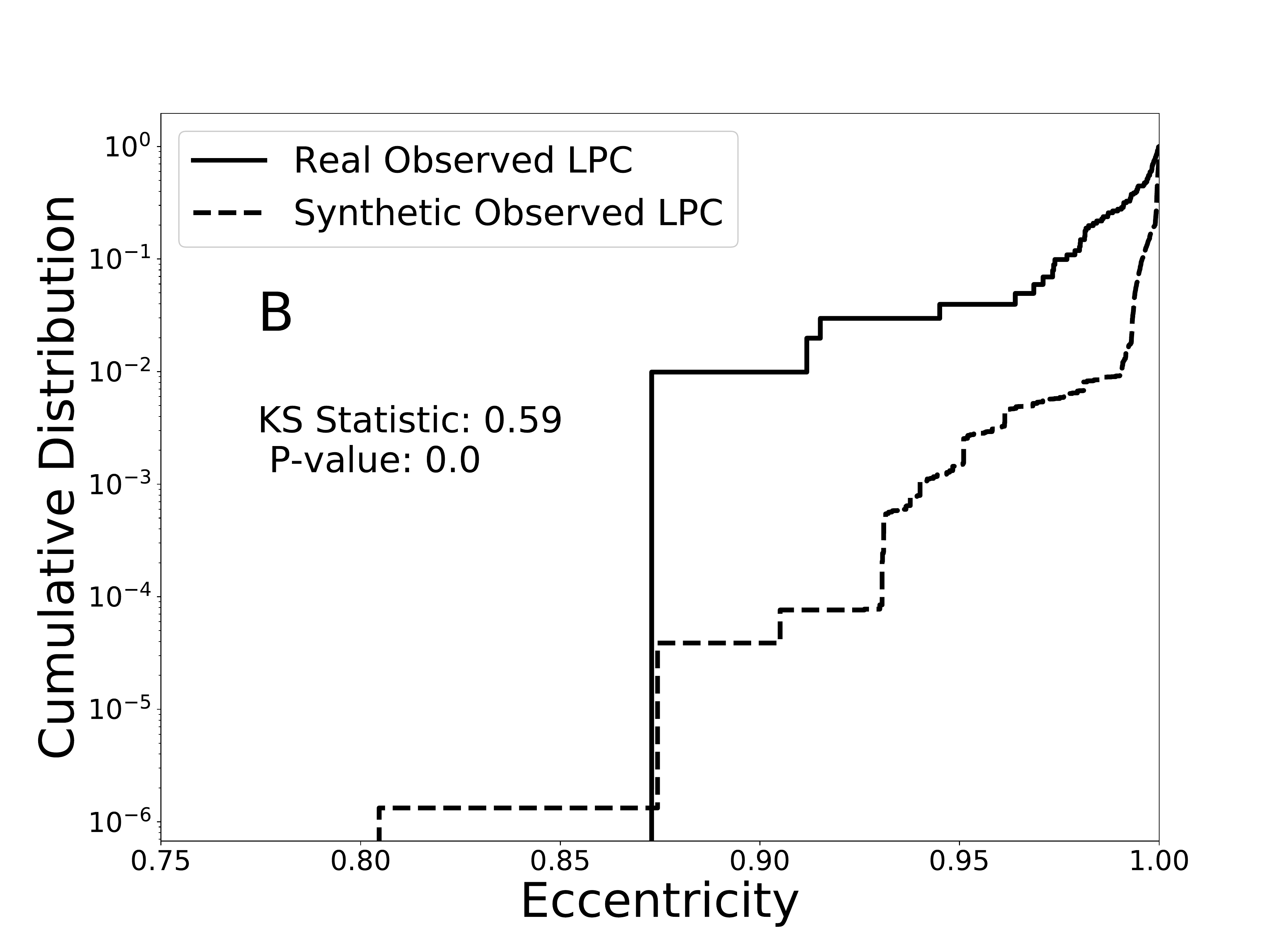}
	\includegraphics[width=0.5\textwidth]{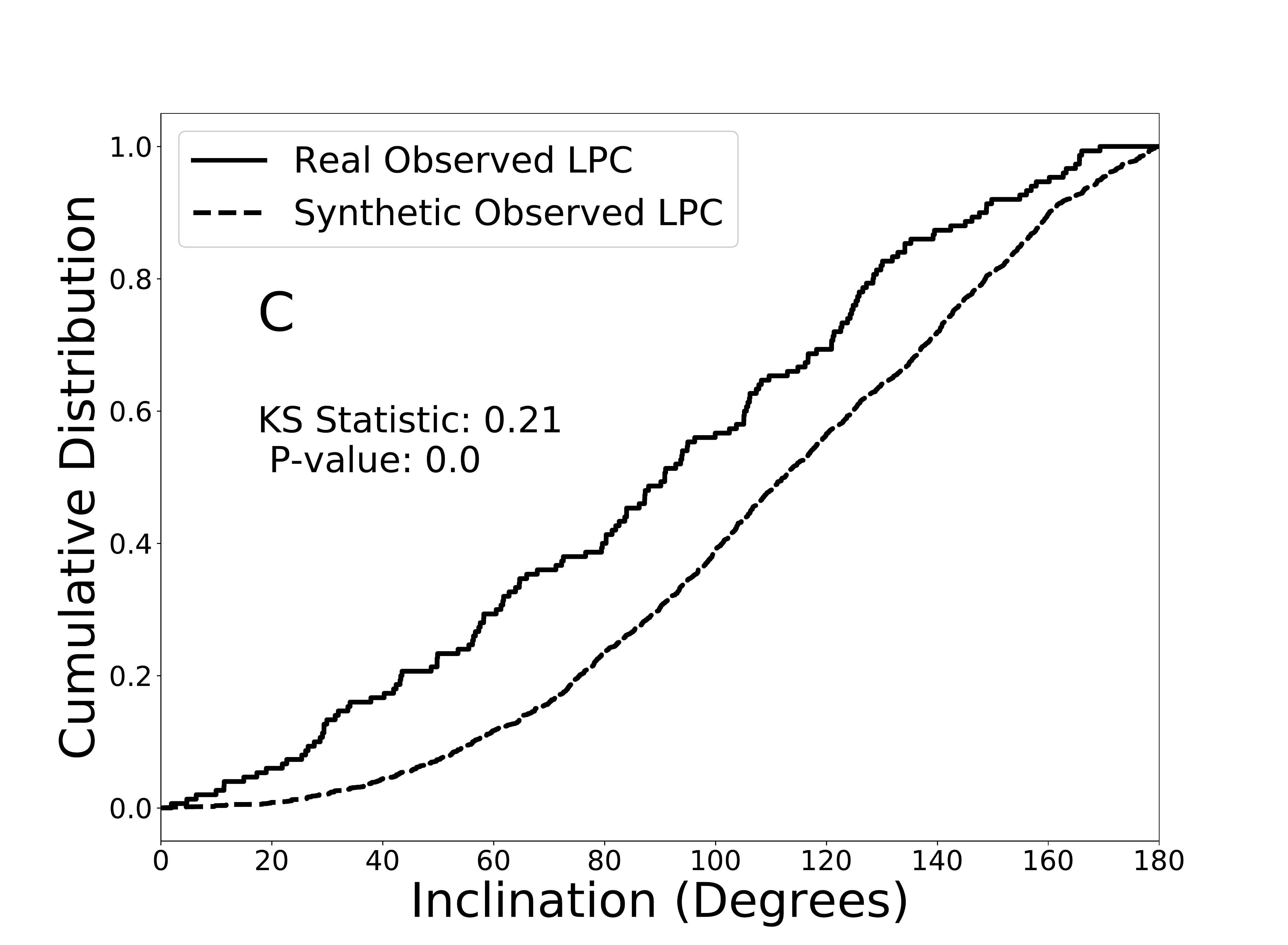}
	\includegraphics[width=0.5\textwidth]{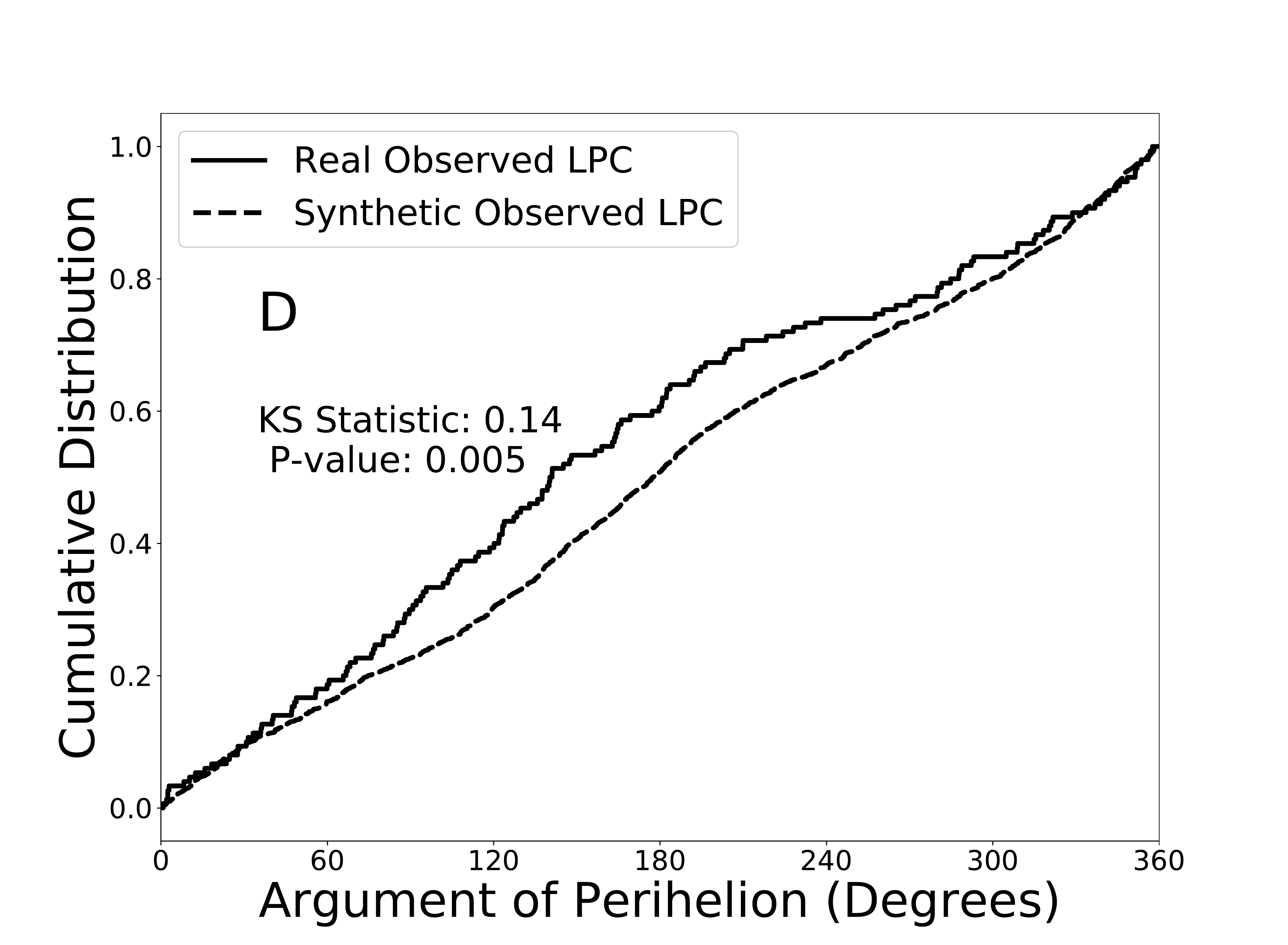}
	\includegraphics[width=0.5\textwidth]{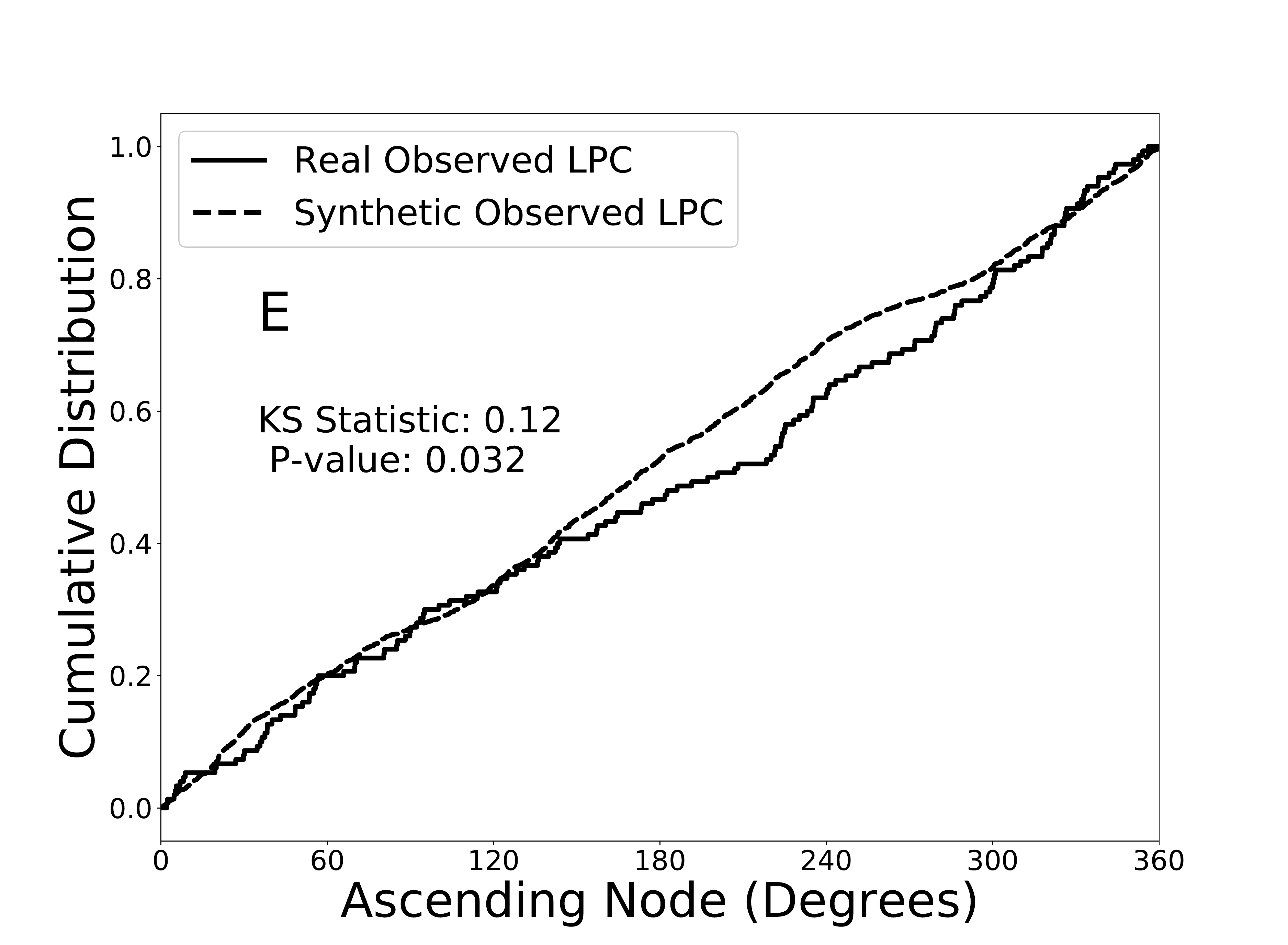}
	\includegraphics[width=0.5\textwidth]{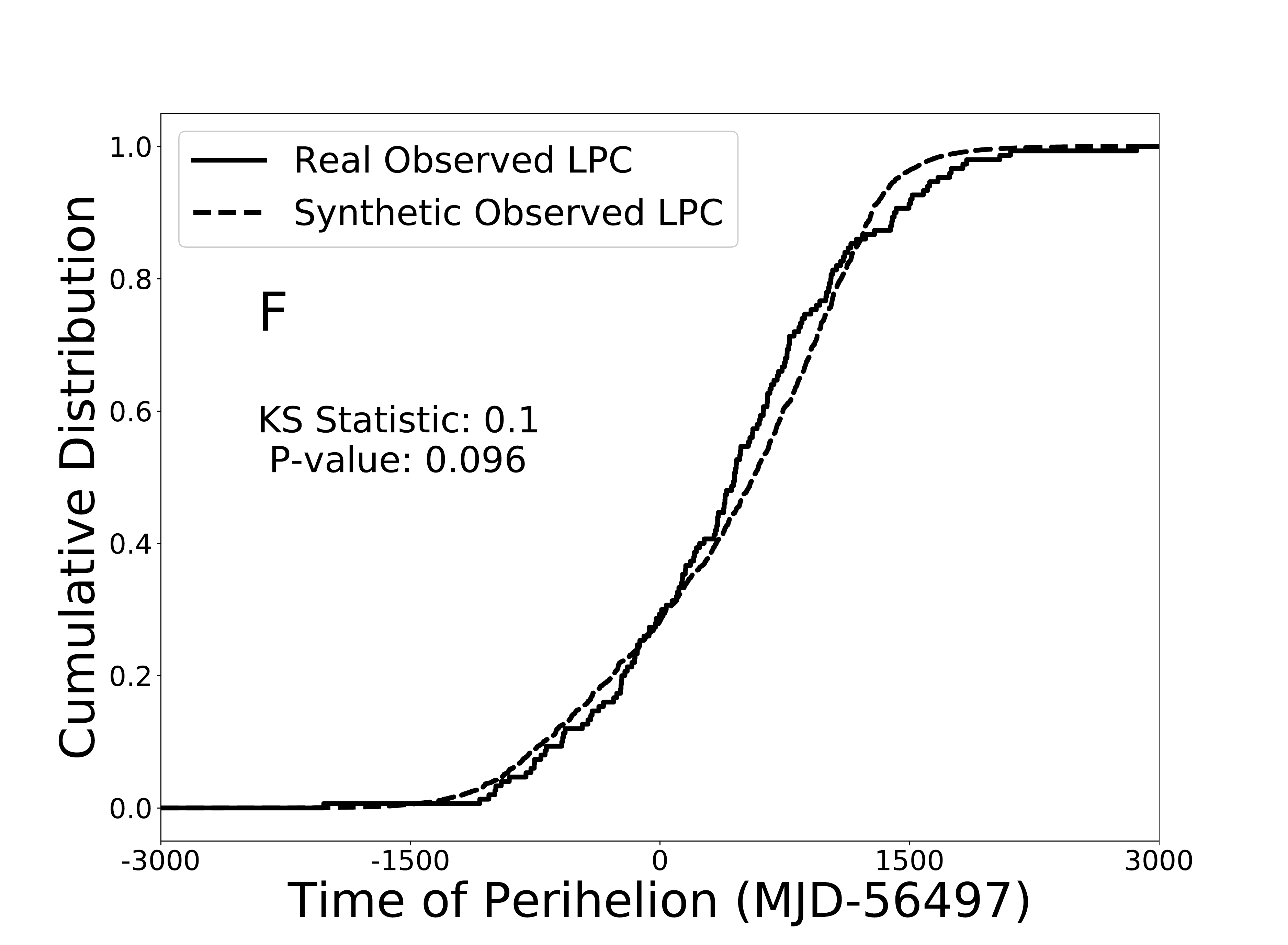}
	
	\caption{Normalized cumulative distributions for each of the six orbital elements for both the real and synthetic LPC population. Each also provides the KS-statistic, representing the maximum difference between the two distributions, and the $p$-value which is the probability that the null hypothesis (that the two populations derive from the same parent population) cannot be rejected. Typically the null hypothesis cannot be rejected if $p > 0.05$.}
	\label{fig.LPC-KS-tests}

\end{figure*}

The KS test on each of the six orbital elements indicates that the real and synthetic observed perihelion and time of perihelion distributions are in formal agreement\footnote{The null hypothesis cannot be rejected if $p > 0.05$} (\fig{fig.LPC-KS-tests}).  We consider the concordance a vindication for the entire simulation because the observed $q$ distribution is the convolution of 1) the actual LPC spatial (orbital) distribution, 2) the details of the sublimation model that is strongly dependent on heliocentric distance, 3) our use of `average' cometary parameters, and 4) the survey simulation, and it is unlikely that we have chosen wrong values in exactly the right combination to deliver the correct perihelion distribution.  Indeed, there are few `tunable' parameters to ensure any agreement at all.  As discussed above, we consider the agreement between the other distributions ($e$, $i$, $\omega$, $\Omega$) to be good, except in inclination (we will discuss that discrepancy below), even though our synthetic distributions are not formally in agreement with the observed ones.

\begin{figure}[htbp]

    \centering
    
    \includegraphics[width=0.45\textwidth]{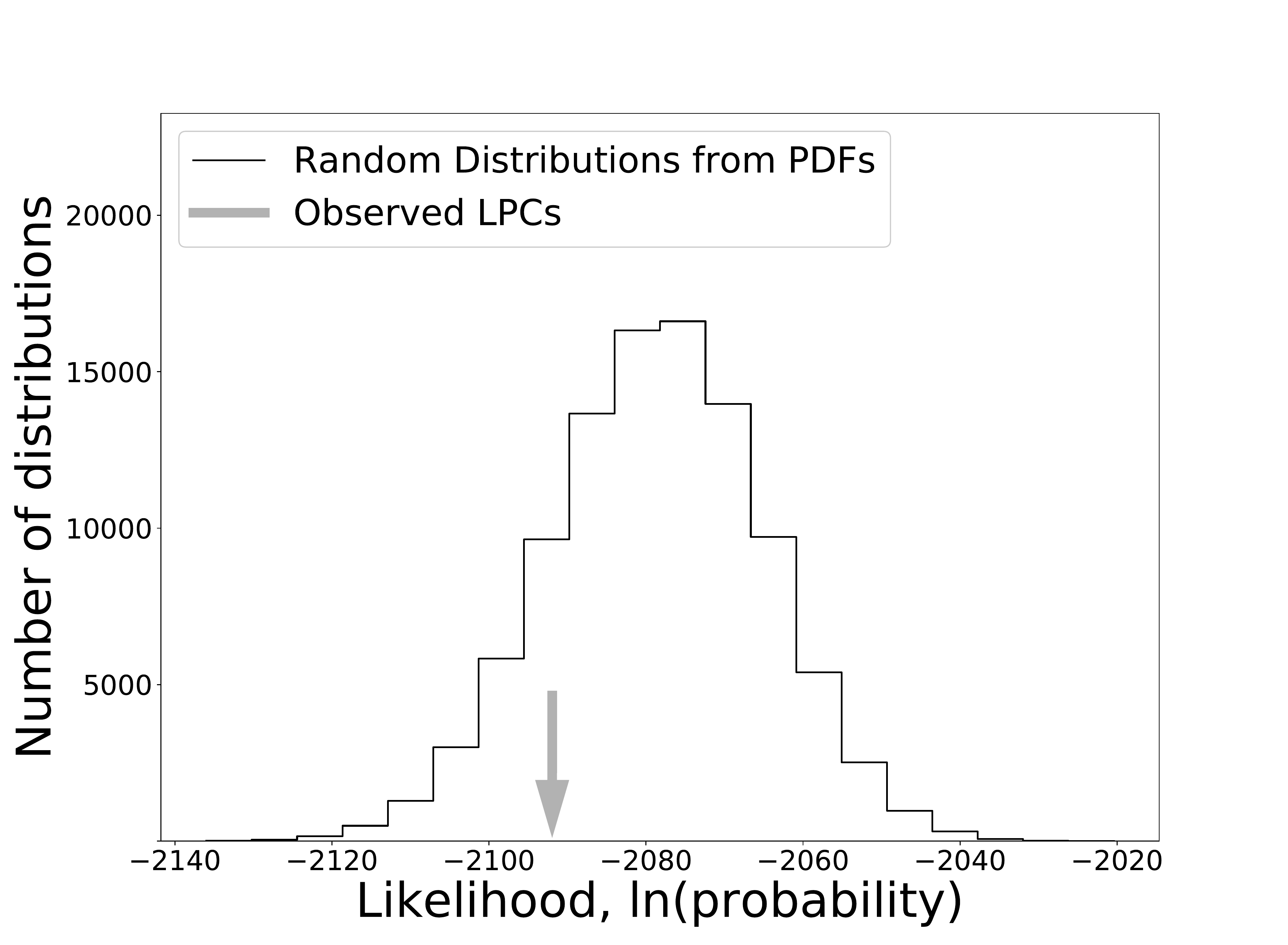}
    \includegraphics[width=0.45\textwidth]{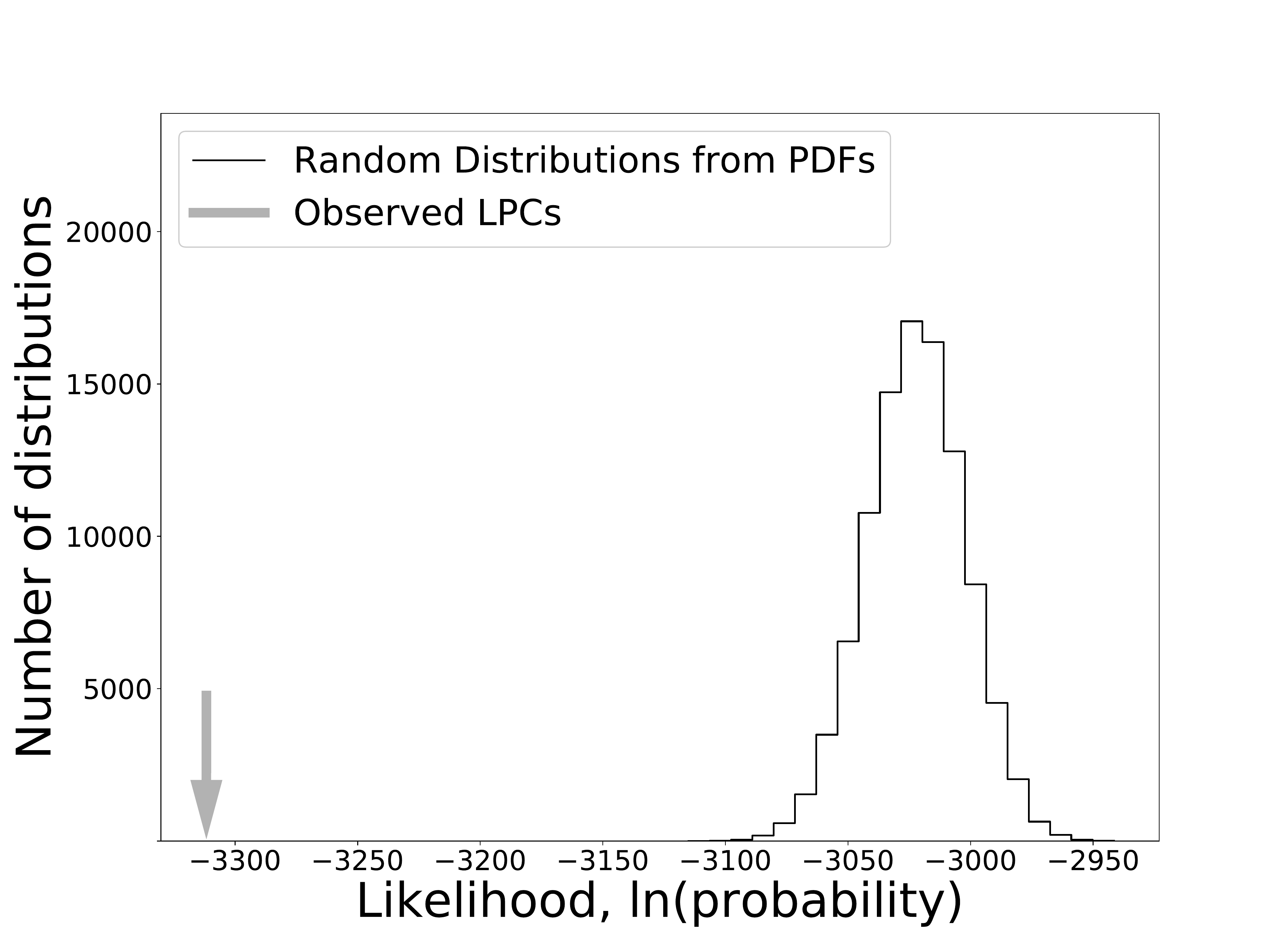}
    
    \caption{
        Log-likelihood of $10^5$ random orbital element distributions generated according to the parametric PDFs of the synthetic detected population with the same number of objects (150) as the observed LPC population (with measured absolute magnitudes). The left panel includes perihelion, time of perihelion, argument of perihelion, ascending node, and the size distribution; the right panel includes the same with the addition of eccentricity, inclination, and aphelion. The log-likelihood of the real observed population given the synthetic model is indicated with a grey arrow in each. 
    }
    
    \label{fig.LogLikelihood}

\end{figure}

To characterize the overall agreement between the real \PSone\ LPC observations and our final synthetic observed model which includes the LPC input model, the \PSone\ detection algorithm and survey simulation, and the sublimation modeling, we calculated the likelihood that the multi-dimensional model agrees with the data.  The synthetic observed orbital element distributions were binned and normalized to create 1-d probability distribution functions (PDFs) that were used to generate $10^5$ different populations of 150 synthetically observed LPCs --- the same number as the real population of LPCs observed by \PSone\ (\ie\ 150 objects have orbital elements).  Letting $P_{ij}$ represent the probability of an object appearing in bin $j$ of the distribution for orbital element $i$, the log-likelihood of each synthetic population is
\begin{equation}
    \mathcal{L}^{syn} = \sum_i \sum_j \ln(P_{ij}n_{ij}^{syn})  \Big{|}_{n_{ij}^{syn} \neq 0}
    \label{eqn.LogLikelihood}
\end{equation}
where $n_{ij}^{syn}$ is the number of synthetic detected objects in the given bin. The sum represents the probability of getting a particular population distribution based on the parametric PDFs, and the distribution of the log-likelihoods for a large sample of synthetic populations provides a quantitative estimate of the expected variation in the sample (\fig{fig.LogLikelihood}).  We then calculate the log-likelihood of the real observed population ($\mathcal{L}^{real}$) given the model letting $n_{ij}^{syn} \rightarrow n_{ij}^{real}$ in \eqn{eqn.LogLikelihood}.  If our overall modeling effort accurately reflects reality we would expect that $\mathcal{L}^{real} \sim \mathcal{L}^{syn}$. We find that the real population and our model agree at about the $1.0-\sigma$ level when we consider only the PDFs for perihelion, time of perihelion, argument of perihelion, ascending node, and the size distribution.  However, $\mathcal{L}^{real}$ is about $\sim 14.3-\sigma$ from the predicted value when we also include eccentricity, aphelion and inclination.  Thus, there is room for improving our overall understanding of the LPC population and our results should be interpreted as a step along that path.

\subsubsection{Corrected LPC size distribution}
\label{sss.CorrectedLPCSizeDistribution}

Our debiased LPC SFD for $13<H_N<25$ (\fig{fig.HEfficiency+HDistribution}) exhibits the long-known transition from a steep slope for large objects to a shallow slope for small objects \citep[\eg\ \eqn{eqn.MeechSFD}; ][]{Meech2004}.  This shape is often fit to a discontinuous function of two lines with different slopes (in $\log$ space) but, following \citet{Granvik2017-NEOSourceRegions}, we introduce a continuous function with similar properties 
\begin{equation}
    n(H_N) = n_1 \; 10^{ (H_N-H_1) \; \int_{-\infty}^{H_N} \alpha(H_N') \; \dif H_N' }
\end{equation}
where
\begin{equation}
    \alpha(H_N) = \alpha_{small}+(\alpha_{big}-\alpha_{small}) \;
                \Biggl[ 1 + \exp\Bigl[ \frac{H_N-H_{trans}}{H_{width}} \Bigr] \Biggr]^{-1}.
\end{equation}
This function has a slope of $\alpha_{big}$ for `big' objects and a slope of $\alpha_{small}$ for small objects where the transition between the two size regimes occurs roughly over the range $H_{width}$ centered at $H_{trans}$.  

We fixed $H_1=17$, corresponding to a $\sim 2.6\km$ diameter LPC with 4\% albedo, so the fitted value of $\log(n_1) = 8.04  \pm 0.04 /\mags$ is the number density at that absolute nuclear magnitude.  We find $H_{trans} = 16.9 \pm 0.3$ and $H_{width} = 0.6 \pm 0.3$ so the transition from `big' to `small' comets occurs at $\sim 2.8\km$ diameter.  Our big LPC SFD slope is $\alpha_{big} = 0.7 \pm 0.1$, much steeper than the asymptotic slope for the smallest LPCs with $\alpha_{small} = 0.07 \pm 0.03$ that is not consistent with a slope of zero.  The `big' LPCs are roughly objects with $H\lesssim 16.3$ (\ie\ $H_{trans}-H_{width}$ or roughly $\gtrsim3.7\km$ diameter) while the `small' LPCs are objects with $H\gtrsim 17.5$ (\ie\ $H_{trans}+H_{width}$ or $\lesssim 2.1\km$ diameter).

The corresponding slopes measured by \citet{Meech2004} for `bare' LPC nuclei are $0.49\pm0.01$ for objects with $2\km \le D < 20\km$ and $0.58\pm0.01$ for $4\km \le D \le 10\km$.  Thus, the entire \citet{Meech2004} size range corresponds to our `big' objects and our slope for objects in that diameter agrees with their values at the $1$-$\sigma$ and $2$-$\sigma$ levels.  Their data also showed an extreme drop in the slope for objects $\lesssim 1\km$ but they were unable to account for observational selection effects in that size range whereas our technique provides us with measurements and detection efficiency for sub-km objects.

A more recent determination of the debiased LPC SFD is significantly shallower than our distribution with $\alpha=0.200\pm0.014$ for $1\km<D<20\km$ \citep{Bauer2017}.  It is difficult to reconcile this value with our slope because, even allowing for the combined statistical+systematic uncertainty in our value (\S\ref{sss.SystematicUncertainties}), the two values  differ by $3$-$\sigma$.  Furthermore, these space-based detections of 35 LPCs using observations in the IR by the NEOWISE mission showed no evidence of a decrease in the number of LPCs at smaller sizes.  This observation might be explained as a consequence of their small statistics in the smallest size bins as we detect a change in the slope for $D\lesssim3\km$, the two smallest bins in their SFD where the debiasing must be strongest.

There is a three-order of magnitude gap between our estimated number of  LPCs at $100\meter$ diameter and the prediction based on a simple extrapolation of the SFD of the large objects to smaller sizes (\fig{fig.HEfficiency+HDistribution}).  This work provides no ability to discriminate between whether 1) these objects are simply inactive or 2) do not exist.  Indeed, the first evidence for inactive objects on LPC-like orbits, colloquially referred to as `Manx'\footnote{The Manx cat is a tailless breed of cat like the tailless LPC-like objects.} objects \citep{Meech2016-Manx}, implies that there does exist a population of not-`active' LPC-like objects that could help to fill the gap between the extrapolation and our estimated LPC population.  \ie\ perhaps cometary activity is even more size-dependent than we have assumed in our simulations.  If smaller Oort cloud objects are less likely to be active than larger ones even after accounting for the size-scaling then it could imply that there is a large population of minimally active Manx-type objects.  Revealing whether this is the case requires a study that models the selection effects for the discovery of these minimally active objects.

The tremendous drop in the slope of the LPC SFD for diameters of $\lesssim 1\km$ is particularly interesting subsequent to the discovery of the $\sim 200 \meter$ diameter inactive interstellar object \Uone\  \citep[1I, \eg][]{Meech2017-U1-Nature}.  Pre-discovery limits on the interstellar object (ISO) number density assumed that they would display cometary activity \citep{Francis2005,Engelhardt2017} as they would have been preserved in the deep-freeze of interstellar space for billions of years and therefore should still contain volatile material \citep{Jewitt2017-U1}.  It was thus a surprise that the first macroscopic ISO discovered passing through our solar system showed no apparent cometary activity \citep[subsequent detailed analysis of the ISO's trajectory suggests that it does exhibit very weak cometary activity;][]{Micheli2018-Nature-Oumuamua}.  Our result (\fig{fig.HEfficiency+HDistribution}) suggests that the LPC SFD is shallow for $D\lesssim 1\km$.  Thus, based on objects from our own solar system we expect there to be relatively few active $100\meter$-scale objects despite the fact that they also spent billions of years in the deep freeze of our own Oort cloud.  Perhaps 1I suggests that the LPC nuclear SFD (\fig{fig.HEfficiency+HDistribution}) is a subset of a much larger population of Oort cloud objects where the small inactive component, like 1I, is yet to be characterized \citep[\eg][]{Fitzsimmons2018-U1-Nature}.

Thus, our LPC SFD suggests that sub-km diameter LPCs behave differently from larger LPCs.  Our analysis assumed that all the parameters governing cometary activity are size-independent and, in this case, \PSone\ would be able to discover and detect $100\meter$-scale diameter LPCs in large numbers.  The fact that \PSone\ does not do so implies either that Oort cloud objects of that size are not active or that they do not exist.  Small inactive Oort cloud objects that enter the inner solar system are far more difficult to detect than LPC with nuclei of the same size (\eg\ \fig{fig.HEfficiency+HDistribution}) so it is not surprising that they have only recently begun to be detected by large-aperture wide-field surveys like \PSone\ \citep{Meech2016-Manx}.  The question then becomes why small Oort cloud objects behave differently from large ones.  It seems unlikely that the small objects would be preferentially volatile-poor and we cannot offer an explanation for why small objects might have surface morphologies that are different from the large objects thereby rendering them inactive.

\subsubsection{The LPC and short period comet size distributions}
\label{sss.LPC+SPC-SizeDistributions}

A comparison of the SFDs of the LPC and short period comets\footnote{Active comets with orbital periods $<200\yr$ including both the Jupiter family comets (JFC) and Halley family comets (HFC).} (SPC) could test current theories of their formation and evolution \citep[\eg][]{Nesvorny2018-review,Brasser+Morbidelli2013-OortCloud}.  Both populations are thought to have formed in the region beyond the snow line in the proto-planetary solar system, in the vicinity of the giant planets that scattered them to more distant regions of the solar system and out to the Oort cloud.  Their similar formation histories, and therefore compositions, but different dynamical and collisional environments, could have left their imprints on their respective SFDs that we observe today.  The problem with the SPCs, as with the LPCs, is that measuring their sizes and debiasing a statistically significant population is difficult due to their cometary activity.

\citet{Davidsson2016} has developed a detailed comet formation model based on the results from the {\it Rosetta} in-situ investigation of the SPC 67P/Churyumov-Gerasimenko.  They conclude that (short period) comet nuclei must be primordial rubble piles and not collisional remnants to be consistent with the low density, low strength, highly porous nuclei studied by space missions, and the presence of super-volatile species (such as \Otwo\ observed by {\it Rosetta}). They argue that comets must have formed from the detritus remaining after the formation of the trans-Neptunian objects and, as a consequence of their model, they expect that cometesimals with $D < 1\km$ should be depleted.  Their model predicts a differential size distribution for SPCs with $D > 1\km$ with a power law index in diameter of 3.5 corresponding to a slope in the absolute magnitude distribution of 0.7, in agreement with our measured value for the LPCs of $\alpha_{big} = 0.7 \pm 0.1$.  Their prediction of a deficit of SPCs with $D<1\km$ is also in agreement with our observations for the LPCs.

The impacting population of small objects on the Jovian planets' satellites provide another indirect means of probing the SFD of small objects in the outer solar system \citep[\eg][]{Bierhaus2009-EuropaCraters,Schenk2007-GalileanCratering,Bierhaus2005-EuropaCraters,Zahnle2003-OuterSolarSystemCratering}.   The impactors are likely dominated by the JFCs but with some contribution from the HFCs and the LPCs as well.  The difficulty in extracting the small impactor SFD from the small crater SFD (those less than a few km diameter) is that secondary craters produced by material ejected by large primary impactors begin to dominate the landscape.  Several attempts to account for the size-dependent secondary craters have led to the realization that the slope of the SFD of the primary craters, and therefore the impacting population, is shallow with $\alpha\sim0.2$ for craters of diameter $<30\km$ diameter \citep{Bierhaus2009-EuropaCraters} but they caution that both larger and smaller slopes are possible.  Given that there are large uncertainties on their reported slope and that our measured slope for the small LPCs is $\alpha_{small} = 0.07 \pm 0.10(stat.+sys.)$ (see \S\ref{sss.SystematicUncertainties} for a discussion of the systematic uncertainties) we consider the values to be in agreement.   

There is also evidence that the trans-Neptunian object (TNO) population, the SPC source population, is depleted in small objects similar to our LPC SFD.  Measurements of the small crater SFD in the Pluto-Charon system \citep{Singer2019,Robbins2017} suggest that the SFD of the sub-$\km$ impacting population of TNOs has a slope of $\alpha\sim0.12$ on Pluto and $\alpha\sim0.28$ on Charon compared with our value of $\alpha_{small} = 0.07 \pm 0.10$ for the same size range.

There are many other measurements of the SPC SFD using different techniques and assumptions.  \citet{Snodgrass2011} utilized a large number of optical observations of JFCs and found $\alpha=0.384\pm0.004$ for $D\ge2.5\km$ while \citet{Fernandez2013} found $\alpha=0.384\pm0.041$ for $2.8\km<D<18\km$ for a set of 98 objects.  \citet{Bauer2017} analyzed the SPC population at the same time as they performed their LPC analysis (\S\ref{sss.CorrectedLPCSizeDistribution}) and found a much steeper JFC slope  of $\alpha=0.462\pm0.004$ for $1\km < D < 20\km$ compared to their debiased LPC population.  \citet{Belton2014} used a population of 161 active JFCs from many sources and found $\alpha=0.564$ for $2\km < D < 20\km$ while he finds $\alpha=0.648$ over the same size range for the TNO population (again, presumably the SPC parent population).  The only conclusions that can be drawn are that the SPC and TNO SFDs are difficult to measure, and they span a wide range of values, but all are numerically less steep than our measured value for the LPCs.

Finally, \citet{Belton2014} found a ``turndown'' in the slope of TNOs with $D\lesssim 2\km$ that could be consistent with our observation of a decrease in the slope of the LPC SFD near $D=2.8\km$.  Given our measurement of a shallow slope for small LPCs, the shallow slope of the primary crater population due to similar diameter impactors on the Jovian satellites and in the Pluto-Charon system, and indications of a turndown in the TNO SFD, it appears that all the outer solar system small body populations may have a shallow SFD at small sizes ($\sim\km$-scale and smaller).

\subsubsection{Oort cloud mass}
\label{sss.OortCloudMass}

There are $(0.46 \pm 0.15) \times 10^9$ LPCs with $q<10\au$ and $1\km \le D < 20\km$ corresponding to a mass of $(3.8 \pm 1.2) \times 10^{-4}\Mearth$ using our measured size distribution\footnote{There are $(2.4\pm0.5) \times 10^9$ objects with $D>100\meter$ corresponding to a mass of $(3.9 \pm 1.2) \times 10^{-4}\Mearth$.} (\fig{fig.HEfficiency+HDistribution}) and our assumed average cometary albedo and density from \tab{tab.CometParameters}.  We extrapolated these values to the entire Oort cloud using two techniques that yield results that differ by about a factor of three.

First, we employed the technique of \citet{Wiegert1996} who calculated that the total number of objects in the Oort cloud ($N_{OC}$) is proportional to the number of dynamically new `visible' comets per year, 
\begin{equation}
    N_{OC} = 1.7 \times 10^{11} \; f_{new} \; N_{vis},
    \label{eqn.N_OC}
\end{equation}
where $N_{vis}$ is the number of `visible' comets ($D\ge1\km$) that reach perihelion with $q\le3\au$ per year and $f_{new}$ is the fraction of those comets that are dynamically new \ie\ on their first passage through the inner solar system.  The fraction of visible comets that are dynamically new LPCs is not well constrained due to the still unknown effects of `fading' but is likely between 1/2 and 1/5 \citep{FestouRickmanWest1993, Wiegert1999} and we adopt $f_{new} = 1/3 \sim 0.33\pm0.17$.  Based on our results (\fig{fig.Synthetic-vs-Corrected-Population}) we calculate that about 6.2  LPCs with $D>1\km$ reach perihelion with $q<3\au$ per year so using this technique there are $(0.35\pm0.18)\times 10^{12}$ objects in the entire Oort cloud in that size range\footnote{The number of objects $\ge 20\km$ diameter is so small that it does not affect this calculation.} and the uncertainty is dominated by $f_{new}$.  \footnote{Extending the calculation to diameters $\ge0.1\km$ we find $N_{vis}(\ge0.1\km) \approx 24$ with a resulting Oort cloud population of $N(D>0.1\km) = (1.3\pm0.7)\times 10^{12}$.}  Thus, using this method, the total Oort cloud mass in LPCs with $D>1\km$ is $f'_{LPC}(q<10\au) ={0.35\times10^{12}}/{0.46\times10^9} \sim 740\times$ the mass of the same size objects with $q\le10\au$, or $0.28\pm0.08\Mearth$.  

Second, we used the \citet{Wiegert1999} LPC orbit model (\S\ref{sss.WiegertLPCOrbitDistribution}) to calculate the size-independent fraction of all Oort cloud objects with $q<10\au$: $f_{LPC}(q<10\au) \sim 1/2250$, about $3\times$ smaller than our first method, which has the effect of increasing our estimate of the number and mass of objects in the Oort cloud by the same factor.  We favor this technique as it is independent of the rather ill-defined constant and terms in \eqn{eqn.N_OC}, but we use the first result to provide a scale for the systematic error of a factor of $3\times$ for our final values reported in \tab{tab.OortCloudStatistics} and \tab{tab.OortCloudStatistics-comparison}.

It is important to remember that our values for the number and mass of Oort cloud objects is valid only for LPCs so they should be considered lower limits given that inactive Oort cloud objects are known to exist \citep[\eg][]{Meech2016-Manx} but the SFD of that population has not yet been measured.  Our range of values for the number of Oort cloud objects overlaps most of the earlier results while our mass estimate range overlaps but is on the low end of previous estimates.  The number of objects at large sizes is small so that extending the diameter range to larger values has no effect on the number.  Extending our fitted SFD to smaller $H$ suggests that the largest Oort Cloud object would have $H\sim5.3$ corresponding to a diameter of about $600\km$ assuming a 4\% albedo.  The mass is dominated by the largest objects due, in part, to the turnover in the slope for $D \lesssim 2.8\km$ (\fig{fig.HEfficiency+HDistribution}) so that including comets down to only $0.1~\km$ increases the total mass by only $\sim2$\%. 

\begin{table}[htbp]
    \centering
    \begin{tabular}{|r|c|c|c|}
    	\hline
    	Diameter range              & $H$ range             & number                & mass \\
    	                            & ($p_V=0.04$)          & ($\times 10^{12}$)    & ($\Mearth$) \\
    	\hline
    	\hline
        $ 20\km \ge D > 1\km$       & $12.5 \le H <19.1$    & $1.5\pm1$             & $0.9\pm0.6$   \\
        $           D > 1\km$       & $         H <19.1$    & $1.5\pm1$             & $1.2\pm0.8$   \\
        $ 20\km \ge D > 100\meter$  & $12.5 \le H <24.1$    & $5.4\pm3.6$           & $0.9\pm0.6$   \\
        $           D > 100\meter$  & $         H <24.1$    & $5.4\pm3.6$           & $1.2\pm0.8$   \\
    	\hline
    \end{tabular}
    \caption{The number and mass of Oort cloud comets (LPC) as derived in this work.  The truncated ranges at large diameters correspond to the range of sizes studied in this work while the values for ranges extending to unlimited size is an extrapolation of our measured SFD.  The uncertainties are dominated by the difference between our two methods of deriving the values.}
    \label{tab.OortCloudStatistics}
\end{table}

\begin{table}[htbp]
    \centering
    \begin{tabular}{|l|c|c|}
    	\hline
    	Source                  & $N_{OC}(D\ge1\km)$       & $M_{OC}(D\ge1\km)$ \\
    	                        & ($\times 10^{12}$)       & $(\Mearth)$ \\
    	\hline
    	\hline
        {\bf this work}         & {\boldm $1.5\pm1$}        & {\boldm $1.3\pm0.8$}   \\
    	\hline
        \citet{Weissman1985}    & 2                         &              \\
    	\hline
    	\citet{Heisler1990}     & 0.5                      &              \\
    	\hline
    	\citet{Weissman1996}    & 1                        & 3.3 to 7     \\
    	\hline
    	\citet{Dones2004}       & 0.3                      &              \\
    	\hline
    	\citet{Francis2005}     & 0.1 to 0.3               & 2 to 40      \\
    	\hline
    	\citet{KaibQuinn2009}   & 1                        &              \\
    	\hline
    \end{tabular}
    \caption{The total number and mass of Oort cloud comets (LPC) larger than $1\km$ diameter as derived in this and several other works.}
    \label{tab.OortCloudStatistics-comparison}
\end{table}

\subsubsection{Corrected LPC orbit distribution}
\label{sss.CorrectedLPCOrbitDistribution}

We think that there is remarkably good agreement between the synthetic LPC model \citep{Wiegert1999} and the debiased \PSone\ distributions (\fig{fig.Synthetic-vs-Corrected-Population}) given all the fixed parameters in our model.  

Of particular interest is that we accurately model the perihelion distribution of the LPCs since this distribution is sensitive to our model of an `average' LPC's behavior and the underlying LPC model (\fig{fig.Synthetic-vs-Corrected-Population}A).  The agreement suggests that all our assumptions are valid since it is unlikely that a set of incorrect assumptions would conspire to yield good agreement in all the distributions.  Interestingly, our corrected perihelion distribution reproduces the dip in the number of objects in the $\sim3\au$ to $\sim5\au$ range predicted by the \citet{Wiegert1999} model.  We have not investigated the details of the origin of the dip except that it is related to a correlation between eccentricity and perihelion in this range that is almost certainly due to Jupiter's influence at $\sim5.2\au$.  Our corrected perihelion distribution is different from the \citet{Fouchard2017b} series of models with the most likely explanation being that it does not include fading of the LPCs with each perihelion passage.

The apparent skew in the eccentricity distribution (\fig{fig.Synthetic-vs-Corrected-Population}B) is exacerbated by the odd binning in that histogram that highlights the fine structure of the distribution as $e \rightarrow 1$.  Our measured distribution is not a good match to either the \citep{Wiegert1999} or the \citet{Fouchard2017b} LPC models. Indeed, our result appear to be a hybrid between the two models, agreeing better with \citet{Fouchard2017b} at smaller eccentricities and better with \citep{Wiegert1999} at the highest eccentricities. The discrepancy could be a clue to details of the LPC injection mechanism and/or their dynamical evolution.

The lack of agreement between the \citet{Wiegert1999} model and our corrected LPC inclination distribution is unsatisfying despite the fact that half the bins agree within $1-\sigma$ and the other half are within $2-\sigma$.  The problem is that the corrected distribution is roughly equally distributed between prograde and retrograde objects whereas the model's inclination distribution is strongly skewed towards retrograde inclinations (as discussed above).  Our debiased distribution is in better agreement with the \citet{Fouchard2017b} model that exhibits the expected $\sin(i)$ distribution based on phase space arguments. 

Both the corrected $\omega$ and $\Omega$ distributions are consistent with being flat and with both LPC models.  Unfortunately, the number of detected objects is not sufficient to resolve the subtle bumps and dips that are predicted by the models.  For instance, the \citet{Fouchard2017b} argument of perihelion distribution appear to be flat but the \citet{Wiegert1999} model seems to have a 10-20\% dip at $\omega\sim60\arcdeg$ and $\sim240\arcdeg$.  Similarly, the \citet{Fouchard2017b} ascending node distribution appears to be at a minimum near $\Omega=150\arcdeg$ and an $\sim2\times$ maximum near $\Omega=330\arcdeg$ while the \citet{Wiegert1999} exhibits $\sim30$\% dips from a flat distribution near $\Omega=90\arcdeg$ and $\Omega=270\arcdeg$.
Given that the primary difference between the two LPC models is the dynamical effect of passing nearby stars, future high statistics comet surveys with good control on their observational selection effects may be able to distinguish between the two models or measure the relative importance of stellar perturbations vs. galactic tides.  It is likely that a high-statistics comparison of the debiased distributions in the galactic rather than ecliptic frames will provide a stronger signal to discriminate between the models.

Finally, the corrected time of perihelion distribution is consistent with being flat, as expected.

\subsubsection{Systematic uncertainties}
\label{sss.SystematicUncertainties}

We performed a study of the systematic impact of our cometary activity model on our results for the LPC SFD (\tab{tab.SystematicUncertainties}).  The nominal value of each model parameter was individually changed by a large but reasonable amount, typically $\pm50$\%, and the complete analysis was re-run to calculate the resulting values of $\Delta\alpha_{big}$, $\Delta\alpha_{small}$, and $\Delta N$.  We did not investigate the systematic impact of the survey simulation parameters because, relative to the cometary activity model's parameters, they are well measured (\eg\ the average Pan-STARRS1 limiting magnitude and the typical radius at which the czar is able to detect cometary activity) and will contribute relatively little to the systematic uncertainty.

The nuclear albedo and the grain radii typically have the largest systematic effects on the fit parameters as might be expected given that they control the total amount of gas and dust and the total reflecting area respectively.  Secondary systematic effects are introduced by our adoptions of the nominal values for the grain density, and active areas for both \HtwoO\ and \COtwo.  All the other parameters that we investigated have minimal effects on the final values of the SFD parameters.  In a few cases the impact of changing the nominal parameters by $\pm50$\% caused both resulting values to change in the same, always small, direction.  We attribute this behavior to small number statistics and the systematic effects are simply related to noise.

The formal quadratic sums of the systematic uncertainties in \tab{tab.SystematicUncertainties} for each parameter are $\Delta\alpha_{big,sys} = _{-0.14}^{+0.15}$, $\Delta\alpha_{small,sys} = \pm0.09$, and $\Delta N_{sys} = _{-3.0}^{+2.1} \times 10^9$.  Since we think the quadratic sum of the $>1\sigma$ systematic uncertainty over-estimates the actual systematic uncertainty, and to not overstate the precision with which we have estimated them, we adopt $\Delta\alpha_{big,sys} =\pm 0.15$, $\Delta\alpha_{small,sys} = \pm 0.09$, and $\Delta N_{sys} = \pm 2.0 \times 10^9$ as the overall systematic error estimate on the SFD parameters.

\begin{table}[htbp]
   \renewcommand*{\arraystretch}{1.4}
   \centering
	\begin{tabular}{ | l | p{2.5cm} | p{2.7cm}  | c       | c      | c |}
		\hline
		\vspace{-0.25cm} 
		{\bf Parameter}           & {\bf Nominal}         & {\bf Systematic}       &  $\Delta\alpha_{big}$  & $\Delta\alpha_{small}$ & $\Delta N \times 10^9$ \\
		                          & {\bf Value}           & {\bf Range}            &                        &                        & $(>100\meter)$        \\ \hline
		Grain radius              &    $1\um$             & $[0.5\um,2.0\um]$      &  $_{-0.07 }^{+0.05 }$  &  $_{-0.05 }^{+0.04 }$  &  $_{-0.78}^{+1.63}$  \\ \hline
		Grain density             & $1,000\kg\meter^{-3}$ & $\pm 50$\%             &  $_{-0.07 }^{+0.04 }$  &  $_{-0.05 }^{+0.02 }$  &  $_{-0.78}^{+0.74}$  \\ \hline
		Nucleus density           &  $400\kg\meter^{-3}$  & $\pm 50$\%             &  $_{+0.002}^{+0.005}$  &  $_{-0.004}^{-0.01}$  &  $_{+0.03}^{+0.01}$  \\ \hline
		Active area (\HtwoO)      &    $4$\%              & $\pm 50$\%             &  $_{-0.0002}^{+0.01}$ &  $_{+0.04 }^{-0.04 }$  &  $_{+1.02}^{-0.36}$  \\ \hline
		Active area (\COtwo)      &  $0.1$\%              & $[0.05,0.2]$\%         &  $_{+0.05 }^{-0.04 }$  &  $_{-0.007}^{-0.05 }$  &  $_{+0.39}^{-0.50}$  \\ \hline
		Nucleus albedo            &    $4$\%  			  & $[2,8]$\%              &  $_{+0.07 }^{-0.13 }$  &  $_{+0.04 }^{-0.05 }$  &  $_{+2.57}^{-0.95}$  \\ \hline
		Emissivity                &    $0.9$              & $\pm 0.05$             &  $_{-0.003}^{+0.01}$  &  $_{+0.002}^{-0.01}$  &  $_{+0.01}^{+0.10}$  \\ \hline
		Nuclear phase coefficient &  $0.04$ mag/deg  	  & $\pm 50$\%             &  $_{-0.0002}^{+0.004}$  &  $_{-0.002}^{-0.004}$  &  $_{+0.001}^{+0.07}$  \\ \hline
		Coma phase coefficient    &  $0.02$ mag/deg    	  & $\pm 50$\%             &  $_{+0.004}^{+0.004}$  &  $_{-0.02 }^{+0.01 }$  &  $_{-0.16}^{+0.30}$  \\ \hline
		$M_{limit,j}$             &  $3,000\Day/n_j$      & $1,500\Day/n_j$        &  $_{-0.04}^{-10^{-7} }$  &  $_{+0.0007 }^{+10^{-8} }$  &  $_{+0.02}^{+0.005}$  \\ \hline
	\end{tabular}
	\caption{Results of a study of the systematic error induced in our primary results on the LPC SFD due to our adoption of the nominal cometary physical parameters.  The rightmost three columns provide the change in the value of $\Delta\alpha_{big}$, $\Delta\alpha_{small}$, and $\Delta N$ induced by changing only the nominal parameter value to the limits of its systematic range.
	}
	\label{tab.SystematicUncertainties}
\end{table}

\section{Conclusions}

We have introduced a new technique to estimate the cometary nuclear size frequency distribution that combines a cometary activity model with a survey simulation model to estimate the size of cometary nuclei.  The technique is intended for use with a well-characterized, high-statistics, comet survey to average over the behavior of many individual comets.  The technique was applied to the 150 long period comets detected by the \PSone\ near-Earth asteroid survey over the past $\sim 7$ years.  

We find that the typical LPC nuclear magnitudes reported by JPL$^{\protect\ref{JPL_H_N}}$ are about $5.5\mags$ brighter than our values, corresponding to an overestimate of the nuclear flux by $\sim160\times$ and an overestimate of the nuclear diameter by $\sim12\times$.

The debiased LPC size-frequency distribution, \ie\ after correcting for observational selection effects, is in agreement with previous estimates for comets with nuclear diameter $\gtrsim 1\km$ but we have extended the size range of the diameter distribution to much smaller sizes approaching only $100\meter$ diameter and we measure a significant drop in the SFD slope for objects with diameters $<1\km$.  Letting $n \propto 10^{\alpha H}$ we find that large objects have a slope $\alpha_{big} = 0.72 \pm 0.09(stat.) \pm 0.15 (sys.)$ while small objects behave as $\alpha_{small} = 0.10 \pm 0.03(stat.) \pm 0.09 (sys.)$.  The large object slope is consistent with comet nucleus models that assume they are low density, low strength, and highly porous.  This work cannot distinguish whether the depletion of small objects indicated by the shallow slope of their SFD is due to their non-existence or an observational selection effect \ie\ it is possible that the deficit is simply due to the small objects being inactive and therefore much more difficult to detect \eg\ the Manx comets.  However, since the slope of our small LPC SFD is consistent with the SFD of small trans-Neptunian objects as derived from the crater SFD in the Pluto-Charon system, and assuming that LPCs and TNOs share a common origin, the lack of small craters in the Pluto-Charon system would indicate that there simply are very few small LPCs instead of implying that the objects are undetected by \PSone\ and other contemporary asteroid surveys.

The total number of objects with diameters $\ge 1\km$ in the Oort cloud that would be considered LPCs if their perihelia evolved to $<10\au$ is $(1.5\pm1)\times10^{12}$ with a combined mass of $1.3\pm0.9\Mearth$.  Our values for both the number of LPCs in the entire Oort cloud and their mass should be considered as lower limits considering that they correspond to the number and mass of LPCs (objects that are or could be active in the inner solar system).  If there is a large and uncharacterized population of inactive Oort cloud objects that are much more difficult to detect then the number and mass of objects could be larger than our values.

The debiased LPC orbit distribution is broadly in agreement with the \citet{Wiegert1999} LPC model but there are interesting hints that the details of the disagreements could point to future opportunities to disentangle the relative importance of stellar perturbations and the galactic tide on producing the LPC population.  We think this method or a similar one could eventually be used to correct for the observational selection effects of the Large Synoptic Survey Telescope and debias the large population of LPCs that will be detected by that system.  Thus, in about ten years we may have a good LPC SFD over a wide range of sizes and a matching dynamical model.

\section*{Acknowledgements}

We offer a special thanks to Eva Lilly for guidance on using MOPS, Serge Chastel for keeping the MOPS servers up and running, Marc Fouchard for graciously providing the data from their LPC model, Giovanni Valsecchi for an historical and dynamical perspective on this work, Bill Bottke for his perspective on the links between this work and the impact crater SFD in the Pluto-Charon system while attending a workshop supported by the Munich Institute for Astro- and Particle Physics (MIAPP) of the DFG cluster of excellence ``Origin and Structure of the Universe", and Beau Bierhaus for help in interpreting the link between the LPC SFD and the impact crater distribution on the Jovian satellites.

We thank the two reviewers, Colin Snodgrass and Luke Dones, for their thorough reading and constructive suggestions for improving the manuscript.

This research has made use of data and/or services provided by the International Astronomical Union's Minor Planet Center and the NASA supported Jet Propulsion Laboratory's Solar System Dynamics Small-Body Database Search Engine. 

The Pan-STARRS1 Surveys (PS1) and the PS1 public science archive have been made possible through contributions by the Institute for Astronomy, the University of Hawaii, the Pan-STARRS Project Office, the Max-Planck Society and its participating institutes, the Max Planck Institute for Astronomy, Heidelberg and the Max Planck Institute for Extraterrestrial Physics, Garching, The Johns Hopkins University, Durham University, the University of Edinburgh, the Queen's University Belfast, the Harvard-Smithsonian Center for Astrophysics, the Las Cumbres Observatory Global Telescope Network Incorporated, the National Central University of Taiwan, the Space Telescope Science Institute, the National Aeronautics and Space Administration under Grants NNX08AR22G, NNX12AR65G, and NNX14AM74G issued through the Planetary Science Division of the NASA Science Mission Directorate, the National Science Foundation Grant No. AST-1238877, the University of Maryland, Eotvos Lorand University (ELTE), the Los Alamos National Laboratory, and the Gordon and Betty Moore Foundation.  K.J.M. was supported by a grant from the National Science Foundation, AST-1413736.

\section*{References}

\bibliographystyle{./model2-names}\biboptions{authoryear}
\bibliography{references}

\newpage
\section*{Appendix}

{\rowcolors{5}{}{mygray}
{\tiny
\begin{longtable}[c]{|l|r|r|r|r|r|r|r|r|}
\caption{The 150 long period comets detected by \PSone\ and used in this analysis.  The LPCs are ordered by the time of first detection by \PSone.  Columns provide the LPC identifier, date of first \PSone\ detection, the apparent measured $V$ magnitude and heliocentric and geocentric distance at the time of first detection, and the object's perihelion distance, eccentricity, and inclination as provided by the JPL Small-Body Database Search Engine (\protect\url{https://ssd.jpl.nasa.gov/sbdb_query.cgi}).}
\label{tab.PS1-LPCs}\\

\hline
                         & date              &     &     &          &          &     &     &     \\
comet designation (name) & (yyyymmdd.dddddd) & $V$ & $r$ & $\Delta$ & $\alpha$ & $q$ & $e$ & $i$ \\
\hline
\endfirsthead

\hline
\multicolumn{9}{|c|}{Continuation of Table \ref{tab.PS1-LPCs}}\\
\hline
                         & date              &     &     &          &          &     &     &     \\
comet designation (name) & (yyyymmdd.dddddd) & $V$ & $r$ & $\Delta$ & $\alpha$ & $q$ & $e$ & $i$ \\
\hline
\endhead

\hline
\endfoot

\hline
\multicolumn{9}{| c |}{End of Table \ref{tab.PS1-LPCs}}\\
\hline\hline
\endlastfoot
                 C/2005L3(McNaught) & 20110314.421070 &  20.5 &   9.660 &   8.880 &   3.800 &   5.595 &   0.998 & 139.469 \\ 
                   C/2010X1(Elenin) & 20110330.366990 &  19.2 &   3.790 &   2.874 &   6.900 &   0.473 &   0.999 &   1.871 \\ 
                   C/2010F3(Scotti) & 20110413.426720 &  20.4 &   5.726 &   4.751 &   2.600 &   5.453 &   0.915 &   4.650 \\ 
                 C/2010L3(Catalina) & 20110514.438200 &  21.1 &   9.934 &   9.293 &   4.700 &   9.893 &   1.000 & 102.500 \\ 
                C/2011L4(PANSTARRS) & 20110521.425250 &  19.6 &   8.068 &   7.064 &   1.000 &   0.302 &   1.000 &  83.958 \\ 
                   C/2011O1(LINEAR) & 20110523.499850 &  18.5 &   5.445 &   4.572 &   5.900 &   3.889 &   0.994 &  76.572 \\ 
                   C/2010R1(LINEAR) & 20110630.372560 &  19.8 &   6.084 &   5.206 &   5.200 &   5.634 &   0.999 & 156.928 \\ 
                    C/2011A3(Gibbs) & 20110702.259320 &  17.5 &   2.974 &   2.818 &  20.000 &   2.332 &   0.997 &  26.081 \\ 
                 C/2009Y1(Catalina) & 20110712.596000 &  16.9 &   3.066 &   2.558 &  18.100 &   2.505 &   0.993 & 107.344 \\ 
                C/2011UF305(LINEAR) & 20110725.318330 &  20.7 &   4.438 &   3.600 &   8.200 &   2.144 &   0.999 &  94.021 \\ 
                 C/2011L3(McNaught) & 20110801.335010 &  18.7 &   1.954 &   0.995 &  13.800 &   1.923 &   0.998 &  87.204 \\ 
                C/2011Q1(PANSTARRS) & 20110817.319750 &  20.3 &   6.788 &   6.050 &   6.200 &   6.766 &   0.998 &  94.889 \\ 
                   C/2009S3(Lemmon) & 20110817.531410 &  21.0 &   6.527 &   6.083 &   8.300 &   6.470 &   0.999 &  60.450 \\ 
    C/2008FK75(Lemmon-SidingSpring) & 20110824.394260 &  19.7 &   5.218 &   4.702 &  10.100 &   4.504 &   1.000 &  61.298 \\ 
                 C/2008S3(Boattini) & 20110909.387500 &  19.3 &   8.036 &   7.155 &   3.700 &   8.008 &   1.000 & 162.696 \\ 
                     C/2012S1(ISON) & 20110930.592270 &  21.1 &   9.339 &   9.626 &   5.800 &   0.013 &   1.000 &  62.767 \\ 
                C/2011U3(PANSTARRS) & 20111024.543840 &  21.8 &   3.306 &   2.389 &   7.900 &   1.072 &   1.000 & 116.712 \\ 
                     C/2010G2(Hill) & 20111128.500330 &  15.7 &   2.240 &   1.256 &   2.300 &   1.981 &   0.979 & 103.761 \\ 
           C/2010FB87(WISE-Garradd) & 20111208.496550 &  17.9 &   4.743 &   3.867 &   6.000 &   2.853 &   0.987 & 107.805 \\ 
                   C/2010S1(LINEAR) & 20111209.243680 &  16.7 &  10.547 &   9.695 &   2.800 &   5.895 &   1.000 & 125.319 \\ 
                C/2012A1(PANSTARRS) & 20120102.453050 &  20.5 &   8.749 &   7.902 &   3.400 &   7.605 &   1.000 & 120.929 \\ 
                   C/2012A2(LINEAR) & 20120115.627170 &  18.8 &   4.413 &   3.924 &  11.800 &   3.535 &   0.997 & 125.904 \\ 
                C/2012F3(PANSTARRS) & 20120119.478650 &  21.8 &   9.938 &   9.236 &   4.100 &   3.457 &   1.000 &  11.354 \\ 
                   C/2012F6(Lemmon) & 20120122.453480 &  21.4 &   5.592 &   4.784 &   6.200 &   0.726 &   0.998 &  82.620 \\ 
                   C/2011F1(LINEAR) & 20120216.653690 &  16.9 &   4.155 &   3.824 &  13.400 &   1.814 &   0.999 &  56.698 \\ 
                C/2012E3(PANSTARRS) & 20120314.597570 &  20.2 &   4.641 &   3.829 &   7.800 &   3.835 &   0.981 & 105.569 \\ 
                   C/2011J2(LINEAR) & 20120316.324450 &  18.8 &   6.556 &   5.583 &   2.000 &   3.442 &   1.000 & 122.795 \\ 
                  C/2013E2(Iwamoto) & 20120319.339590 &  21.4 &   4.503 &   3.540 &   3.700 &   1.406 &   0.994 &  21.873 \\ 
                  C/2009P1(Garradd) & 20120426.253770 &  14.0 &   7.200 &   7.684 &   6.800 &   1.559 &   1.000 & 106.203 \\ 
                 C/2009UG89(Lemmon) & 20120427.348610 &  20.7 &   5.734 &   5.152 &   8.700 &   3.935 &   1.000 & 130.099 \\ 
                   C/2012K5(LINEAR) & 20120506.578840 &  18.7 &   3.095 &   2.707 &  18.500 &   1.145 &   0.998 &  92.818 \\ 
                   C/2006S3(LONEOS) & 20120517.467350 &  15.9 &   5.144 &   4.211 &   4.800 &   5.136 &   1.000 & 166.023 \\ 
                C/2012K1(PANSTARRS) & 20120517.545420 &  19.1 &  13.722 &  12.998 &   3.000 &   1.054 &   1.000 & 142.402 \\ 
                   C/2012L1(LINEAR) & 20120607.443830 &  19.0 &   3.170 &   3.099 &  18.600 &   2.263 &   0.996 &  87.237 \\ 
                   C/2012K8(Lemmon) & 20120806.377360 &  20.5 &  11.024 &  10.013 &   0.400 &   6.466 &   0.999 & 106.119 \\ 
                C/2012S3(PANSTARRS) & 20120927.298833 &  20.2 &   4.238 &   3.931 &  13.400 &   2.307 &   0.999 & 112.955 \\ 
                C/2012S4(PANSTARRS) & 20120928.414945 &  19.3 &   4.865 &   4.256 &  10.000 &   4.342 &   1.000 & 126.560 \\ 
             C/2013A1(SidingSpring) & 20121004.592670 &  19.7 &   7.931 &   7.726 &   7.200 &   1.400 &   1.000 & 129.026 \\ 
                C/2012U1(PANSTARRS) & 20121018.316523 &  21.0 &   6.991 &   6.018 &   1.900 &   5.270 &   0.998 &  56.349 \\ 
                   C/2012T5(Bressi) & 20121021.559037 &  19.8 &   2.477 &   1.602 &  13.700 &   0.324 &   1.000 &  72.249 \\ 
                C/2012V1(PANSTARRS) & 20121103.340655 &  20.4 &   3.658 &   2.714 &   5.500 &   2.088 &   0.999 & 157.836 \\ 
                 C/2013F2(Catalina) & 20121228.587855 &  20.0 &   6.269 &   5.784 &   8.100 &   6.230 &   0.994 &  61.779 \\ 
                 C/2012J1(Catalina) & 20130210.229250 &  16.7 &   3.222 &   3.532 &  16.000 &   3.161 &   1.000 &  34.094 \\ 
                 C/2011R1(McNaught) & 20130228.643600 &  16.8 &   2.583 &   2.271 &  22.400 &   2.083 &   1.000 & 116.161 \\ 
                 C/2013F1(Boattini) & 20130408.508730 &  19.6 &   2.428 &   1.641 &  17.800 &   1.864 &   1.000 &  79.588 \\ 
                C/2013G3(PANSTARRS) & 20130410.421660 &  20.7 &   6.204 &   5.213 &   1.500 &   3.852 &   1.000 &  64.683 \\ 
                C/2013G8(PANSTARRS) & 20130414.602790 &  20.1 &   5.388 &   5.455 &  10.600 &   5.136 &   0.997 &  27.638 \\ 
                   C/2012L3(LINEAR) & 20130508.441330 &  21.5 &   4.361 &   3.375 &   3.200 &   3.041 &   0.990 & 134.136 \\ 
                 C/2013G7(McNaught) & 20130509.413590 &  19.2 &   5.288 &   4.278 &   0.300 &   4.681 &   0.997 & 105.109 \\ 
                 C/2013H2(Boattini) & 20130515.506300 &  19.5 &   7.862 &   6.941 &   3.200 &   7.503 &   0.999 & 128.388 \\ 
             C/2012OP(SidingSpring) & 20130604.431290 &  19.7 &   3.957 &   3.054 &   7.600 &   3.606 &   0.995 & 114.834 \\ 
                     C/2012E1(Hill) & 20130609.319090 &  21.3 &   8.681 &   8.517 &   6.700 &   7.506 &   0.993 & 122.576 \\ 
                  C/2013G9(Tenagra) & 20130618.335750 &  19.4 &   6.733 &   5.880 &   5.000 &   5.341 &   1.000 & 146.253 \\ 
                C/2013P2(PANSTARRS) & 20130726.519190 &  19.5 &   3.504 &   2.874 &  14.400 &   2.837 &   0.996 & 125.619 \\ 
    C/2013PE67(Catalina-Spacewatch) & 20130803.531930 &  20.9 &   2.410 &   1.449 &  10.000 &   1.873 &   0.964 & 116.672 \\ 
               C/2013US10(Catalina) & 20130814.498740 &  19.9 &   8.922 &   8.079 &   3.800 &   0.823 &   1.000 & 148.879 \\ 
                  C/2013P3(Palomar) & 20130831.492330 &  20.3 &   9.030 &   8.210 &   3.900 &   8.644 &   1.000 &  93.919 \\ 
                C/2013X1(PANSTARRS) & 20131017.592120 &  21.4 &   9.252 &   9.116 &   6.200 &   1.315 &   1.000 & 163.232 \\ 
                 C/2013V4(Catalina) & 20131023.481500 &  19.2 &   7.339 &   6.474 &   4.100 &   5.186 &   1.000 &  67.860 \\ 
                 C/2013S1(Catalina) & 20131104.429620 &  20.9 &   2.975 &   2.015 &   5.900 &   2.822 &   0.999 &  83.957 \\ 
                C/2013UQ4(Catalina) & 20131109.486210 &  18.9 &   3.424 &   2.673 &  12.300 &   1.088 &   0.981 & 145.035 \\ 
               C/2013V5(Oukaimeden) & 20131204.432370 &  18.9 &   4.362 &   3.516 &   7.400 &   0.627 &   0.998 & 154.866 \\ 
                 C/2013B2(Catalina) & 20131204.610890 &  20.8 &   3.983 &   3.993 &  14.200 &   3.742 &   1.000 &  43.463 \\ 
                C/2013Y2(PANSTARRS) & 20131230.512600 &  18.4 &   2.748 &   2.012 &  15.900 &   1.921 &   0.992 &  29.399 \\ 
                C/2014A5(PANSTARRS) & 20140104.424720 &  21.4 &   5.083 &   4.134 &   3.300 &   4.801 &   0.945 &  31.973 \\ 
               C/2014AA52(Catalina) & 20140104.584900 &  20.5 &   4.933 &   4.289 &   9.300 &   2.001 &   1.000 & 105.201 \\ 
              C/2013TW5(Spacewatch) & 20140105.564210 &  19.4 &   6.045 &   5.178 &   4.800 &   5.835 &   0.980 &  31.405 \\ 
                 C/2013J3(McNaught) & 20140110.624590 &  21.2 &   4.824 &   4.853 &  11.700 &   3.995 &   0.996 & 118.204 \\ 
                 C/2012K6(McNaught) & 20140227.609830 &  18.3 &   4.238 &   3.439 &   8.800 &   3.360 &   0.998 & 135.221 \\ 
                C/2014G1(PANSTARRS) & 20140405.574800 &  20.5 &   5.577 &   5.004 &   8.900 &   5.476 &   0.991 & 165.634 \\ 
                C/2014G3(PANSTARRS) & 20140410.542430 &  19.9 &   5.205 &   4.706 &  10.100 &   4.701 &   0.912 & 156.071 \\ 
                     C/2014F1(Hill) & 20140419.583220 &  19.8 &   3.922 &   3.291 &  12.500 &   3.500 &   0.998 & 108.260 \\ 
              C/2014H1(Christensen) & 20140504.366320 &  19.4 &   2.150 &   1.164 &   7.600 &   2.142 &   0.984 &  99.938 \\ 
               C/2014OE4(PANSTARRS) & 20140508.473640 &  21.0 &   8.842 &   7.861 &   1.600 &   6.244 &   1.000 &  81.348 \\ 
                C/2015H2(PANSTARRS) & 20140519.279740 &  21.2 &   7.970 &   7.258 &   5.400 &   4.967 &   1.000 &  33.705 \\ 
                  C/2013R1(Lovejoy) & 20140603.360570 &  16.7 &   2.694 &   1.681 &   1.000 &   0.813 &   0.998 &  63.928 \\ 
                C/2014M1(PANSTARRS) & 20140624.489280 &  20.8 &   6.381 &   5.530 &   5.400 &   5.580 &   1.000 & 160.184 \\ 
                C/2012LP26(Palomar) & 20140627.433890 &  19.8 &   7.104 &   6.192 &   3.900 &   6.536 &   1.000 &  25.377 \\ 
                C/2014N2(PANSTARRS) & 20140702.548510 &  19.6 &   2.462 &   1.703 &  19.000 &   2.190 &   0.999 & 132.875 \\ 
                 C/2012C1(McNaught) & 20140727.478180 &  19.8 &   6.332 &   5.403 &   4.000 &   4.835 &   0.998 &  96.239 \\ 
                  C/2014N3(NEOWISE) & 20140815.595930 &  18.7 &   4.283 &   3.685 &  11.800 &   3.881 &   0.999 &  61.644 \\ 
                C/2014Q1(PANSTARRS) & 20140816.547460 &  18.4 &   4.893 &   4.044 &   7.100 &   0.315 &   1.000 &  43.109 \\ 
               C/2014QU2(PANSTARRS) & 20140816.593330 &  19.8 &   2.265 &   1.482 &  20.200 &   2.220 &   0.991 & 124.816 \\ 
                 C/2014M3(Catalina) & 20140819.403640 &  19.7 &   2.516 &   1.552 &   8.900 &   2.432 &   0.982 & 164.904 \\ 
                C/2014Q6(PANSTARRS) & 20140831.349080 &  19.7 &   4.360 &   3.485 &   7.400 &   4.221 &   0.999 &  49.794 \\ 
                C/2014S2(PANSTARRS) & 20140902.450480 &  21.4 &   5.265 &   4.376 &   5.700 &   2.092 &   0.991 &  64.610 \\ 
                C/2014R3(PANSTARRS) & 20140906.328560 &  20.1 &   8.514 &   8.208 &   6.600 &   7.276 &   1.000 &  90.839 \\ 
                C/2014S1(PANSTARRS) & 20140919.563260 &  21.3 &   8.370 &   7.763 &   5.700 &   8.135 &   1.000 & 123.807 \\ 
                 C/2009F4(McNaught) & 20140920.575830 &  21.7 &   8.817 &   8.285 &   5.700 &   5.453 &   0.999 &  79.440 \\ 
                C/2014S3(PANSTARRS) & 20140922.524320 &  21.5 &   2.104 &   1.566 &  27.000 &   2.047 &   0.977 & 169.323 \\ 
                    C/2014R4(Gibbs) & 20141002.619990 &  20.2 &   1.834 &   1.981 &  30.100 &   1.819 &   0.999 &  42.405 \\ 
                   C/2014A4(SONEAR) & 20141108.465630 &  17.3 &   4.888 &   4.197 &   9.000 &   4.179 &   1.000 & 121.368 \\ 
                C/2014W2(PANSTARRS) & 20141117.353070 &  18.8 &   5.343 &   4.398 &   3.400 &   2.671 &   0.998 &  82.001 \\ 
                C/2014Y1(PANSTARRS) & 20141117.553490 &  20.2 &   4.959 &   4.432 &  10.200 &   2.242 &   1.000 &  14.928 \\ 
                C/2014W3(PANSTARRS) & 20141118.466070 &  20.5 &   6.340 &   5.764 &   7.600 &   6.069 &   1.000 &  90.131 \\ 
                C/2014W8(PANSTARRS) & 20141122.423410 &  21.5 &   5.503 &   4.535 &   2.200 &   5.024 &   0.973 &  41.943 \\ 
                 C/2014W6(Catalina) & 20141202.624510 &  19.9 &   3.256 &   3.404 &  16.800 &   3.089 &   1.000 &  53.564 \\ 
               C/2014XB8(PANSTARRS) & 20141215.372300 &  20.4 &   3.204 &   2.316 &   9.000 &   3.009 &   0.997 & 149.788 \\ 
                C/2015B2(PANSTARRS) & 20150106.544190 &  21.0 &   5.483 &   4.964 &   9.200 &   3.370 &   1.000 & 105.086 \\ 
                C/2013U2(Holvorcem) & 20150109.565660 &  18.7 &   5.170 &   4.503 &   8.600 &   5.139 &   0.993 &  43.236 \\ 
                  C/2015V2(Johnson) & 20150111.388820 &  19.6 &   8.792 &   7.963 &   3.600 &   1.637 &   1.000 &  49.875 \\ 
              C/2015ER61(PANSTARRS) & 20150121.564370 &  22.4 &   8.842 &   8.148 &   4.700 &   1.042 &   0.997 &   6.349 \\ 
                    C/2015X7(ATLAS) & 20150128.232430 &  19.9 &   5.952 &   5.883 &   9.500 &   3.685 &   1.000 &  57.570 \\ 
                C/2015D3(PANSTARRS) & 20150219.490140 &  20.4 &   8.567 &   7.700 &   3.400 &   8.149 &   1.000 & 128.506 \\ 
                C/2015J2(PANSTARRS) & 20150321.573660 &  20.9 &   4.542 &   4.046 &  11.600 &   4.321 &   0.983 &  17.279 \\ 
                C/2015J1(PANSTARRS) & 20150514.335830 &  20.4 &   6.418 &   5.475 &   3.500 &   6.023 &   1.000 &  94.987 \\ 
                C/2015K2(PANSTARRS) & 20150518.334810 &  21.8 &   1.487 &   0.562 &  25.900 &   1.455 &   0.995 &  29.116 \\ 
                C/2015K4(PANSTARRS) & 20150524.468620 &  19.0 &   2.032 &   1.076 &  13.000 &   2.008 &   1.000 &  80.259 \\ 
                  C/2013V2(Borisov) & 20150530.330690 &  18.7 &   4.073 &   3.650 &  13.700 &   3.510 &   1.000 &  37.852 \\ 
               C/2015LC2(PANSTARRS) & 20150607.339800 &  19.9 &   5.898 &   5.100 &   6.600 &   5.893 &   1.000 &  93.710 \\ 
                  C/2015F4(Jacques) & 20150613.536070 &  15.7 &   1.816 &   0.982 &  25.100 &   1.643 &   0.986 &  48.717 \\ 
                  C/2014R1(Borisov) & 20150615.360230 &  19.3 &   3.068 &   2.066 &   3.800 &   1.347 &   0.993 &   9.927 \\ 
                C/2015M1(PANSTARRS) & 20150620.451770 &  19.8 &   2.134 &   1.158 &  10.300 &   2.091 &   0.995 &  57.304 \\ 
                C/2015M3(PANSTARRS) & 20150629.429770 &  20.1 &   3.590 &   3.170 &  15.800 &   3.552 &   0.974 &  65.942 \\ 
                  C/2015F2(Polonia) & 20150703.506670 &  20.0 &   1.561 &   0.912 &  38.300 &   1.210 &   0.969 &  28.702 \\ 
                C/2015O1(PANSTARRS) & 20150719.470140 &  19.6 &   8.216 &   7.212 &   1.200 &   3.730 &   1.000 & 127.210 \\ 
             C/2011KP36(Spacewatch) & 20150728.430840 &  16.4 &   5.350 &   4.398 &   4.200 &   4.883 &   0.873 &  18.986 \\ 
                   C/2015K1(MASTER) & 20150730.530150 &  18.7 &   3.910 &   3.380 &  13.700 &   2.556 &   0.986 &  29.384 \\ 
                     C/2015P3(SWAN) & 20150810.253150 &  16.4 &   0.765 &   0.910 &  73.900 &   0.715 &   0.997 &  58.182 \\ 
                C/2015R3(PANSTARRS) & 20150912.578510 &  22.3 &   6.559 &   5.964 &   7.400 &   4.902 &   0.999 &  83.634 \\ 
                C/2015T2(PANSTARRS) & 20151009.471070 &  21.8 &   7.474 &   6.687 &   5.000 &   6.935 &   1.000 & 124.545 \\ 
                C/2015T4(PANSTARRS) & 20151014.533890 &  19.2 &   3.512 &   3.032 &  15.400 &   2.296 &   0.974 &  87.920 \\ 
                C/2015TQ209(LINEAR) & 20151022.361560 &  19.5 &   4.095 &   3.117 &   3.000 &   1.413 &   0.999 &  11.395 \\ 
                C/2015WZ(PANSTARRS) & 20151025.469459 &  21.3 &   2.707 &   1.901 &  14.700 &   1.377 &   0.993 & 134.144 \\ 
                C/2015V3(PANSTARRS) & 20151102.272290 &  21.8 &   4.240 &   3.501 &   9.900 &   4.235 &   0.994 &  86.232 \\ 
                C/2015V1(PANSTARRS) & 20151102.604680 &  20.2 &   7.429 &   7.072 &   7.300 &   4.266 &   1.000 & 139.231 \\ 
                   C/2015G2(MASTER) & 20151204.512300 &  18.4 &   3.110 &   2.257 &  10.700 &   0.781 &   1.000 & 147.578 \\ 
                    C/2015W1(Gibbs) & 20151206.511800 &  20.1 &   2.883 &   2.232 &  16.800 &   2.232 &   1.000 &  87.315 \\ 
                C/2015X5(PANSTARRS) & 20151206.553640 &  21.3 &   8.380 &   7.997 &   6.400 &   6.802 &   1.000 & 124.278 \\ 
   C/2015VL62(Lemmon-Yeung-PANSTARR & 20151216.273080 &  19.9 &   6.397 &   5.479 &   3.400 &   2.720 &   1.000 & 165.614 \\ 
                C/2016A5(PANSTARRS) & 20151217.580897 &  21.0 &   3.403 &   2.945 &  15.800 &   2.947 &   0.998 &  40.240 \\ 
                C/2016A1(PANSTARRS) & 20160101.539920 &  19.3 &   7.309 &   7.027 &   7.500 &   5.328 &   1.000 & 121.180 \\ 
                  C/2015XY1(Lemmon) & 20160101.615210 &  20.2 &   9.436 &   8.616 &   3.400 &   7.928 &   1.000 & 148.836 \\ 
                   C/2015Y1(LINEAR) & 20160112.478180 &  18.5 &   2.841 &   1.981 &  11.600 &   2.514 &   0.991 &  71.213 \\ 
                C/2016A6(PANSTARRS) & 20160113.444340 &  21.0 &   2.527 &   1.632 &  11.600 &   2.413 &   0.988 & 120.919 \\ 
                C/2016C1(PANSTARRS) & 20160212.495940 &  19.9 &   8.465 &   7.648 &   3.900 &   8.461 &   1.000 &  56.180 \\ 
                C/2016E1(PANSTARRS) & 20160303.544070 &  21.1 &   8.623 &   8.124 &   5.900 &   8.177 &   1.000 & 131.891 \\ 
                  C/2014Q2(Lovejoy) & 20160411.557620 &  19.1 &   5.337 &   5.109 &  10.700 &   1.291 &   0.998 &  80.277 \\ 
                   C/2016K1(LINEAR) & 20160616.584340 &  19.4 &   2.310 &   2.010 &  26.000 &   2.291 &   1.000 &  90.844 \\ 
                C/2016M1(PANSTARRS) & 20160622.466780 &  19.7 &   7.700 &   7.686 &   7.600 &   2.211 &   0.999 &  90.996 \\ 
                C/2016N6(PANSTARRS) & 20160627.404660 &  20.8 &   7.337 &   6.786 &   6.900 &   2.670 &   0.998 & 105.831 \\ 
                C/2016P4(PANSTARRS) & 20160731.427470 &  21.7 &   5.914 &   4.906 &   1.300 &   5.889 &   0.982 &  29.902 \\ 
                   C/2016N4(MASTER) & 20160801.558990 &  19.5 &   4.958 &   4.258 &   9.200 &   3.199 &   1.000 &  72.560 \\ 
                C/2016Q2(PANSTARRS) & 20160826.341270 &  21.6 &  12.592 &  11.604 &   1.000 &   7.206 &   0.980 & 109.652 \\ 
                C/2016R2(PANSTARRS) & 20160830.625230 &  18.9 &   6.370 &   6.051 &   8.900 &   2.602 &   0.997 &  58.216 \\ 
                  C/2016T1(Matheny) & 20161008.257310 &  20.0 &   2.638 &   2.472 &  22.200 &   2.298 &   0.980 & 129.820 \\ 
                C/2016T3(PANSTARRS) & 20161010.380040 &  20.1 &   4.229 &   3.349 &   7.200 &   2.650 &   0.981 &  22.674 \\ 
                 C/2010U3(Boattini) & 20161017.532420 &  19.4 &   9.912 &   9.379 &   5.000 &   8.446 &   1.000 &  55.511 \\ 
                   C/2016X1(Lemmon) & 20161126.505860 &  21.4 &   9.305 &   8.326 &   0.800 &   7.601 &   0.987 &  26.526 \\ 
                C/2017A1(PANSTARRS) & 20170102.333005 &  19.6 &   2.730 &   2.015 &  16.500 &   2.291 &   0.971 &  49.794 \\ 
\end{longtable}
}  
}  

\end{document}